\documentclass[12pt]{article}
\usepackage{amssymb,amsmath,latexsym}
\usepackage[english]{babel}
\usepackage[T1]{fontenc}
\usepackage{color,hyperref}
\newtheorem{thm}{Theorem}
\newtheorem{prop}{Proposition}

\newtheorem{lem}{Lemma}

\newtheorem{ex}{Example}
\numberwithin{ex}{section}
\numberwithin{rem}{section}
\numberwithin{equation}{section}
\numberwithin{thm}{section}
\numberwithin{lem}{section}
\numberwithin{coro}{section}

\def\1g{1\hskip -3pt \mbox{l}}

\small\normalsize

\title{Corrected portmanteau tests for VAR models with time-varying variance}

\author{
{\sc Valentin Patilea\footnote{CREST (Ensai) \& IRMAR (UEB), ENSAI - Campus de Ker-Lann,
Rue Blaise Pascal - BP 37203,
35172 BRUZ cedex, France.
E-mail: valentin.patilea@insa-rennes.fr},
\hspace{1em} and Hamdi Ra\"{i}ssi\footnote{IRMAR-INSA, 20 avenue des buttes de Coësmes, CS 70839, F-35708 Rennes Cedex 7, France. E-mail:
hamdi.raissi@insa-rennes.fr}}}

\begin{document}

\maketitle  \noindent {\em Abstract:} The problem of test of fit for Vector AutoRegressive (VAR) processes
with unconditionally heteroscedastic errors is studied. The
volatility structure is deterministic but time-varying and allows
for changes that are  commonly observed in economic or financial
multivariate series such as breaks or smooth transitions. Our
analysis is based on the residual autocovariances and
autocorrelations obtained from Ordinary Least Squares (OLS),
Generalized Least Squares (GLS) and Adaptive Least Squares (ALS)
estimation of the autoregressive parameters. The OLS residuals are
the standards estimates of the VAR model errors. To build the GLS
residuals we use the GLS estimate of the VAR coefficients to
estimate the model errors that we further standardize by the
time-varying volatility. Hence, the GLS estimates require the
knowledge of the variance structure. The ALS approach is the GLS
approach adapted to the \emph{unknown} time-varying volatility that
is then estimated by kernel smoothing. The properties of the three
types of residual autocovariances and autocorrelations are derived.
In particular it is shown that the ALS and GLS residual
autocorrelations are asymptotically equivalent. It is also found
that the asymptotic distribution of the OLS residual
autocorrelations can be quite different from the standard chi-square
asymptotic distribution obtained in a correctly specified VAR model
with iid innovations. As a consequence the standard portmanteau
tests are unreliable in our framework. The correct critical values
of the standard portmanteau tests based on the OLS residuals are
derived. Moreover, modified portmanteau statistics based on ALS
residual autocorrelations are introduced and their asymptotic
critical values are obtained. The finite sample
properties of the goodness-of-fit tests we consider are investigated
by Monte Carlo experiments. The theoretical results are also
illustrated using a U.S. economic data set.

\vspace*{.7cm}
\noindent

\vspace*{.7cm} \noindent {\em Keywords:} VAR model; Unconditionally heteroscedastic errors; Residual autocorrelations; Portmanteau tests.\\
%\textit{JEL Classification:} C01; C32

\newpage
\section {Introduction}
\label{S1}

In the econometric analysis numerous tools are routinely used in the framework of VAR (Vector AutoRegressive) modeling
of time series variables (see L\"{u}tkepohl (2005) and references therein). Nevertheless
it is well known that these tools are in general noticeably affected by the adjusted
autoregressive order. For instance Thornton and Batten (1985),
Stock and Watson (1989) or Jones (1989)
discussed the importance of well specified VAR model for the test of linear Granger causality in mean. Therefore the checking of goodness-of-fit
of the autoregressive order is commonly performed in applied works before proceeding to the analysis of the dynamics of time series.
%as a first step for the statistical inference of time series.
The dominant tests
for the adequacy of the autoregressive order are the portmanteau tests introduced in the VAR framework by Chitturi (1974) and Hosking (1980).
The properties of the tests based on the residual autocorrelations are well explored in the case of stationary processes (see e.g. Francq, Roy
and Zakoïan (2005) in the univariate case,
Francq and Raïssi (2007) or Boubacar Mainassara (2010) in the multivariate case). Duchesne (2005), Br\"{u}ggemann, L\"{u}tkepohl and Saikkonen (2006) and Raïssi (2010) developed tests for residual autocorrelation in a cointegrated framework with stationary innovations.

However many applied studies pointed out the presence of non stationary volatility in economic time series. For instance Ramey and Vine (2006) found a declining volatility in the U.S. automobile industry. Watson (1999) noted a declining volatility of short-term U.S. interest rates and increasing volatility for long-term U.S. interest rates. Sensier and van Dijk (2004) considered 214 U.S. macroeconomic variables and found that approximately 80\% of these variables have a volatility that changes in time. These findings stimulated an interest on the effects of non-stationary volatility in time series analysis amongst econometricians (see e.g. Kim, Leybourne and Newbold (2002) or Cavaliere and Taylor (2007)).

The present paper is motivated by the need of reliable tools for testing the adequacy of the autoregressive order of VAR models with non stationary volatility. On one hand, we show that in such cases the use of standard procedures for testing the adequacy of the autoregressive order can be
quite misleading. On the other hand, valid portmanteau tests based on Ordinary Least Squares (OLS) and Adaptive Least Squares (ALS) residual autocovariances are proposed for testing the goodness-of-fit tests of non-stationary but stable VAR processes.
More precisely we consider the VAR model of order $p\geq 0$ and dimension $d\geq 1$
\begin{eqnarray}\label{VAR}
&&{X}_t={A}_{01}{X}_{t-1}+\dots+{A}_{0p}{X}_{t-p}+u_t,\\&&
u_t=H_t\epsilon_t,\qquad t=1,2, ... \nonumber
\end{eqnarray}
where $X_t$ are random vectors of dimension $d$ and the $d\times d-$matrices $ {A}_{0i}$, $i\in\{1,\dots,p\}$, are such that the process $(X_t)$ is stable, that means $\det(A(z))\neq 0$ for all $|z|\leq 1$, with ${A}(z)=I_d-\sum_{i=1}^{p}{A}_{0i} z^i$. Here,  $H_t$ is an \emph{unknown} $d\times d$ matrix-valued deterministic function of time and  $(\epsilon_t)$ is an innovation process of unit variance that could be serially dependent. Phillips and Xu (2005), Xu and Phillips (2008) already studied the problem of estimation of such univariate stable autoregressive processes. Patilea and Raïssi (2010) investigated the estimation and the test of parameter restrictions of multivariate stable autoregressive
processes like in (\ref{VAR}).

The usual way to check the adequacy of a stable VAR(p) model, implemented in any specialized software, is to assume that the error term $u_t$ is second order stationary, to fix an integer $m>0$ and to test
\begin{equation}\label{hyp0}
\mathcal{H}_0: \,\,\mbox{Cov}(u_t,u_{t-h})=0, \,\, \text{for all } \,0 < h \leq m,
\end{equation}
using a classical (Box-Pierce or Ljung-Box) portmanteau test statistic and chi-square type critical values. The errors $u_t$ are approximated using the OLS type estimates of the coefficients $A_{0i}$. With the volatility structure we assumed in model (\ref{VAR}), the variance of $u_t$ depends on $t$ and the usual chi-square type critical values are in general inaccurate, that is the asymptotic distribution of the classical portmanteau statistics under $\mathcal{H}_0$ is no longer of chi-square type.

Here we propose two ways to correct this problem. First, we derive the correct asymptotic distribution of the classical portmanteau test statistics under $\mathcal{H}_0$ and the conditions of model (\ref{VAR}). This asymptotic distribution is a weighted sum of $d^2m$ independent chi-square distributions. Next, we indicate how the correct critical values can be approximated.

To explain our second approach, let us notice that  $\mbox{Cov}(u_t,u_{t-h})=0$ is equivalent to $\mbox{Cov}(\epsilon_t,  \epsilon_{t-h})=0$ and the variance of $\epsilon_t$ does not depend on the time $t$. Thus an alternative idea for checking the adequacy of a model like (\ref{VAR}) is to test
\begin{equation}\label{hyp0bis}
\mathcal{H}_0^\prime : \,\,\mbox{Cov}(\epsilon_t,\epsilon_{t-h})=0, \,\, \text{for all } \,0 < h \leq m.
\end{equation}
The values $\epsilon_t$ are approximated by residuals built using a nonparametric estimate of the deterministic function $H_t$ and Adaptive Least Squares (ALS) type estimates of the coefficients $A_{0i}$ that take into account the volatility structure. More precisely, to build the ALS residual vector at time $t$ we use the ALS estimate of the VAR coefficients to estimate the VAR model error vector at time $t$ that we further standardize by the nonparametric estimate of time-varying volatility $H_t$. Next, we build classical portmanteau test statistics using the estimates of $\epsilon_t$ and we derive the asymptotic distribution under $\mathcal{H}_0$. The asymptotic distribution is again a weighted sum of $d^2m$ independent chi-square distributions and the weights can be easily estimated from the data. In some important particular cases, including the univariate (i.e. $d=1$) autoregressive models,
we retrieve the standard chi-squared
%these weights are all equal to 1 and thus the (pas le cas pour $p>0$)!!
asymptotic distribution.
%becomes a chi-square.

The remainder of the paper is organized as follows. In section \ref{S2} we specify the framework of our study and state the
asymptotic behavior of the OLS and the Generalized Least Squares (GLS) estimators of the VAR coefficients. The asymptotic
normality of the OLS and the infeasible GLS residual autocovariances and autocorrelations is established in section \ref{S3}.
The GLS residuals are defined as the standardized (by the true volatility $H_t$) estimates of the model error vector obtained with the GLS estimates of the VAR coefficients. In section \ref{S4} we highlight the unreliability of the chi-square type critical values for standard portmanteau statistics and we derive their correct critical values in our framework. In section \ref{S5} the ALS residual autocovariances and autocorrelations are introduced.
Since the GLS residual autocovariances and autocorrelations are infeasible, we investigate the relationship between the GLS and ALS residual autocovariances and autocorrelations and we show that, in some sense, they are asymptotically equivalent. This result is used to introduce portmanteau tests based on the ALS residuals that have the same critical values like those based on the infeasible GLS residuals. In section \ref{modifstats} we propose suitably modified quadratic forms of OLS and ALS residual autocovariances in order to obtain alternative test statistics with chi-square asymptotic distributions under the null hypothesis. Such modified statistics are nothing but Wald type test statistics for testing the nullity of a vector of autocovariances. In section \ref{S5b} some theoretical comparisons of the asymptotic power, in the Bahadur sense, are carried out:  classical Box-Pierce portmanteau test vs. modified quadratic forms of OLS residual autocorrelations based test; and
ALS vs. OLS residual autocorrelations based portmanteau tests. A possible extension of our findings on testing the order of a VAR model to the case of heteroscedastic co-integrated variables is briefly described in section \ref{Scointegration}. The finite sample properties of the
different tests considered in this paper are studied by mean of Monte Carlo experiments in section \ref{S6}. In section \ref{S7} applications
to U.S. economic real data sets are used to illustrate the theoretical results:
the U.S. balance on services and balance on merchandise trade data, and the U.S. energy-transport consumer price indexes.
The summary of our finding and some concluding remarks are given in section \ref{S8}. The proofs and the tables and figures are
relegated in the appendices.

\section {Parameters estimation}
\label{S2}

In  the following weak convergence
is denoted by $\Rightarrow$ while $\stackrel{P}{\rightarrow}$ stands for convergence in probability. The symbol $\otimes$ denotes the usual Kronecker product for matrices and $A^{\otimes 2}$ stands for $A\otimes A$. The symbol $\mbox{vec}(\cdot)$ is used for the column vectorization operator. We denote by $[a]$ the integer part of a real number $a$. For a squared matrix $A$, $\mbox{tr}(A)$ denotes the trace.
For a random variable $x$ we define $\parallel x\parallel_r=(E\parallel x\parallel^r)^{1/r}$,
where $\parallel x\parallel$ denotes the Euclidean norm.  We also
define the $\sigma-$field $\mathcal{F}_{t}=\sigma(\epsilon_s:s\leq t)$. The following conditions on the innovations process $(u_t)$ are assumed to hold.

\quad

\textbf{Assumption A1:}\:(i) The $d\times d$ matrices $H_{t}$  are  positive definite and
the components $\{g_{ij}(r): 1\leq k,l \leq d\}$ of the matrix $G(r)$
are measurable deterministic functions on the interval $(0,1]$, such that $H_{t}=G(t/T)$ and,
$\forall \,1\leq k,l\leq d$, $ \sup_{r\in(0,1]}|g_{k,l}(r)|<\infty$ and $g_{k,l}(\cdot)$ satisfies a Lipschitz condition piecewise on a finite number of some sub-intervals that partition $(0,1]$ (the partition may depend on $k,l$).
The matrix $\Sigma(r)=G(r)G(r)'$ is assumed positive definite for all $r$.\\
(ii) The process $(\epsilon_t)$ is $\alpha$-mixing and such that
$E(\epsilon_t\mid \mathcal{F}_{t-1})=0$,
$E(\epsilon_t\epsilon_t'\mid \mathcal{F}_{t-1})=I_d$ and
$\sup_t\parallel\epsilon_{it}\parallel_{4\mu}<\infty$ for some
$\mu>1$ and all $i\in\{1,\dots,d\}$.\\

The second approach we propose for checking the adequacy of a VAR(p) model requires the estimation of the innovations $\epsilon_t$, and hence we will need an identification condition for $G(r)$ and an estimate of the matrix $H_t$. The condition $H_{t}$  is positive definite matrix identifies $G(r)$ as the square root of $\Sigma(r)$ and this is a convenient choice for the mathematical proofs. Nevertheless one can notice from the following that our results could be stated using alternative conditions, like for instance  $H_t$ is a lower triangular matrix with diagonal components restricted to be positive. The conditions on the unknown volatility function $G(r)$ are general
and allow for a large set of dynamics for the innovation variance as
for instance abrupt shifts or piecewise affine functions. This
assumption generalizes to a multivariate framework the specification of the innovation variance
considered in Xu and Phillips (2008). The conditional homoscedasticity of $(\epsilon_t)$ imposed in (ii) ensures the identifiability of $\Sigma(r)$. We call a model like in (\ref{VAR}) with the innovation process  $(u_t)$ satisfying Assumption \textbf{A1} a stable VAR(p) model with time-varying variance.

To introduce the OLS and GLS estimators of the autoregressive parameters, set the observations $X_{-p+1},\dots,X_0$ equal to the null vector of $\mathbb{R}^d$ (or any other initial values) and
denote by $\theta_0 = (\mbox{vec}\:(A_{01})' \dots \mbox{vec}\:(A_{0p})')'\in \mathbb{R}^{pd^2}$ the vector of true parameters. The equation (\ref{VAR}) becomes
\begin{eqnarray}\label{VARvec}
&& X_t=(\tilde{X}_{t-1}'\otimes I_d)\theta_0+u_t,\quad t=1,2,\dots \\&&
u_t=H_t\epsilon_t,\nonumber
\end{eqnarray}
with $\tilde{X}_{t-1}=(X_{t-1}^\prime,\dots,X_{t-p}^\prime)^\prime$. Then  the OLS estimator is
\begin{equation*}
\hat{\theta}_{OLS}=\hat{\Sigma}_{\tilde{X}}^{-1}\mbox{vec}\:\left(\hat{\Sigma}_{X}\right),
\end{equation*}
where
$$\hat{\Sigma}_{\tilde{X}}=T^{-1}\sum_{t=1}^T\tilde{X}_{t-1}\tilde{X}_{t-1}'\otimes
I_d\quad\mbox{and}\quad\hat{\Sigma}_X=T^{-1}
\sum_{t=1}^TX_t\tilde{X}_{t-1}'.$$
%Let us define $\Sigma_t:=H_tH_t'$.

Multiplying by  $H_t^{-1}$ on the left in equation (\ref{VARvec}) we obtain
\begin{eqnarray*}%\label{VARvec}
&& H_t^{-1}X_t=H_t^{-1}(\tilde{X}_{t-1}'\otimes
I_d)\theta_0+\epsilon_t,
\end{eqnarray*}
and then the GLS estimator is
\begin{equation}\label{GLS}
\hat{\theta}_{GLS}=\hat{\Sigma}_{\tilde{\underline{X}}}
^{-1}\mbox{vec}\:\left(\hat{\Sigma}_{\underline{X}}\right),
\end{equation}
with
$$\hat{\Sigma}_{\tilde{\underline{X}}}=T^{-1}\sum_{t=1}^T\tilde{X}_{t-1}\tilde{X}_{t-1}'
\otimes\Sigma_t^{-1}, \quad\hat{\Sigma}_{\underline{X}}=T^{-1}
\sum_{t=1}^T\Sigma_t^{-1}X_t\tilde{X}_{t-1}'.$$
In general, the GLS estimator is infeasible since it involves the true volatility matrix.

Due to the stability condition, we can write
$X_t=\sum_{i=0}^{\infty}\psi_{i}u_{t-i},$
where $\psi_{0}=I_d$ and the components of the $\psi_i$'s are
absolutely summable $d\times d-$matrices. Then \label{pg5}
\begin{equation*}\label{MA2}
\tilde{X}_t=\sum_{i=0}^{\infty}\tilde{\psi}_{i}u_{t-i}^p,
\end{equation*}
where $u_t^p$ is given by $u_t^p=\mathbf{1}_p\otimes u_t$,
$\mathbf{1}_p$ is the vector of ones of dimension $p$, and
\[
\tilde{\psi}_{i}= diag\{ \psi_i, \psi_{i-1}, \dots, \psi_{i-p+1} \},
\]
taking $\psi_j=0$ for $j<0$.

Let $\mathbf{1}_{p\times p}$ stand for the $p\times p-$matrix with all components equal to one.
Patilea and Raïssi (2010) proved that under {\bf A1}
\begin{equation}\label{res1}
T^{\frac{1}{2}}(\hat{\theta}_{GLS}-\theta_0)\Rightarrow
\mathcal{N}(0,\Lambda_1^{-1}),
\end{equation}
where $$\Lambda_1=\int_0^1
\sum_{i=0}^{\infty}\left\{\tilde{\psi}_i(\mathbf{1}_{p\times
p}\otimes\Sigma(r))\tilde{\psi}_i'\right\}\otimes\Sigma(r)^{-1}dr,$$
and
\begin{equation}\label{res2}
T^{\frac{1}{2}}(\hat{\theta}_{OLS}-\theta_0)\Rightarrow
\mathcal{N}(0,\Lambda_3^{-1}\Lambda_2\Lambda_3^{-1}),
\end{equation}
with $$\Lambda_2=\int_0^1
\sum_{i=0}^{\infty}\left\{\tilde{\psi}_i(\mathbf{1}_{p\times
p}\otimes\Sigma(r))\tilde{\psi}_i'\right\}\otimes\Sigma(r)dr,$$
$$\Lambda_3=\int_0^1
\sum_{i=0}^{\infty}\left\{\tilde{\psi}_i(\mathbf{1}_{p\times
p}\otimes\Sigma(r))\tilde{\psi}_i'\right\}dr\otimes I_d.$$
Moreover, they showed that $\Lambda_3^{-1}\Lambda_2\Lambda_3^{-1}-\Lambda_1^{-1}$ is positive semi-definite.

%%%%%%%% I took off the part on Reinsel that can be found on 15mars version

\section {Asymptotic behavior of the residual autocovariances}
\label{S3}

Let us define the OLS-based estimates of $u_t$ and the GLS-based
estimates of $\epsilon_t$
$$\hat{u}_t=X_t-(\tilde{X}_{t-1}'\otimes
I_d)\hat{\theta}_{OLS}\quad\text{and}\quad
\hat{\epsilon}_t=H_t^{-1}X_t-H_t^{-1}(\tilde{X}_{t-1}'\otimes
I_d)\hat{\theta}_{GLS}.$$ The corresponding residual autocovariances
are defined as
\[\hat{\Gamma}_{OLS}^u(h)=T^{-1}\sum_{t=h+1}^T\hat{u}_t\hat{u}_{t-h}'
\quad\mbox{and}\quad\hat{\Gamma}_{GLS}^\epsilon(h)=T^{-1}\sum_{t=h+1}^T\hat{\epsilon}_t\hat{\epsilon}_{t-h}'.\]
In general the estimated residuals $\hat{\epsilon}_t$ as well as the
autocovariances $\hat{\Gamma}_{GLS}^\epsilon(h)$ are not computable
since they depend on the unknown matrices $H_t$ and the infeasible
estimator $\hat{\theta}_{GLS}$.

For any fixed integer $m\geq 1 $, the estimates of the first $m$ residual autocovariances are defined by
$$\hat{\gamma}_m^{u, OLS}
\! = \! \mbox{vec}\!\left\{\! \left( \hat{\Gamma}^u_{OLS}(1),\dots,
\hat{\Gamma}^u_{OLS}(m)\right)\! \right\}, \,\,\, \,
\hat{\gamma}_m^{\epsilon, GLS}\!=\!\mbox{vec}\!\left\{\!\left(
\hat{\Gamma}^\epsilon_{GLS}(1),\dots,
\hat{\Gamma}^\epsilon_{GLS}(m)\right)\!\right\}.$$ To state the
asymptotic behavior of $\hat{\gamma}_m^{u,OLS}$ and
$\hat{\gamma}_m^{\epsilon, GLS}$   let us define\label{pg6}
$$K=\left(
            \begin{array}{cccc}
              A_{01} & \dots & A_{0p-1} & A_{0p} \\
              I_d & 0 & \dots & 0 \\
                & \ddots & \ddots & \vdots \\
              0 &  & I_d & 0 \\
            \end{array}
          \right).$$
Note that if $\tilde{u}_t=(u_t',0\dots,0)'$,  $\tilde{X}_t=K\tilde{X}_{t-1}+\tilde{u}_t.$ Now, let $\Sigma_{G}=\int_0^1\Sigma(r)dr$, $\Sigma_{G^{\otimes2}}=\int_0^1\Sigma(r)^{\otimes2}dr$ and
\begin{equation}\label{estim0}
\Phi ^u_ {m}=\sum_{i=0}^{m-1}\left\{e_m(i+1)e_p(1)'\otimes
\Sigma_{G}\otimes I_d\right\}  \left\{ K^{i \,\prime} \otimes I_d\right\},
\end{equation}
\begin{equation}\label{estim1}
\Lambda^{u,\theta}_{m}=\sum_{i=0}^{m-1}\left\{e_m(i+1)e_p(1)'\otimes
\Sigma_{G^{\otimes2}}\right\}\left\{K^{i\,\prime}\otimes I_d\right\},
\end{equation}
\begin{equation}\label{estim12}
\Lambda^{\epsilon,\theta}_{m}= \sum_{i=0}^{m-1}\left\{e_m(i+1)e_p(1)'\otimes
\int_0^1
%\Sigma(r)^{1/2}
G(r)^\prime
\otimes
%\Sigma(r)^{- 1/2}
G(r)^{-1}
dr\right\}\left\{K^{i\,\prime} \otimes
I_d\right\},
\end{equation}
\begin{equation}\label{estim4}
\Lambda^{u,u}_m=I_m\otimes\Sigma_{G^{\otimes2}},
\end{equation}
where $e_m(j)$ is the vector of dimension $m$ such that the $j$th component is
equal to one and zero elsewhere.\footnote{Recall that our identification condition for $H_t$ implies
$G(r)=\Sigma(r)^{1/2}$.}

\begin{prop}\label{propostun}  If model (\ref{VAR}) is correct and Assumption {\bf A1} holds true, we have
\begin{equation}\label{gamols}
T^{\frac{1}{2}}\hat{\gamma}_m^{u,OLS}\Rightarrow\mathcal{N}(0,\Sigma^{u,OLS}),
\end{equation}
where
\begin{equation}\label{olsmat}\Sigma^{u,OLS}=\Lambda^{u,u}_m
- \Lambda^{u,\theta}_m\Lambda_3^{-1}\Phi_m^{u\, \prime}-\Phi^u_m\Lambda_3^{-1}\Lambda^{u,\theta\, \prime}_{m }
+\Phi_m^u \Lambda_3^{-1} \Lambda_2 \Lambda_3^{-1} \Phi_m^{u\,\prime},
\end{equation}
\begin{equation}\label{gamgls}
T^{\frac{1}{2}}\hat{\gamma}_m^{\epsilon,
GLS}\Rightarrow\mathcal{N}(0,\Sigma ^{\epsilon,GLS}),
\end{equation}
where
\begin{equation}\label{glsmat}\Sigma^{\epsilon,GLS}=I_{d^2m}
-\Lambda^{\epsilon,\theta}_m\Lambda_1^{-1}\Lambda^{\epsilon,\theta\,^\prime}_{m}.
\end{equation}
In the particular case $p=0$, $\Sigma^{u,OLS}=\Lambda^{u,u}_m$ and  $\Sigma^{\epsilon,GLS}=I_{d^2m}$.
\end{prop}

\quad

Let us discuss the conclusions of Proposition \ref{propostun} in some particular situations.
In the case where $\Sigma(\cdot) = \sigma^2(\cdot) I_d$ for some  positive scalar function $\sigma(\cdot)$, we have
\begin{equation*}\label{eqsimp}
\Lambda^{\epsilon,\theta}_{m}=\! \sum_{i=0}^{m-1}\!\left\{e_m(i+1)e_p(1)'\otimes I_d\otimes I_d
\right\}\left\{K^{i\,\prime} \otimes
I_d\right\}\!,\quad\Lambda_1\! =
\! \sum_{i=0}^{\infty} \left\{ \tilde{\psi}_i(\mathbf{1}_{p\times
p}\otimes I_d)\tilde{\psi}_i' \right\}\otimes I_d,
\end{equation*}
so that in this case the asymptotic distribution of the $\epsilon_t$
autocovariances estimates $\hat{\gamma}_m^{\epsilon, GLS}$ do not
depend on the volatility function $\Sigma(\cdot)$. Meanwhile, the
(asymptotic) covariance matrix $\Sigma^{u,OLS}$ still depends on the
volatility function.

If we suppose that $(u_t)$ have a time-constant variance $\Sigma(r) \equiv \Sigma_u$,
we obtain
$$\Lambda_1=\!E\!\left[\tilde{X}_t\tilde{X}_t'\right]\otimes\Sigma_u^{-1},\:\:
\Lambda^{\epsilon,\theta}_m=\!E\!\left[\epsilon_{t}^m\tilde{X}_t'\right]\!\otimes G_u^{-1},\:
\Lambda^{u,u}_m=I_m\otimes\Sigma_u^{\otimes2},\:\:
\Lambda_3=\!E\!\left[\tilde{X}_t\tilde{X}_t'\right]\!\otimes I_d,$$  where $\Sigma_u = G_u G_u^\prime$, and
$$\Lambda^{u,\theta}_m=E\left[u_{t}^m\tilde{X}_t'\right]\!\otimes\Sigma_u,
\:\:\Lambda_2=E\left[\tilde{X}_t\tilde{X}_t'\right]\!\otimes\Sigma_u,
\:\:\Phi^u_m=E\left[u_{t}^m\tilde{X}_t'\right]\otimes I_d,$$ where
$u_t^m\!=(u_t',\dots,u_{t-m}')'$ and
$\epsilon_t^m\!=(\epsilon_t',\dots,\epsilon_{t-m}')'$. By
straightforward computations
\begin{equation}\label{bruges}
\Sigma^{u,OLS}=I_m\otimes\Sigma_u^{\otimes2}-E\left[u_{t}^m\tilde{X}_t'\right]
E\left[\tilde{X}_t\tilde{X}_t'\right]^{-1}
E\left[u_{t}^m\tilde{X}_t'\right]'\otimes\Sigma_u,
\end{equation}
\begin{equation}\label{gand}
\Sigma^{\epsilon,GLS}=I_{d^2m}-E\left[\epsilon_{t}^m\tilde{X}_t'\right]
E\left[\tilde{X}_t\tilde{X}_t'\right]^{-1}
E\left[\epsilon_{t}^m\tilde{X}_t'\right]'\otimes I_d.
\end{equation}
Formula (\ref{bruges}) (resp. (\ref{gand})) corresponds to the
(asymptotic) covariance matrix obtained in the standard case with an
i.i.d. error process of variance $\Sigma_u$ (resp. $I_d$), see
L\"{u}tkepohl (2005), Proposition 4.5. Herein, some dependence of
the error process is allowed.  In particular, equation (\ref{gand})
indicates that the homoscedastic (time-constant variance) case is
another situation where $\Sigma^{\epsilon,GLS}$ does not depend on
error process variance $\Sigma_u$.

Proposition \ref{propostun} shows that in general VAR models with
time-varying variance the  covariance matrix $\Sigma^{\epsilon,GLS}$
 depends on  $\Sigma(\cdot)$. For the sake of
simpler notation, hereafter we write $\hat \Gamma_{OLS}(h)$ (resp.
$\hat \gamma^{OLS}_m$) (resp. $\Sigma^{OLS}$) instead of $\hat
\Gamma_{OLS}^u(h)$ (resp. $\hat \gamma^{u,OLS}_m$) (resp.
$\Sigma^{u,OLS}$). Similar notation simplification will be applied
for $\hat \Gamma_{\epsilon, GLS}(h)$, $\hat \gamma^{\epsilon,
GLS}_m$ and $\Sigma^{\epsilon,GLS}$.

The following example  shows that when the error process is
heteroscedastic, the covariance matrices $\Sigma^{OLS}$ and
$\Sigma^{GLS}$  can be quite different and far from the covariance
matrices obtained in the stationary case.

\begin{figure}[h]\!\!\!\!\!\!\!\!\!\!
\vspace*{4.3cm} \hspace*{8.5cm}$\tau_1$
\hspace*{11.7cm}$\sigma_{21}$ \vspace*{-5.3 cm}

\vspace*{4.3cm} \hspace*{2.15cm}$\tau_1$ \hspace*{2.3cm}$\sigma_{21}$ %\hspace*{3.8cm}$\tau_1$
\vspace*{0.2 cm}

\protect \includegraphics{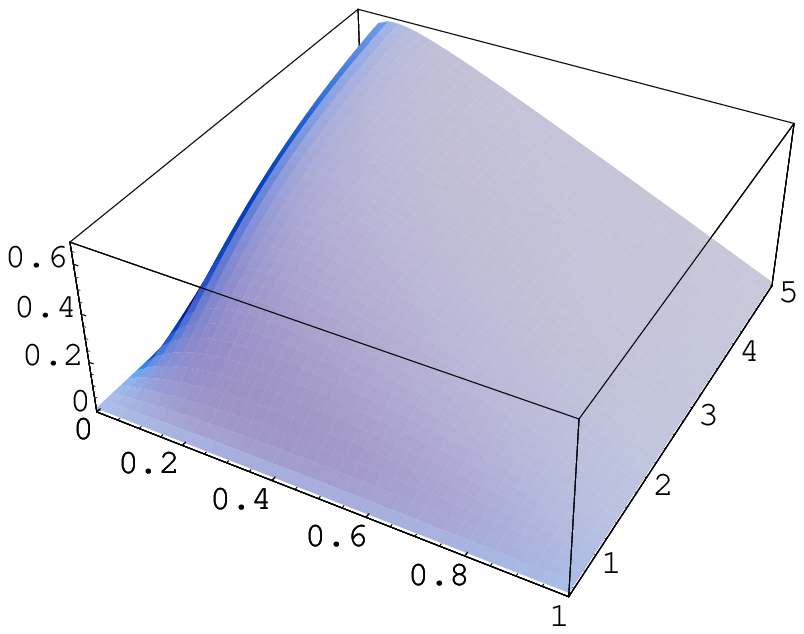} \protect \includegraphics{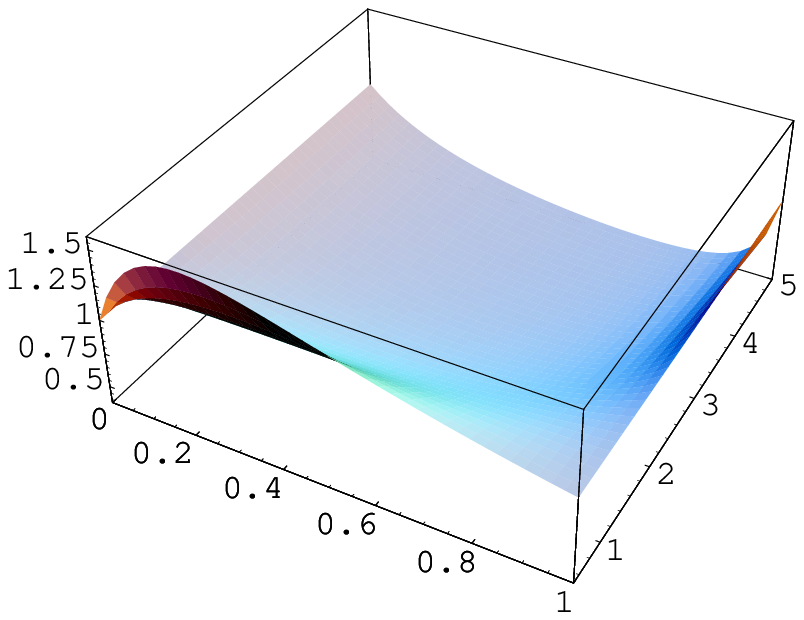} \caption{\label{fig1}
{\footnotesize The asymptotic variance $\Sigma_{GLS}(2,2)$ on the left and the ratio $\Sigma_{OLS}(6,6)/\Sigma_{OLS}^S(6,6)$ on the right for $\tau_1=\tau_2$ in Example \ref{ex1}.}}
\end{figure}

\begin{figure}[h]\!\!\!\!\!\!\!\!\!\!
\vspace*{4.3cm} \hspace*{8.5cm}$\tau_1$
\hspace*{11.7cm}$\tau_2$ \vspace*{-5.9 cm}

\vspace*{5.3cm} \hspace*{2.15cm}$\tau_1$ \hspace*{2.5cm}$\tau_2$ \vspace*{0.2 cm}

\protect \includegraphics{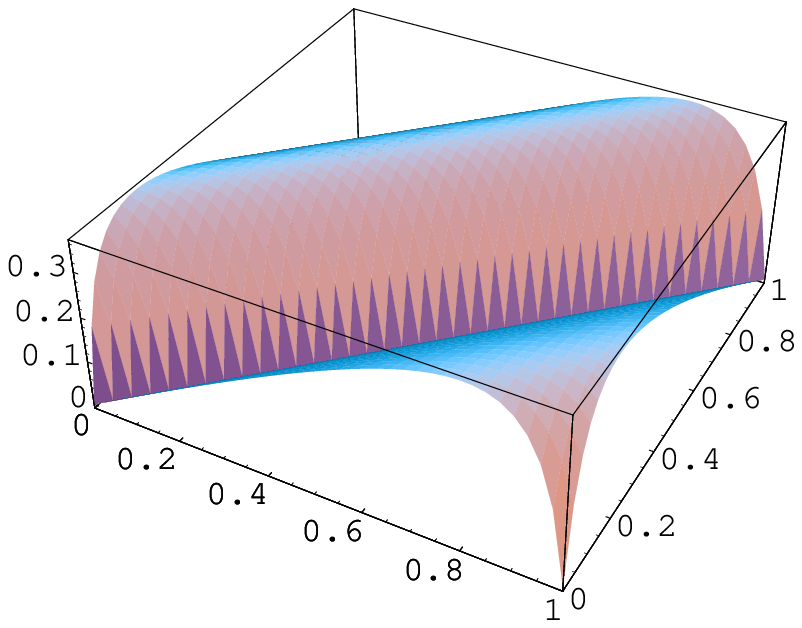} \protect \includegraphics{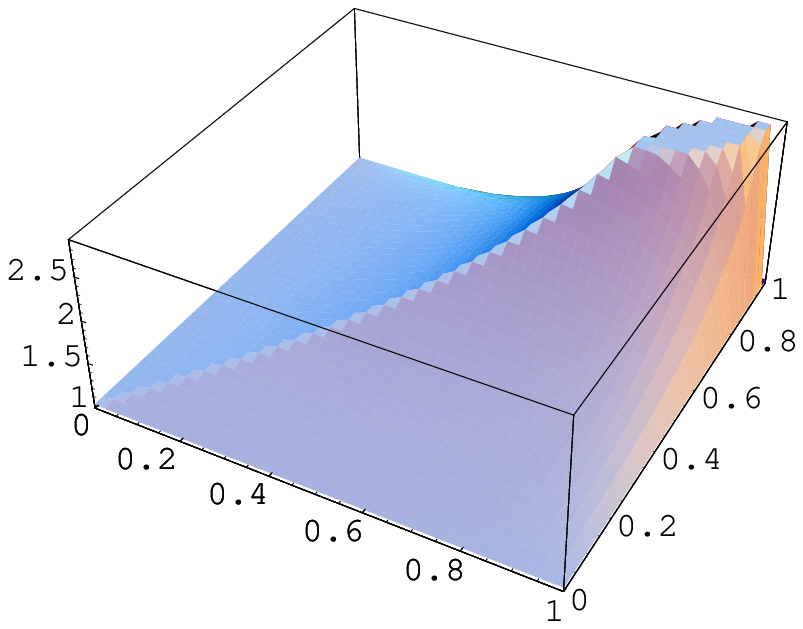} \caption{\label{fig2}
{\footnotesize The same as in Figure \ref{fig1} but for $\tau_1\neq\tau_2$ in general.}}
\end{figure}

\begin{ex}\label{ex1}
{\em Consider a bivariate $AR(1)$ model $X_t=A_0X_{t-1} + u_t$ with
true parameter $A_0$ equal to the zero $2\times2-$matrix.  One can
use such a model to study linear Granger causality in mean between
uncorrelated variables.  However in practice one has first to check
that the error process is a white noise. If we assume for simplicity
that
$$\Sigma(r)=\left(
             \begin{array}{cc}
               \Sigma_1(r) & 0 \\
               0 & \Sigma_2(r) \\
             \end{array}
           \right),$$
we obtain diagonal covariance matrices
$\Sigma^{OLS}=diag\{0_{4\times4},I_{m-1}\otimes\breve{\Sigma}^{OLS}\}$
and $\Sigma^{GLS}=diag\{\breve{\Sigma}^{GLS},I_{4(m-1)}\}$,
with
$${\breve{\Sigma}^{OLS}=\left(
           \begin{array}{cccc}
             \int_0^1\Sigma_1(r)^2 dr & 0 & 0 & 0 \\
             0 & \int_0^1\Sigma_1(r)\Sigma_2(r) dr & 0 & 0 \\
             0 & 0 & \int_0^1\Sigma_1(r)\Sigma_2(r) dr & 0 \\
             0 & 0 & 0 & \int_0^1\Sigma_2(r)^2 dr \\
           \end{array}
         \right)}$$
and
$${\breve{\Sigma}^{GLS}=\left(
                 \begin{array}{ccccc}
                   0 & 0 & 0 & 0  \\
                   0 & 1-\frac{(\int_0^1\Sigma_1(r)^{\frac{1}{2}}\Sigma_2(r)^{-\frac{1}{2}} dr)^2}{\int_0^1\Sigma_1(r)\Sigma_2(r)^{-1} dr} & 0 & 0  \\
                    0& 0 & 1-\frac{(\int_0^1\Sigma_1(r)^{-\frac{1}{2}}\Sigma_2(r)^{\frac{1}{2}} dr)^2}{\int_0^1\Sigma_2(r)\Sigma_1(r)^{-1} dr} & 0  \\
                   0 & 0 & 0 & 0  \\
                 \end{array}
               \right)}.
$$
We denote by $0_{q\times q}$ the null matrix of dimension $q\times q$. Note that the matrix $I_{4(m-1)}$ which appears in the expression
of $\Sigma^{GLS}$ is a consequence of the assumption $A_0=0_{2\times 2}$. If we
suppose that the errors are homoscedastic, that is $\Sigma(r)$ is constant, equal to some $\Sigma_u$, we obtain
$\Sigma^{OLS}=diag\{0_{4\times4},I_{m-1}\otimes\Sigma_u^{\otimes2}\}$
and $\Sigma^{GLS}=diag\{0_{4\times4},I_{4(m-1)}\}$. Therefore in the OLS approach and if the innovations variance is spuriously assumed constant, the asymptotic spurious covariance matrix $\Sigma_S^{OLS}=diag\{0_{4\times4},I_{m-1}\otimes\Sigma_{u,S}^{\otimes2}\}$ is used with $$\Sigma_{u,S}=\left(
             \begin{array}{cc}
               \int_0^1\Sigma_1(r)dr & 0 \\
               0 & \int_0^1\Sigma_2(r)dr \\
             \end{array}
           \right).
$$
Now we illustrate the difference between the covariance matrices obtained if we take into account the unconditional heteroscedasticity of the process and the case where the process is spuriously supposed homoscedastic. We take
\begin{equation}\label{evident}
\Sigma_1(r)=\sigma_{10}^2+(\sigma_{11}^2-\sigma_{10}^2)\times
\mathbf{1}_{\{r\geq\tau_1\}}(r)
\end{equation}
and
\begin{equation}\label{evident_b}
\Sigma_2(r)=\sigma_{20}^2+(\sigma_{21}^2-\sigma_{20}^2)\times \mathbf{1}_{\{r\geq\tau_2\}}(r),
\end{equation}
where $\tau_i\in[0,1]$ with $i\in\{1,2\}$. This specification
of the volatility function is inspired by Example 1 of Xu and
Phillips (2008) (see also Cavaliere (2004)). In Figure \ref{fig1},
we take $\tau_1=\tau_2$, $\sigma_{10}^2=\sigma_{20}^2=1$ and
$\sigma_{11}^2=0.5$, so that only the break dates and $\sigma_{21}^2$
vary freely. In figure \ref{fig2} only the break dates vary with
$\tau_1\neq\tau_2$ in general, and $\sigma_{10}^2=\sigma_{20}^2=1$,
$\sigma_{11}^2=\sigma_{21}^2=4$. In Figure \ref{fig1} and \ref{fig2}
we plot in the left graphics the second component
$\Sigma^{GLS}(2,2)$ on the diagonal of $\Sigma^{GLS}$ and in the
right graphics the ratio of $\Sigma^{OLS}(6,6)/\Sigma^{OLS}_S(6,6)$.

From the left graphic of Figure \ref{fig1} it turns out that
$\Sigma^{GLS}(2,2)$ could be far from zero for larger values of
$\sigma_{21}$ and when the breaking point $\tau_1$ is located early in the sample. From the right graphic of Figure \ref{fig1} we can see that the ratio of
$\Sigma^{OLS}(6,6)/\Sigma^{OLS}_S(6,6)$ can be far from 1, however
the relation between this ratio and the variations of $\tau_1$, $\sigma_{21}$ is not clear. From the left graphic of Figure
\ref{fig2}, it appears that $\Sigma^{GLS}(2,2)$ can be far from zero. According to the right graphic of Figure \ref{fig2}, the
relative difference between $\Sigma^{OLS}(6,6)$ and $\Sigma^{OLS}_S(6,6)$ is
significantly larger when the breaking points $\tau_1$ and $\tau_2$ are located at the end of
the sample. This example shows that the
standard results for the analysis of the autocovariances can be
quite misleading when the unconditional homoscedasticity assumption on the
innovations process does not hold.}
\end{ex}

We also consider the vector of residual autocorrelations: for a
given integer $m\geq 1$, define
$$\hat{\rho}_m^{OLS}
=\mbox{vec}\:\left\{\left( \hat{R}_{OLS}(1),\dots,
\hat{R}_{OLS}(m)\right)\right\}\quad\mbox{where}\quad\hat{R}_{OLS}(h)
=\hat{S}_u^{-1}\hat{\Gamma}_{OLS}(h)\hat{S}_u^{-1} $$
with $\hat{S}_u^2=\mbox{Diag}\{\hat{\sigma}_u^2(1),\dots,\hat{\sigma}_u^2(d)\}$,
$\hat{\sigma}_u^2(i)=T^{-1}\sum_{t=1}^T\hat{u}_{it}^2$, and
$$\hat{\rho}_{a,m}^{GLS}
=\mbox{vec}\:\left\{\left( \hat{R}_{GLS}(1),\dots,
\hat{R}_{GLS}(m)\right)\right\}\quad\mbox{where}\quad\hat{R}_{GLS}(h)
=\hat{S}_{\epsilon}^{-1}\hat{\Gamma}_{GLS}(h)\hat{S}_{\epsilon}^{-1},$$
with
$\hat{S}_{\epsilon}^2=\mbox{Diag}\{\hat{\sigma}_{\epsilon}^2(1),\dots,$
$\hat{\sigma}_{\epsilon}^2(d)\}$,
$\hat{\sigma}_{\epsilon}^2(i)=T^{-1}\sum_{t=1}^T\hat{{\epsilon}}_{it}^2$.
Since $\epsilon_t$ has identity variance matrix, we can also define
$$\hat{\rho}_{b,m}^{GLS} = \hat{\gamma}_m^{GLS}.$$

\begin{prop}\label{propostun_b} If model (\ref{VAR}) is correct and Assumption {\bf A1} holds true,
we have
\begin{equation}\label{rhools}
T^{\frac{1}{2}}\hat{\rho}_m^{OLS}\Rightarrow\mathcal{N}(0,\Psi^{OLS}),
\end{equation}
where
$$\Psi^{OLS}=\{I_m\otimes(S_u\otimes S_u)^{-1}\}\Sigma^{OLS}\{I_m\otimes(S_u\otimes S_u)^{-1}\},$$
where $S_u^2=\mbox{Diag}\{\Sigma_{G, 11} ,\dots, \Sigma_{G, dd}\}$. Moreover,
\begin{equation}\label{rhogls}
T^{\frac{1}{2}}\hat{\rho}_{m}^{GLS}\Rightarrow\mathcal{N}(0,\Sigma^{GLS}),
\end{equation}
where  $\hat{\rho}_{m}^{GLS}$ stands for any of $\hat{\rho}_{a,m}^{GLS}$ or $\hat{\rho}_{b,m}^{GLS}$.
\end{prop}

Using Proposition \ref{propostun_b}, $\hat{S}_u$  and a consistent estimator of $\Sigma^{OLS}$ (that can build in a similar way to that of $\Delta_m^{OLS}$, see Section \ref{S4}, p. \pageref{qsqs}), one can easily build a consistent estimate of $\Psi^{OLS}$ and confidence intervals for the OLS residual autocorrelations.

\section{Modified portmanteau tests based on OLS estimation}
\label{S4}
Corrected portmanteau tests based on the OLS residual autocorrelations are proposed below. We use the standard Box-Pierce  statistic, Box and Pierce (1970), introduced in the VAR framework by Chitturi
(1974)
\begin{eqnarray}
Q_m^{OLS}&=&T\sum_{h=1}^m\mbox{tr}\left(\hat{\Gamma}_{OLS}'(h)\hat{\Gamma}_{OLS}^{-1}(0)
\hat{\Gamma}_{OLS}(h)\hat{\Gamma}_{OLS}^{-1}(0)\right)\nonumber\\&=&T\hat{\gamma}^{OLS'}_m\left(I_m\otimes
\hat{\Gamma}_{OLS}^{-1}(0)\otimes
\hat{\Gamma}_{OLS}^{-1}(0)\right)\hat{\gamma}^{OLS}_m.\label{olstatport}
\end{eqnarray}
We also consider the Ljung-Box statistic (Ljung and Box (1978))
introduced in the VAR framework by Hosking (1980)
\begin{eqnarray*}
\tilde{Q}_m^{OLS}&=&T^2\sum_{h=1}^m(T-h)^{-1}\mbox{tr}\left(\hat{\Gamma}_{OLS}'(h)\hat{\Gamma}_{OLS}^{-1}(0)
\hat{\Gamma}_{OLS}(h)\hat{\Gamma}_{OLS}^{-1}(0)\right).
\end{eqnarray*}
The
following result, a direct consequence of Proposition \ref{propostun} equation (\ref{rhools}), provides the asymptotic distribution of
$Q_m^{OLS}$ and $\tilde{Q}_m^{OLS}$.

\begin{thm} \label{th_4_1} If model (\ref{VAR}) is correct and Assumption {\bf A1} holds true, the statistics $Q_m^{OLS}$ and $\tilde{Q}_m^{OLS}$ converge in law to
\begin{equation}\label{truedist}
U(\delta_m^{OLS})=\sum_{i=1}^{d^2m}\delta^{ols}_iU_i^2,
\end{equation}
as $T\to\infty$, where
$\delta_m^{OLS}=(\delta^{ols}_1,\dots,\delta^{ols}_{d^2m})'$ is the vector of
the eigenvalues of the matrix
$$\Delta_m^{OLS}=(I_m\otimes\Sigma_G^{-1/2}\otimes\Sigma_G^{-1/2})
\Sigma^{OLS}(I_m\otimes\Sigma_G^{-1/2}\otimes\Sigma_G^{-1/2}),$$
$\Sigma_{G}=\int_0^1\Sigma(r)dr$ and the $U_i$'s are independent $\mathcal{N}(0,1)$ variables.
\end{thm}
 When the error process is homoscedastic i.i.d. and $m$ is large, it is well known that the asymptotic distribution of the statistics $Q_m^{OLS}$ and $\tilde{Q}_m^{OLS}$ under the null hypothesis $\mathcal{H}_0$  can be approximated by a chi-square law with $d^2(m-p)$ degrees of freedom, see Box and Pierce (1970). In our framework, even for large $m$, the limit distribution in (\ref{truedist}) can be very different from a chi-square law. The following example illustrate this point.

\begin{ex}\label{ex2}
{\em Consider the bivariate process in Example
\ref{ex1}. Then
\begin{equation*}
\Delta_m^{OLS}=diag\{0_{4\times4},I_{m-1}\otimes\breve{\Delta}_{OLS}\}
\end{equation*}
with
$$\breve{\Delta}_{OLS}=\left(
           \begin{array}{cccc}
             \frac{\int_0^1\Sigma_1(r)^2 dr}{(\int_0^1\Sigma_1(r)dr)^2} & 0 & 0 & 0 \\
             0 & \frac{\int_0^1\Sigma_1(r)\Sigma_2(r) dr}{\int_0^1\Sigma_1(r)dr\int_0^1\Sigma_2(r)dr} & 0 & 0 \\
             0 & 0 & \frac{\int_0^1\Sigma_1(r)\Sigma_2(r) dr}{\int_0^1\Sigma_2(r)dr\int_0^1\Sigma_1(r)dr} & 0 \\
             0 & 0 & 0 & \frac{\int_0^1\Sigma_2(r)^2 dr}{(\int_0^1\Sigma_2(r)dr)^2} \\
           \end{array}
         \right).$$
If we suppose that $\Sigma(r)$ is constant and  $A_0=0_{2\times 2}$, we obtain
$\breve{\Delta}_{OLS}=I_4$, so that the asymptotic distribution of
$Q_m^{OLS}$ and $\tilde{Q}_m^{OLS}$ is $\chi^2(d^2(m-p))$ with $p=1$ and $d=2$. However
it is easy to see that the $d^2(m-p)$ non zero diagonal elements in
$\Delta_m^{OLS}$ can be far from one if the error process is
heteroscedastic. From the Jensen inequality the components $\breve{\Delta}_{OLS}(1,1)$ and $\breve{\Delta}_{OLS}(4,4)$ are greater or equal than one.
For illustration, in the right graphics of Figures
\ref{fig1} and \ref{fig2} we present the second diagonal element of
$\breve{\Delta}_{OLS}$ when the volatility function is like in (\ref{evident})-(\ref{evident_b}).}
\end{ex}

Estimates of the weights which appear in (\ref{truedist}) can be
obtained as follows. First, let us recall the following results   proved by Patilea and Raïssi (2010):
\begin{equation}\label{ge}
\hat{\Sigma}_{G^{\otimes2}}:=T^{-1}\sum_{t=2}^T\hat{u}_{t-1}\hat{u}_{t-1}'\otimes\hat{u}_t
\hat{u}_t'=\Sigma_{G^{\otimes2}}+o_p(1),
\end{equation}
\begin{equation}\label{gege}
\hat{\Sigma}_{G}:=T^{-1}\sum_{t=1}^T\hat{u}_t\hat{u}_t'=\Sigma_{G}+o_p(1),
\end{equation}
\begin{equation}\label{lam2}
\hat{\Lambda}_2:=T^{-1}\sum_{t=1}^T\tilde{X}_{t-1}\tilde{X}_{t-1}'\otimes\hat{u}_t\hat{u}_t'
=\Lambda_2+o_p(1),
\end{equation}
and
\begin{equation}\label{lam3}
\hat{\Lambda}_3:=\hat{\Sigma}_{\tilde{X}}=\Lambda_3+o_p(1).
\end{equation}
A consistent estimator of $\Phi^u_{m}$\label{qsqs} and $\Lambda^{u,\theta}_{m}$ given in (\ref{estim0}) and (\ref{estim1}) is easily obtained by
replacing $A_{01},\dots,A_{0p}$ with their OLS estimators in $K$ and using (\ref{ge}) and (\ref{gege}).
Thus from this and the equations (\ref{ge}) to (\ref{lam3}),
one can easily define a consistent estimator of $\Delta_m^{OLS}$. Denote the
estimated eigenvalues of $\Delta_m^{OLS}$ by
$\hat{\delta}_m^{OLS}=(\hat{\delta}_1^{ols},\dots,\hat{\delta}_{d^2m}^{ols})'$.

We are now ready to introduce the OLS residuals-based corrected versions of the Box-Pierce (resp. Ljung-Box) portmanteau tests for testing the order of the VAR model  (\ref{VAR}). With at hand a vector $\hat{\delta}_m^{OLS}$, at the asymptotic level $\alpha$,
the Box-Pierce (resp. Ljung-Box) procedure consists in rejecting the null
hypothesis (\ref{hyp0}) of uncorrelated innovations when
\begin{equation*}\label{pvalue}
P(Q_m^{OLS} > U_{OLS}(\hat{\delta}_m^{OLS})\mid X_1,\dots,X_T)<\alpha
\end{equation*}
(resp.\:
$P(\tilde{Q}_m^{OLS} > U_{OLS}(\hat{\delta}_m^{OLS})\mid X_1,\dots,X_T)<\alpha).$
The $p$-values  can be evaluated using the Imhof
algorithm (Imhof, 1961) or the saddle point method, see e.g. Kuonen
(1999).

Let us end this section with some remarks on the particular case $\Sigma(\cdot) = \sigma^2(\cdot) I_d$ (that includes the univariate AR(p) models with time-varying variance). In this case
\begin{equation}\label{prt_part1}
\Delta_m^{OLS} = \left[ \int_0^1\sigma^2(r)dr \right]^{-2} \Sigma^{OLS} = \left[ \int_0^1\sigma^2(r)dr \right]^{-2} \left[\int_0^1\sigma^4(r)dr \right] \Sigma^{GLS} = : c_\sigma  \Sigma^{GLS},
\end{equation}
and clearly, $c_\sigma \geq 1$. If in addition $p=0$, by Proposition
\ref{propostun} we have $\Sigma^{GLS} = I_{d^2m}$ and hence
$\delta_m^{OLS} = c_\sigma (1,\cdots, 1)^\prime$.

\section{Adaptive portmanteau tests}
\label{S5}

An alternative way to build portmanteau tests for VAR(p) models with time-varying variance we consider herein is to use approximations of the innovation $\epsilon_t$. A nonparametric estimate of the volatility function is needed for building such approximations. For this purpose we generalize the approach of Xu and Phillips (2008) to the multivariate case, see also Patilea and Ra\"{i}ssi (2010). Let us denote by $A\odot B$ the Hadamard  (entrywise) product of two matrices of same dimension $A$ and $B$. Define the \emph{symmetric} matrix
$$\check{\Sigma}_t^0=\sum_{i=1}^T w_{ti} \odot \hat{u}_i\hat{u}_i',$$
where, as before the $\hat{u}_i$'s are the OLS residuals and the $kl-$element, $k\leq l$, of the $d\times d$ matrix of weights $w_{ti}$ is given by
$$w_{ti}(b_{kl})= \left(\sum_{i=1}^TK_{ti}(b_{kl})\right)^{-1} K_{ti}(b_{kl}),$$
with $b_{kl}$ the bandwidth and
$$K_{ti} (b_{kl}) =\left\{
              \begin{array}{c}
                K(\frac{t-i}{Tb_{kl}})\quad \mbox{if}\quad t\neq i,\\
                0  \quad\mbox{if}\quad t=i.\\
              \end{array}
            \right.$$
The kernel function $K(z)$ is bounded nonnegative and such that $\int_{-\infty}^\infty K(z)dz=1$. For all $1\leq k\leq l\leq d$ the bandwidth $b_{kl}$ belongs to a range $\mathcal{B}_T = [c_{min} b_T, c_{max} b_T]$ with $c_{min}, c_{max}>0$ some constants and $b_T \downarrow 0$ at a suitable rate that will be specified below.

When using the same bandwidth $b_{kl}\in \mathcal{B}_T$ for all the
cells of $\check{\Sigma}_t^0$, since $\hat{u}_i$, $i=1,...,T$ are
almost sure linear independent each other, $\check{\Sigma}_t^0$ is
almost sure positive definite provided $T$ is sufficiently large.
When using several bandwidths $b_{kl}$ a regularization of
$\check{\Sigma}_t^0$ could be necessary in order to ensure positive
definiteness. Let us consider
\[
\check{\Sigma}_t = \left\{\left(\check{\Sigma}_t^0\right)^2  + \nu_T I_d\right\}^{1/2}
\]
where $\nu_T >0$, $T\geq 1$, is a sequence of real numbers
decreasing to zero at a suitable rate that will be specified below.
Our simulation experiments indicate that in applications with
moderate and large samples $\nu_T$ could be even set equal to 0.

In practice the bandwidths $b_{kl}$ can be chosen by minimization of
a cross-validation criterion like
$$\sum_{t=1}^T\parallel\check{\Sigma}_t-\hat{u}_t\hat{u}_t'\parallel^2,$$
with respect to all $b_{kl}\in\mathcal{B}_T$, $1\leq k\leq l\leq d$,
where $\parallel \cdot \parallel $ is some norm for a square matrix,
for instance the Frobenius norm that is the square root of the sum
of the squares of matrix elements. Like in Patilea and Raïssi
(2010), the theoretical results below are obtained uniformly with
respect to the bandwidths $b_{kl}\in\mathcal{B}_T$ and this provides
a justification for the common cross-validation bandwidth selection
approach in the framework we consider.

Let us now  introduce the following  adaptive least squares (ALS) estimator
\begin{equation*}%\label{ALS}
\hat{\theta}_{ALS}=\check{\Sigma}_{\tilde{\underline{X}}}
^{-1}\mbox{vec}\:\left(\check{\Sigma}_{\underline{X}}\right),
\end{equation*}
with
$$\check{\Sigma}_{\tilde{\underline{X}}}=T^{-1}\sum_{t=1}^T\tilde{X}_{t-1}\tilde{X}_{t-1}'
\otimes\check{\Sigma}_t^{-1},\quad\mbox{and}\quad\check{\Sigma}_{\underline{X}}=T^{-1}
\sum_{t=1}^T\check{\Sigma}_t^{-1}X_t\tilde{X}_{t-1}'.$$
The ALS
residuals, proxies of the infeasible GLS residuals, are defined as
$\check{\epsilon}_t=\check{H}_t^{-1}X_t-\check{H}_t^{-1}(\tilde{X}_{t-1}'\otimes
I_d)\hat{\theta}_{ALS},$ and the adaptive autocovariances and
autocorrelations
$$\hat{\Gamma}_{ALS}(h)=\hat{\Gamma}_{ALS}^\epsilon (h)=T^{-1}\sum_{t=h+1}^T\check{\epsilon}_t\check{\epsilon}_{t-h}',
\quad\hat{R}_{ALS}(h)=\check{S}_{\epsilon}^{-1}\hat{\Gamma}_{ALS}(h)\check{S}_{\epsilon}^{-1},$$
where
$\check{S}_{\epsilon}=\mbox{Diag}\{\check{\sigma}_{\epsilon}(1),\dots,$
$\check{\sigma}_{\epsilon}(d)\}$,
$\check{\sigma}_{\epsilon}^2(i)=T^{-1}\sum_{t=1}^T\check{{\epsilon}}_{it}^2$,
and $\check{H}_t$ is the nonparametric estimator obtained from
$\check{\Sigma}_t$ and the identification condition on $H_t$ (see Assumption \textbf{A1}(i)), that is $\check{H}_t=\check{\Sigma}_t^{1/2}$.

Let $\hat \gamma_m^{ALS} =
\mbox{vec}\!\{( \hat{\Gamma}_{ALS}(1),\dots,
\hat{\Gamma}_{ALS}(m))\} $. Following the notation of the previous section, for a given integer
$m\geq 1$, define the residual autocorrelations
$$\hat{\rho}_{a,m}^{ALS}
=\mbox{vec}\left\{\:\left( \hat{R}_{ALS}(1),\dots,
\hat{R}_{ALS}(m)\right)\!\right\}\quad\mbox{and}\quad
\hat{\rho}_{b,m}^{ALS} =\hat \gamma_m^{ALS}.
$$
The main result of this section  shows  that $\hat \gamma_m^{ALS}$
and $\hat{\rho}_{a,m}^{ALS}$ are asymptotic equivalent to $\hat
\gamma_m^{GLS}$ and $\hat{\rho}_{a,m}^{GLS}$. This will allow us to
define new portmanteau statistics based on the ALS residuals. For
this purpose, we need the following assumptions.

\vspace{0.3 cm}

\textbf{Assumption A1':} Suppose that all the conditions in Assumption \textbf{A1}(i) hold true.  In addition:

  (i) $\inf_{r\in(0,1]} \lambda_{min}(\Sigma(r)) >0$ where for any symmetric matrix $A$ the real value $ \lambda_{min}(A)$ denotes its smallest eigenvalue.

 (ii) $\sup_{t} \| \epsilon_{kt} \|_8 <\infty$ for all $k\in\{1,...,d  \}$.

\vspace{0.3 cm}

\textbf{Assumption A2:} \, (i) The kernel $K(\cdot)$ is a bounded density function defined on the real line such that $K(\cdot)$ is nondecreasing on $(-\infty, 0]$ and decreasing on $[0,\infty)$ and $\int_\mathbb{R} v^2K(v)dv < \infty$. The function $K(\cdot)$ is differentiable except a finite number of points and the derivative $K^\prime(\cdot)$  is an integrable function.
Moreover, the Fourier Transform $\mathcal{F}[K](\cdot)$ of $K(\cdot)$ satisfies $\int_{\mathbb{R}}  \left| s \mathcal{F}[K](s) \right|ds <\infty$.

(ii) The bandwidths $b_{kl}$, $1\leq k\leq l\leq d$, are taken in the range $\mathcal{B}_T = [c_{min} b_T, c_{max} b_T]$ with $0< c_{min}< c_{max}< \infty$ and $b_T + 1/Tb_T^{2+\gamma} \rightarrow 0$ as $T\rightarrow \infty$, for some $\gamma >0$.

(iii) The sequence $\nu_T$ is such that $T\nu_T^2 \rightarrow 0.$

\vspace{0.3 cm}

Below, we say that a sequence of random matrices $A_T$, $T\geq 1$ is $o_p(1)$ uniformly with respect to (w.r.t.) $b_{kl}\in\mathcal{B}_T$ as $T\rightarrow \infty$ if $\sup_{1\leq k\leq l\leq d} \sup_{b_{kl}\in\mathcal{B}_T} \|\mbox{vec}\:\left(A_T\right)\| \stackrel{P}{\longrightarrow} 0$. The following proposition gives the asymptotic behavior of variances, autocovariances and autocorrelations estimators based on the ALS estimator of $\theta_0$ and the nonparametric estimate of the time-varying variance structure $\Sigma_t$. The results are uniformly w.r.t the bandwidths.

\begin{prop}\label{lemALS}
If model (\ref{VAR}) is correct and Assumptions {\bf A1'} and {\bf
A2} hold, uniformly w.r.t. $b\in\mathcal{B}_T$
\begin{equation}\label{ademain}
T^{-1}\sum_{t=1}^T\check{H}_t^{\prime}\otimes\check{H}_t^{-1}=
\int_0^1 G(r)^{\prime}\otimes G(r)^{-1}dr+o_p(1),
\end{equation}
\begin{equation}\label{sigtilde}
\check{\Sigma}_{\tilde{\underline{X}}}=\Lambda_1+o_p(1).
\end{equation}
Moreover, given any $m\geq 1$,
\begin{equation}\label{equivalent}
T^{\frac{1}{2}}\left\{\hat{\gamma}^{ALS}_m - \hat{\gamma}^{GLS}_m
\right\}=o_p(1) \quad\text{and}\quad
T^{\frac{1}{2}}\left\{\hat{\rho}^{ALS}_m-\hat{\rho}^{GLS}_m\right\}=o_p(1),
\end{equation}
where $\hat{\rho}^{ALS}_m$ (resp. $\hat{\rho}^{GLS}_m$) stands for any of
$\hat{\rho}^{ALS}_{a,m}$ and $\hat{\rho}^{ALS}_{b,m}$ (resp. $\hat{\rho}^{GLS}_{a,m}$ and $\hat{\rho}^{GLS}_{b,m}$).
\end{prop}

This asymptotic equivalence result allows us to propose portmanteau
test statistics adapted to the case of time-varying variance.
Consider the Box-Pierce type statistic
\begin{eqnarray*}
Q_{a,m}^{ALS}&=&T\sum_{h=1}^m\mbox{tr}\left(\hat{\Gamma}_{ALS}'(h)\hat{\Gamma}_{ALS}^{-1}(0)
\hat{\Gamma}_{ALS}(h)\hat{\Gamma}_{ALS}^{-1}(0)\right)\\&=&T\hat{\gamma}^{ALS'}_m\left(I_m\otimes
\hat{\Gamma}_{ALS}^{-1}(0)\otimes
\hat{\Gamma}_{ALS}^{-1}(0)\right)\hat{\gamma}^{ALS}_m,
\end{eqnarray*}
and
\[
Q_{b,m}^{ALS}=T\hat{\rho}^{ALS'}_{b,m}\hat{\rho}^{ALS}_{b,m}.
\]
Consider also the Ljung-Box type statistics
\[
\tilde{Q}_{a,m}^{ALS}=T^2\sum_{h=1}^m(T-h)^{-1}\mbox{tr}\left(\hat{\Gamma}_{ALS}'(h)\hat{\Gamma}_{ALS}^{-1}(0)
\hat{\Gamma}_{ALS}(h)\hat{\Gamma}_{ALS}^{-1}(0)\right)
\]
and
\[
\tilde{Q}_{b,m}^{ALS}=T^2\sum_{h=1}^m(T-h)^{-1}\mbox{tr}\left(\hat{\Gamma}_{ALS}'(h)
\hat{\Gamma}_{ALS}(h)\right).
\]
The following theorem is a direct consequence of (\ref{gamgls}) and
Proposition \ref{lemALS} and hence the proof is omitted.

\begin{thm}\label{th_5_1} Under the assumptions of Proposition \ref{lemALS}, the statistics $Q_{a,m}^{ALS}$, $Q_{b,m}^{ALS}$ and $\tilde{Q}_{a,m}^{ALS}$,
$\tilde{Q}_{b,m}^{ALS}$
converge in distribution to
\begin{equation}\label{truedistadapt}
U(\delta_m^{ALS})=\sum_{i=1}^{d^2m}\delta^{als}_iU_i^2,
\end{equation}
as $T\to\infty$, where
$\delta_m^{ALS}=(\delta^{als}_1,\dots,\delta^{als}_{d^2m})'$ is the
vector of the eigenvalues of $\Sigma^{GLS},$ and the $U_i$'s are
independent $\mathcal{N}(0,1)$ variables.
\end{thm}

To compute the critical values of the adaptive portmanteau tests,  we
first obtain a consistent estimator of $\Lambda^{\epsilon,\theta}_{m}$
given in (\ref{estim12}) by replacing $A_{01},\dots,A_{0p}$ by their
ALS estimators in $K$ and using (\ref{ademain}). Next we consider the estimate of $\Lambda_1$ given in (\ref{sigtilde}).
Plugging these estimates into the formula (\ref{glsmat}), we obtain a consistent estimator of  $\Sigma^{GLS}$ with eigenvalues
$\hat{\delta}_m^{ALS}=(\hat{\delta}_1^{als},\dots,\hat{\delta}_{d^2m}^{als})'$ that consistently estimate $\delta_m^{ALS}$.

There are several important particular cases that could be
mentioned. In the case of a VAR(0) model (i.e., the process $(u_t)$
is observed), $\Sigma^{GLS}= I_{d^2 m}$ (see Proposition
\ref{propostun}) and hence the asymptotic distribution of the four
test statistics in Theorem \ref{th_5_1} would be $\chi^2_{d^2m}$,
that means independent of the variance structure given by
$\Sigma(\cdot)$. In the general case $p \geq 1$ where the autoregressive coefficients $A_{0i}$, $i=1,...,p$ have to be estimated, the matrix $I_{d^2 m} - \Sigma^{GLS}$ being positive semi-definite,  the eigenvalues $\delta_1^{als}, ..., \delta_{d^2m}^{als}$ are smaller or equal to 1. Since, in some sense, the unconditional heteroscedasticity is removed in the ALS residuals, one could expect that the $\chi^2_{d^2(m-p)}$ asymptotic approximation is reasonably accurate for the ALS tests. Example \ref{ex1} indicates that this is may not the case, the asymptotic distribution we obtain for the ALS portmanteau statistics can be very different from the $\chi^2_{d^2(m-p)}$ approximation when the errors are heteroscedastic.
Finally note that Patilea and Raïssi (2010) pointed out that using the adaptive estimators of autoregressive parameters instead of the OLS estimators lead to a gain of efficiency, so that it is advisable to compute the kernel smoothing estimator of the variance function $\Sigma(\cdot)$ at the estimation stage. Therefore since the kernel estimator of the variance $\Sigma_t$ is available for the validation stage, the ALS tests are not more complicated than the OLS tests to implement.

 Let us also point out that the eigenvalues $\delta_1^{als}, ..., \delta_{d^2m}^{als}$ will
not depend on the variance structure when $\Sigma(\cdot) = \sigma^2(\cdot) I_d$
(in particular in the univariate case), whatever the value of $p$ is.
Moreover, using the arguments of Box and Pierce (1970), see also Brockwell and Davis (1991, pp. 310--311), one can easily show that for large values of $m$, the law of $U(\delta_m^{ALS})$ is accurately approximated by a $\chi^2_{d^2 (m-p)}$ distribution. However, in the general the multivariate setup the asymptotic distribution in (\ref{truedistadapt}) depend on the variance function $\Sigma(\cdot)$.

\section{Modified portmanteau statistics with standard chi-square asymptotic distributions}
\label{modifstats}

In the previous sections we considered portmanteau tests for which the asymptotic critical values are given by weighted sums of chi-squares in the general VAR(p) case. Using a suitable change of our quadratic forms one can propose alternative portmanteau test statistics with chi-squared asymptotic distributions under the null hypothesis. This type of modification was already proposed in the recent time series literature but in different contexts.

First note that as remarked above when testing that the observed process is uncorrelated ($p=0$) and using the standard portmanteau statistic (\ref{olstatport}) we obtain a non standard asymptotic distribution. Then following the approach of Lobato, Nankervis and Savin (2002) we consider the modified portmanteau test statistic
$$\underline{Q}_m^{OLS}=T\hat{\gamma}^{OLS'}_m\left(\hat{\Lambda}_m^{u,u}\right)^{-1}\hat{\gamma}^{OLS}_m,$$
where $\hat{\Lambda}_m^{u,u}=I_m\otimes\hat{\Sigma}_{G^{\otimes2}}$ with $\hat{\Sigma}_{G^{\otimes2}}$ defined in equation (\ref{ge}). The invertibility of $\hat{\Lambda}_m^{u,u}$ is guaranteed asymptotically by our assumptions. In view of Proposition \ref{propostun} it is clear that under the null hypothesis of uncorrelated observed process, the asymptotic distribution of the $\underline{Q}_m^{OLS}$ statistic is $\chi^2_{d^2m}$. Recall that this kind of statistic correction is not necessary to obtain a standard asymptotic distribution for the adaptive portmanteau tests when the non correlation of the observed process is tested.

This approach can be generalized to the case of VAR($p$) models with possibly $p>0$ using the approach of Katayama (2008) for building tests with standard asymptotic distributions. In this part we take $p<m<T$. Let us introduce
$$D_m^{OLS}=\Phi_m\left\{\Phi_m'\left(\Lambda_m^{u,u}\right)^{-1}\Phi_m\right\}^{-1}\Phi_m'\left(\Lambda_m^{u,u}\right)^{-1}$$ $$D_m^{GLS}=\Lambda_m^{\epsilon,\theta}\left\{\Lambda_m^{\epsilon,\theta'}\Lambda_m^{\epsilon,\theta}\right\}^{-1}\Lambda_m^{\epsilon,\theta'}$$
so that $(I_{d^2m}-D_m^{OLS})\Phi_m=0$ and $(I_{d^2m}-D_m^{GLS})\Lambda_m^{\epsilon,\theta}=0$. From the proof of Proposition \ref{propostun} (equation (\ref{petitop})), it is easy to see that
$$(I_{d^2m}-D_m^{OLS})T^{\frac{1}{2}}\hat{\gamma}_m^{OLS}=(I_{d^2m}-D_m^{OLS})T^{\frac{1}{2}}c_m^u+o_p(1)$$
where $T^{1/2}c_m^u$ is asymptotically normal of mean 0 and variance $\Lambda_m^{u,u}$. Deduce that
$$(I_{d^2m}-D_m^{OLS})T^{\frac{1}{2}}\hat{\gamma}_m^{OLS}\Rightarrow\mathcal{N}(0,V),$$where $V=(I_{d^2m}-D_m^{OLS})\Lambda_m^{u,u}(I_{d^2m}-D_m^{OLS})^\prime$.
Now, notice that
$$(I_{d^2m}-D_m^{OLS})\Lambda_m^{u,u}=\Lambda_m^{u,u}-\Phi_m\left\{\Phi_m'\left(\Lambda_m^{u,u}\right)^{-1}\Phi_m\right\}^{-1}\Phi_m^\prime= \Lambda_m^{u,u}(I_{d^2m}-D_m^{OLS})^\prime .$$ From this and the fact that $I_{d^2m}-D_m^{OLS}$ is a projector, deduce that the matrix $AV$ is idempotent, where
$$A=(I_{d^2m}-D_m^{OLS})^\prime(\Lambda_m^{u,u})^{-1}(I_{d^2m}-D_m^{OLS}). $$
Moreover, since $\Phi_m$ is of full column rank $d^2p$, it is easy to see that the rank of $A$ is $d^2(m-p)$. A classical result in multivariate data analysis implies
\begin{equation}\label{portdiststdols}
T\hat{\gamma}_m^{OLS'}(I_{d^2m}-D_m^{OLS})^\prime(\Lambda_m^{u,u})^{-1}(I_{d^2m}-D_m^{OLS})
\hat{\gamma}_m^{OLS}\Rightarrow\chi_{d^2(m-p)}^2.
\end{equation}
Similarly we obtain
\begin{equation}\label{katayama1}
(I_{d^2m}-D_m^{GLS})T^{\frac{1}{2}}\hat{\gamma}_m^{GLS}\Rightarrow\mathcal{N}(0,I_{d^2m}-D_m^{GLS})
\end{equation}
and we deduce
\begin{equation}\label{portdiststdgls}
T\hat{\gamma}_m^{GLS'}(I_{d^2m}-D_m^{GLS})
\hat{\gamma}_m^{GLS}\Rightarrow\chi_{d^2(m-p)}^2.
\end{equation}

The matrices $D_m^{OLS}$ and $\Lambda_m^{u,u}$ could be estimated as suggested in Section \ref{S4}, see comments after equation (\ref{lam3}), and hence a modified portmanteau test statistic based on the OLS estimates $\hat{\gamma}_m^{OLS}$ and having standard chi-square critical values could be derived from equation  (\ref{portdiststdols}). On the other hand, using a nonparametric estimate of $H_t$ one could easily estimate $D_m^{GLS}$, see Proposition \ref{lemALS} and the comments after Theorem \ref{th_5_1}. Moreover, Proposition \ref{lemALS} allows us to replace $\hat{\gamma}_m^{GLS}$ with $\hat{\gamma}_m^{ALS}$ and thus to introduce an adaptive portmanteau test with a modified statistic and standard chi-square critical values based on equation (\ref{portdiststdgls}). Clearly one can consider similar modification for Ljung-Box type statistics.

The chi-square critical values are certainly more convenient for portmanteau tests. Moreover, in section \ref{S5b} we provide evidence that the test based on the statistic (\ref{portdiststdols}) could be more powerful, in the Bahadur slope sense, than the OLS estimates based test based on the $Q_m^{OLS}$ statistic investigated  in Theorem  \ref{VAR}. However, it is not necessarily true that the modified procedures presented in this section are preferable in applications. Indeed, the empirical evidence presented in Section \ref{S6} shows that test statistics like in (\ref{portdiststdols}) and (\ref{portdiststdgls}) are unstable and induce bad levels even with series of few hundred observations.

\section{Testing for autocorrelation in heteroscedastic series: some theoretical power comparisons}
\label{S5b}

In this part we carry out some theoretical power comparisons for the tests we considered above in the important case where the non correlation of the observed
process $X_t=u_t$ is tested. The case $p\geq 1$ will be considered elsewhere.
%the order $p$ in the model (\ref{VAR}).
On one hand we compare the classical Box-Pierce portmanteau test and modified quadratic forms of OLS residual autocorrelations based test introduced in section \ref{modifstats}. On the other hand we compare ALS and OLS residual autocorrelations based portmanteau tests. For this purpose we use the Bahadur slope approach that we briefly recall here. Let  $Q_A$ denote a test statistic  and, for any $x>0$, define $q_A(x) =  -\log P_{0} (Q_A > x)$ where $P_{0} $ stands for the limit distribution of $Q_A$ under the null hypothesis. Following Bahadur (1960) (see also van der Vaart 1998, chapter 14),  consider the \emph{(asymptotic) slope} $ c_A(\varrho) = 2\lim_{T\rightarrow\infty} T^{-1} q_A(Q_A) $ under a \emph{fixed} alternative $\mathcal{H}_1$ such that the limit exists in probability.  The asymptotic relative efficiency of the test based on $Q_A$ with respect to a competing test based on a test statistic $Q_B$ is then defined as the ratio $ARE_{A,B}(\varrho) = c_A(\varrho)/c_B(\varrho)$. A relative efficiency $ARE_{A,B}(\varrho)\geq 1$ suggests that the test given by $Q_A$ is better suited to detect $\mathcal{H}_1 $ because the associated $p-$values wanes faster or equally faster compared to the $p-$values of the test based on $Q_B$.

For the sake of simplicity we restrict our attention to the BP statistics and
%VAR(0) models, the case $p\geq 1$ will be considered elsewhere.
%That iswe
consider the case where one tests the non correlation of the observed process, while the underlying true process is the autoregressive process of order 1

\begin{equation}
u_t=Bu_{t-1}+\tilde{H}_t\epsilon_t,\label{eqdeux}
\end{equation}
where $\det(I_d-Bz)\neq0$ for all $|z|\leq1$ and $B\neq0$.
We keep the notation $E(X_tX_t')=E(u_tu_t')=\Sigma_t$ and we introduce $E(\tilde{H}_t\epsilon_t\epsilon_t'\tilde{H}_t')=\tilde{H}_t\tilde{H}_t':=\tilde{\Sigma}_t$.
Under the alternative hypothesis (\ref{eqdeux}) we have the relationship
\begin{equation}\label{sigtosig}
\Sigma_t=\sum_{i=0}^{\infty}B^i\tilde{\Sigma}_{t-i}B^{i'}.
\end{equation}
Using similar arguments to that of the proofs of Lemmas \ref{lem1} to \ref{lemtcllln} in the Appendix, deduce that
\begin{equation}\label{autools}
\hat\Gamma_{OLS} (h) = T^{-1}\sum_{t=h+1}^Tu_tu_{t-h}'=B^h\int_0^1\Sigma(r)dr+o_p(1)
\end{equation}
and
\begin{equation}\label{autoals}
\hat\Gamma_{GLS} (h) = T^{-1}\sum_{t=h+1}^TH_t^{-1}u_tu_{t-h}'H_{t-h}^{-1'}=\int_0^1G(r)^{-1}B^hG(r)dr+o_p(1).
\end{equation}
Using basic properties of the vec($\cdot$) operator and the Kronecker product we obtain
$$T^{-1}Q_m^{OLS}=\mathcal{B}'\left\{I_m\otimes\int_0^1\Sigma(r)dr
\otimes\left(\int_0^1\Sigma(r)dr\right)^{-1}\right\}\mathcal{B}+o_p(1)$$
$$T^{-1} \underline{Q}_m^{OLS}\! = \!\mathcal{B}' \!\left\{ \!
I_m\otimes \left(\! \int_0^1 \!\!\! \Sigma(r)dr
\!\otimes\! I_d \! \right) \!\! \left(\!\int_0^1 \!\! \!\Sigma(r)\!\otimes \!\Sigma(r)dr \!\! \right)^{\!\! -1} \!\! \left(\!\int_0^1 \!\!\! \Sigma(r)dr
\!\otimes\! I_d \! \right)
\!\! \right\}\mathcal{B} +   o_p(1)
$$
and
$$T^{-1}Q_{i,m}^{ALS}=\mathcal{B}'\left\{
I_m\otimes\left(\int_0^1G(r)'\otimes G(r)^{-1}dr\right)^2
%\left(\int_0^1\Sigma(r)^{-\frac{1}{2}}\otimes\Sigma(r)^{\frac{1}{2}}dr\right)
\right\}\mathcal{B}+o_p(1),$$
with $i\in\{a,b\}$ and $\mathcal{B}=\mbox{vec}\left\{(B^1,\dots,B^m)\right\}$.

\begin{prop}\label{bahadur_comp}
(i) If Assumption {\bf A1} holds true and the observations follow the model (\ref{eqdeux}), the asymptotic relative efficiency of the portmanteau test based on $\underline{Q}_m^{OLS}$ with respect to the portmanteau tests based on $Q_m^{OLS}$ is larger or equal to 1.

(ii) Suppose that $\Sigma(\cdot) = \sigma^2(\cdot) I_d$ where $\sigma(\cdot)$ is some positive scalar function. Suppose that Assumptions {\bf A1'} and {\bf A2} holds true and the observations follow the model (\ref{eqdeux}). Then asymptotic relative efficiencies of the portmanteau test based on $Q_m^{ALS}$ with respect to the portmanteau tests based on $Q_m^{OLS}$ or $\underline{Q}_m^{OLS}$ are larger or equal to 1.
\end{prop}

In the first part of Proposition \ref{bahadur_comp} the result is obtained without additional restriction on $\Sigma(\cdot) $ while in the second part we impose  $\Sigma(\cdot) = \sigma^2(\cdot) I_d$ which is for instance true in the univariate case ($d=1$). In the general multivariate case the portmanteau test based on ALS residual autocorrelations does not necessarily outperforms, in the sense of the Bahadur slope, the tests based on the OLS residual autocorrelations considered above.

\section{Extending the scope: testing the order of a heteroscedastic co-integration model}
\label{Scointegration}
Consider the case of a unit root multivariate process $(X_t)$ with time-varying volatility, see for instance Cavaliere, Rahbek and Taylor (2010) or Boswijk (2010).
With $Z_{0t}:=X_t-X_{t-1}$  the model (\ref{VAR}) can be rewritten  in its error correction form
\begin{eqnarray}\label{VECM}
&&Z_{0t}=\Pi_{01} X_{t-1}+\sum_{i=2}^{p}\Pi_{0i}Z_{0t-i+1}+u_t\\&&
u_t=H_t\epsilon_t.\nonumber
\end{eqnarray}
The matrices $\Pi_{0i}$ are functions of the $A_i$'s, and such that the assumptions of the Granger representation theorem hold (see for instance Assumption 1 of Cavaliere, Rahbek and Taylor (2010)), $\Pi_{01}=\alpha_0\beta_0'$ where the $d\times s$-dimensional matrices $\alpha_0$ and $\beta_0$ are identified in some appropriate way (see e.g. Johansen 1995, p. 72, for the identification problem). If $p=1$ the sum in (\ref{VECM}) vanishes. In this section we follow Cavaliere, Rahbek and Taylor (2010) and we slightly strengthen {\bf A1} assuming  that $(\epsilon_t)$ is iid. Then it follows from their Lemma 1 that $(X_t)$ have a random walk behavior and also that $(\beta_0'X_t)$ is stable. By analogy with the homoscedastic case, the number $s$ of independent linear stable combinations in $(\beta_0'X_t)$ is the cointegrating rank (see section 2.3 of Cavaliere, Rahbek and Taylor 2010, for a detailed discussion on the concept of cointegration in our framework).
If $s=0$ the process $(X_t)$ is not cointegrated and the procedures described in the previous sections apply directly to the process $(Z_{0t})$. Many contributions in the literature that considered the standard homoscedastic  framework pointed out that the choice of the lag length is important for the contegrating rank analysis, see e.g. Boswijk and Franses (1992, section 4). It seems reasonable to imagine that a similar  remark remains true with a time-varying variance.

To describe the estimation procedure of (\ref{VECM}), let us define $Z_{1t}(\beta)=(X_{t-1}'\beta,Z_{0t-1}',$ $\dots,Z_{0t-p+1}')'$ for any $d\times s$-matrix $\beta$ and
rewrite model (\ref{VECM}) under the form
$$Z_{0t}=(Z_{1t}(\beta_0)'\otimes I_d)\vartheta_0+u_t,$$
where $\vartheta_0=\mbox{vec}\left\{\left(\alpha_0,\Pi_{02},\dots,\Pi_{0p}\right)\right\}$. The estimator of the long run relationships $\hat{\beta}$ can be obtained using the reduced rank regression method introduced by Anderson (1951). Cavaliere \textit{et al} (2010) showed that in our framework
\begin{equation}\label{betaconsistent}
T(\hat{\beta}-\beta)=O_p(1).
\end{equation}
Now, let us define
\begin{equation*}
\hat{\vartheta}_{OLS}(\beta)=\hat{\Sigma}_{Z_1}^{-1}(\beta)\mbox{vec}\:\left(\hat{\Sigma}_{Z_0}(\beta)\right),
\end{equation*}
where
$$\hat{\Sigma}_{Z_1}(\beta)=T^{-1}\sum_{t=1}^TZ_{1t}(\beta)Z_{1t}(\beta)'\otimes
I_d\quad\mbox{and}\quad\hat{\Sigma}_{Z_0}(\beta)=T^{-1}
\sum_{t=1}^TZ_{0t}Z_{1t}(\beta)',$$
and similarly to (\ref{GLS}) let us introduce
\begin{equation*}
\hat{\vartheta}_{GLS}(\beta)=\hat{\Sigma}_{\underline{Z}_1}(\beta)
^{-1}\mbox{vec}\:\left(\hat{\Sigma}_{\underline{Z}_0}(\beta)\right),
\end{equation*}
with
$$\hat{\Sigma}_{\underline{Z}_1}=T^{-1}\sum_{t=1}^TZ_{1t}(\beta)Z_{1t}(\beta)'
\otimes\Sigma_t^{-1}, \quad\hat{\Sigma}_{\underline{Z}_0}=T^{-1}
\sum_{t=1}^T\Sigma_t^{-1}Z_{0t}Z_{1t}(\beta)',$$
where the volatility $\Sigma_t$ is assumed known. Next, for any fixed $\beta$,  let us define the estimated residuals
$$\hat{u}_t(\beta)=Z_{0t}-(Z_{1t}(\beta)'\otimes I_d)\hat{\vartheta}_{OLS}(\beta),$$
$$\hat{\epsilon}_t(\beta)=H_t^{-1}Z_{0t}-H_t^{-1}(Z_{1t}(\beta)'\otimes I_d)\hat{\vartheta}_{GLS}(\beta)$$
and the corresponding estimated autocovariance matrices
$$\hat{\Gamma}_{OLS}(h,\beta)=T^{-1}\sum_{t=h+1}^T\hat{u}_t(\beta)\hat{u}_{t-h}(\beta)'\quad\mbox{and}\quad
\hat{\Gamma}_{GLS}(h,\beta)=T^{-1}\sum_{t=h+1}^T\hat{\epsilon}_t(\beta)\hat{\epsilon}_{t-h}(\beta)'.$$
From (\ref{betaconsistent}) we obviously have
$$\hat{\Gamma}_{OLS}(h,\hat{\beta})=\hat{\Gamma}_{OLS}^u(h,\beta_0)+o_p(T^{-\frac{1}{2}})
\quad\mbox{and}\quad\hat{\Gamma}_{GLS}(h,\hat{\beta})=\hat{\Gamma}_{GLS}^u(h,\beta_0)+o_p(T^{-\frac{1}{2}}).$$
Defining
$$\check{\Sigma}_t^0(\beta)=\sum_{i=1}^Tw_{ti}\odot\hat{u}_i(\beta)\hat{u}_i(\beta)',$$
it is also obvious that
$$\check{\Sigma}_t^0(\hat{\beta})=\check{\Sigma}_t^0(\beta_0)+o_p(T^{-\frac{1}{2}}).$$
It is clear now that one can treat $\beta_0$ as known and, following the lines of the previous sections, one can use $\hat{\Gamma}_{OLS}(h,\hat{\beta})$ and the ALS version of $\hat{\Gamma}_{GLS}(h,\hat{\beta})$ to build portmanteau tests  for checking the order $p$  of the model (\ref{VECM}).

\section{Monte Carlo experiments}
\label{S6}

In the sequel  $LB_m^{OLS}$ and $LB_m^{ALS}$ will denote the Ljung-Box type
portmanteau tests based on the adaptive approach with non standard distributions (\ref{truedist}) and (\ref{truedistadapt}). For the sake of brevity, only the results obtained with the test statistic $\tilde Q_{a, m}^{ALS}$ will be reported. %those obtained with $\tilde Q_{yy, m}^{ALS}$ were quite close.
Moreover, since we found similar results for
the $BP$ and $LB$ tests, we only report on  $LB$ tests.
The $LB$ tests
based on modified statistics which are built using the results in Section \ref{modifstats} are denoted by $\widetilde{LB}_m^{OLS}$
and $\widetilde{LB}_m^{ALS}$. If we assume that
the volatility function is known, one can also build portmanteau
tests using the result in (\ref{rhogls}) in a similar way to the adaptive
portmanteau tests. These infeasible tests denoted by $LB_m^{GLS}$
and $\widetilde{LB}_m^{GLS}$ will serve as a benchmark for comparison with the
ALS tests $LB_m^{ALS}$ and $\widetilde{LB}_m^{ALS}$. It is clear that the asymptotic critical values of the
ALS tests are the same as the critical values of the GLS tests. In
this section we investigate by simulations the finite sample
properties of the ALS and GLS portmanteau tests and we compare them
with the OLS estimation-based tests. In the next section we study the model adequacy of two real data sets: the US energy and transportation price indexes for all urban consumers on one hand and the US balances on merchandise trade and on services on the other hand.

\subsection{Empirical size}

Our Monte Carlo experiments are based on the following Data
Generating Process (DGP) specification
\begin{equation}\label{dgpmc}
\left(
    \begin{array}{c}
      X_{1t} \\
      X_{2t} \\
    \end{array}
  \right)
=\left(\begin{array}{cc}
0.3 & -0.3 \\
0 & -0.1 \\
\end{array}
\right)\left(
    \begin{array}{c}
      X_{1t-1} \\
      X_{2t-1} \\
    \end{array}
  \right)+\left(
\begin{array}{cc}
\mathfrak{a} & 0 \\
0 & \mathfrak{a} \\
\end{array}
\right)\left(
    \begin{array}{c}
      X_{1t-2} \\
      X_{2t-2} \\
    \end{array}
  \right)+\left(
            \begin{array}{c}
              u_{1t} \\
              u_{2t} \\
            \end{array}
          \right),
\end{equation}
where $\mathfrak{a}=0$ in the empirical size part of the study and $\mathfrak{a}=-0.3$ in the empirical power part. The autoregressive parameters are such that the stability condition hold and are inspired from the ALS estimation obtained for the U.S. balance on services and merchandise trade data (see Table \ref{estimates} below). In the case of smooth
variance structure we consider
\begin{equation}\label{smoothspe}
\Sigma(r)=\left(
        \begin{array}{cc}
          (1+\pi_1 r)(1+\varpi^2) & \varpi(1+\pi_1 r)^{\frac{1}{2}}(0.1+\pi_2 r)^{\frac{1}{2}} \\
          \varpi(1+\pi_1 r)^{\frac{1}{2}}(0.1+\pi_2 r)^{\frac{1}{2}} & (0.1+\pi_2 r) \\
        \end{array}
      \right),
\end{equation}
where we take $\pi_1=250$ and $\pi_2=5$. In order to investigate the properties of the tests when a volatility break is present we also consider the following specification

\begin{equation}\label{breakspe}
\Sigma(r)=\left(
        \begin{array}{cc}
          (6+f_1(r))(1+\varpi^2) & \varpi(6+f_1(r))^{\frac{1}{2}}(0.5+f_2(r))^{\frac{1}{2}} \\
          \varpi(0.5+f_2(r))^{\frac{1}{2}}(6+f_1(r))^{\frac{1}{2}} & (0.5+f_2(r))(1+\rho^2) \\
        \end{array}
      \right),
\end{equation}
with $f_1(r)=54\times\mathbf{1}_{(r\geq1/2)}(r)$ and $f_2(r)=3\times\mathbf{1}_{(r\geq1/2)}(r)$. In this case we have a common volatility break at the date $t=T/2$. In all experiments we fix $\varpi=0.2$. These volatility specifications are inspired by the real data studies we consider in the next section.
For instance to fix the $(1,1)-$component of the specification (\ref{smoothspe}) we noticed that the last estimated variances with the balance services and merchandise trade data are all greater than 200. In the energy-transport price indexes data some of the last estimated variances of the first component are even greater than 300. The amplitudes of the functions $f_1(\cdot)$ and $f_2(\cdot)$ in the volatility specification with an abrupt break defined in (\ref{breakspe}) were calibrated close to the means of the first $T/2$ estimated volatilities and of the $T/2$ last volatilities for the balance services and merchandise trade data.
To assess the finite sample properties of the tests under comparison
when the errors are stationary, we also considered standard i.i.d.
Gaussian error processes. For each experiment $N=1000$ independent
trajectories are simulated using DGP (\ref{dgpmc}).
Samples of length $T=50$, $T=100$ and $T=200$ are simulated.

We first study the empirical size of the tests taking $\mathfrak{a}=0$, and adjusting a VAR(1) model to the simulated processes.
The portmanteau tests for the non correlation of the error terms are applied using $m=5$ and $m=15$ at the asymptotic nominal level 5\%.
The results
%for the test of the null hypothesis $p=1$
are given in Tables \ref{tab1}, \ref{tab2} and \ref{tab3}. Since $N=1000$
replications are performed and assuming that the finite sample size
of the tests is $5\%$, the relative rejection frequencies should be
between the significant limits 3.65\% and 6.35\% with probability
0.95. Then the relative rejection frequencies are displayed in bold
type when they are outside these significant limits.  Note that the
distributions (\ref{truedist}) and (\ref{truedistadapt}) are given
for fixed $m$, while the $\chi_{d^2(m-p)}^2$ approximation (see discussion after
Theorem  \ref{th_4_1} above) should be accurate only for large  $m$.
%Therefore the choice of $m$ in
%finite samples is discussed for the ALS, GLS and OLS
%portmanteau tests.
Therefore we only comment the results for small and
moderate samples ($T=50$ and $T=100$) for the standard portmanteau
tests with $m=5$, while the results for large samples ($T=200$) are considered when $m=15$ is taken.

In Table \ref{tab1} i.i.d. standard Gaussian errors are used, that means $\Sigma(\cdot) \equiv I_2$. In this
simple case the relative rejection frequencies of the tests converge
to the asymptotic nominal level. In general we do not remark a major
loss of efficiency of the $LB_m^{ALS}$ test when compared to the standard
test. Therefore one can use the $LB_m^{ALS}$ test in case of doubt of the
presence of unconditional heteroscedasticity. The same remark can be
made for the $LB_m^{OLS}$ test when $m$ is small. However we note that the $LB_m^{OLS}$ is oversized for small samples and when $m$ is large. It also appears that the tests based on modified statistics are oversized when $m$ is large. This can be explained by the fact that the matrices $\Lambda_m^{\epsilon,\theta'}\Lambda_m^{\epsilon,\theta}$ and $\Phi_m'(\Lambda_m^{u,u})^{-1}\Phi_m$ are difficult to invert in such situations.

In Table \ref{tab2} heteroscedastic errors with an abrupt volatility break are considered, while the trending specification (\ref{smoothspe}) is used for the volatility in Table \ref{tab3}. In line with the theoretical
the relative rejection frequencies of the ALS, GLS and OLS tests
converge to the asymptotic nominal level. As expected the standard
portmanteau test is not valid. In general it emerges from this part that the
$LB_m^{ALS}$ and $LB_m^{GLS}$ tests have similar results. Then it seems that
estimating the volatility entails no major loss of efficiency when
building this kind of test. We did not found clear advantage for the $LB_m^{ALS}$ when compared to the $LB_m^{OLS}$ in the presented experiments. However note that the asymptotic distribution (\ref{truedist}) of the standard portmanteau statistic seems estimated with less precision than the asymptotic distribution of the ALS portmanteau statistic. For instance we see in Table \ref{weights} that for $m=5$ the standard deviations of the ALS weights are lower or equal to the standard deviations of the OLS weights. We found that this difference is more marked for $m=15$ which may lead to problems for the control of the error of first kind for the $LB_m^{OLS}$ as already noted in the homoscedastic case (see Table \ref{tab1} for $m=15$). Other simulation results not reported here show that the $LB_m^{OLS}$ test can be oversized when $m$ is large as in the homoscedastic case with $m=15$ and we noted that the estimation of the weights requires a relatively large number of
observations for the OLS approach. A possible explanation is also that  $\delta_i^{als}\in[0,1]$ while $\delta_i^{ols}\in[0,\infty)$ for $i\in\{1,\dots,d^2m\}$, which may proceed more instable estimation for $\delta_i^{ols}$ in many cases. In addition the energy-transportation price indexes example below show that the estimation of the weights may be problematic in the OLS case. Therefore we recommend to choose small $m$ when the samples are small and use large $m$ only when the samples are large
despite the asymptotic results hold
true for fixed $m$ when the $LB_m^{OLS}$ is considered. We again note that the tests with modified statistics are in general clearly oversized. A possible explanation is that White (1980) type correction matrices are inverted in the modified portmanteau statistics which may lead to oversized tests as pointed out in Vilasuso (2001) in the stationary case. We can conclude that with a data generating process close to our simulation design the $LB_m^{ALS}$ test controls reasonably well the error of the first kind in all the situations we considered.

\subsection{Empirical power}

\begin{figure}[h]\!\!\!\!\!\!\!\!\!\!
\vspace*{1.5cm} \hspace*{0.3cm}
\hspace*{5.2cm}

\vspace*{1.5cm} \hspace*{6.3cm} \hspace*{6.0cm}
\vspace*{1.4cm}

\protect \includegraphics{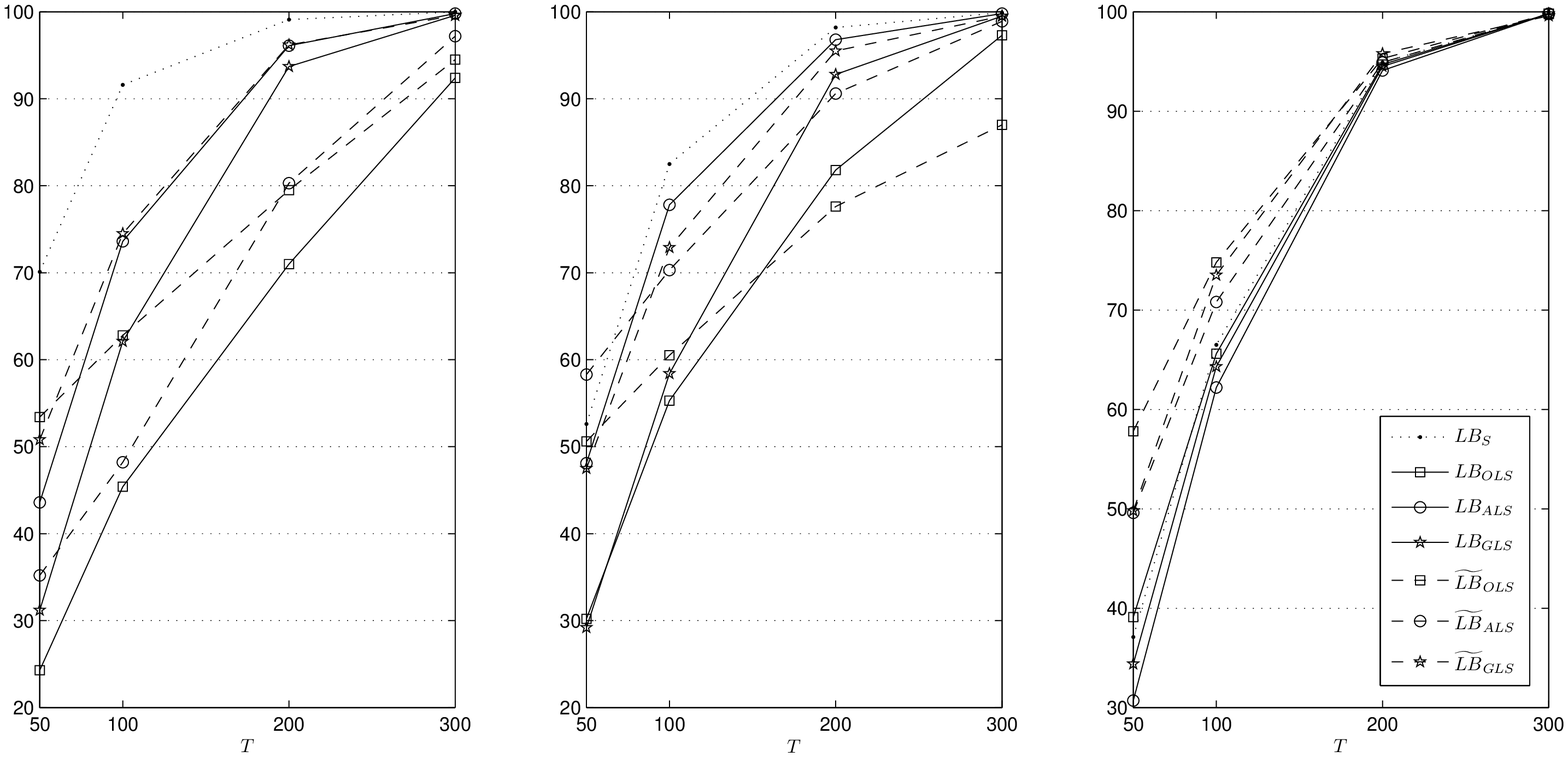} \caption{\label{powhomo}
{\footnotesize Empirical power (in \%) of the portmanteau tests with $m=10$. The adequacy of a VAR(1) model to VAR(2) processes is tested. The innovations are
homoscedastic on the right. The variance exhibits a break at $T/2$ on the left and have a trending behavior in the middle.}}
\end{figure}

In the empirical power part of this section, we examine the ability
of the different tests to detect underspecified autoregressive
order. The power investigation is realized in the Bahadur sense, that is the sample size is increased while the alternative is kept fixed. More precisely we set $\mathfrak{a}=-0.3$, and we adjusted a
VAR(1) model to the simulated VAR(2) processes with $T=50,100,200,300$. We simulated $N=1000$ independent trajectories using DGP (\ref{dgpmc}) with standard Gaussian innovations and heteroscedastic volatility specifications (\ref{smoothspe}) and (\ref{breakspe}). The non correlation of the error process is again tested at the asymptotic nominal level 5\%,
but taking $m=10$ in all the experiments. In Figure \ref{powhomo} we consider the homoscedastic case, errors where an abrupt volatility shift is observed and errors with smoothly varying variance.

When the variance is constant it appears that the tests with non standard distribution have similar power to the standard test when the errors are homoscedastic. Therefore we again note that there is no major loss of efficiency when the tests with modified distribution are used while the variance is constant. The tests with standard distribution may seem more powerful when the samples are small, but this mainly comes from the fact that these tests are oversized. When the errors are heteroscedastic the standard test seems more powerful than the other tests. However the standard test is oversized and in this case and the comparison is again not fair. A similar comment can be made when the tests with modified distribution are compared to the tests with standard distribution. It emerges that the $LB_m^{ALS}$ test is more powerful than the $LB_m^{OLS}$. The relation between the powers of $\widetilde{LB}_m^{OLS}$ and $\widetilde{LB}_m^{ALS}$ is not clear. Finally we can remark that in the presented experiments the GLS type tests are not necessarily more powerful than the ALS tests.

\begin{figure}[h]\!\!\!\!\!\!\!\!\!\!
\vspace*{1.0cm} \hspace*{0.3cm}
\hspace*{5.2cm}

\vspace*{1.5cm} \hspace*{6.3cm} \hspace*{6.0cm}
\vspace*{1.4cm}

\protect \includegraphics{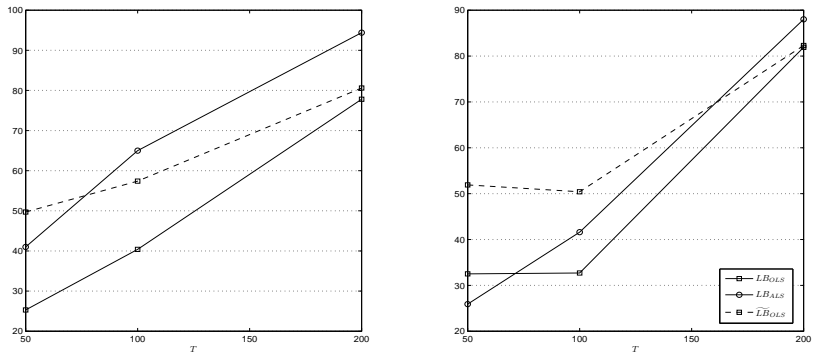} \caption{\label{powhomouncorr}
{\footnotesize Empirical power (in \%) of the portmanteau tests with $m=10$. The non correlation of VAR(1) processes is tested. The variance have a trending behavior on the left and exhibits an abrupt shift on the right.}}
\end{figure}

The results of section \ref{S5b} are also illustrated. To this aim $N=1000$ independent trajectories are simulated using a VAR(1) DGP with $A_{01}=-0.3I_2$.
We consider heteroscedastic errors with trending behavior volatility specification:

$$\Sigma(r)=\left(
              \begin{array}{cc}
                (1+\pi_1 r) & 0 \\
                0 & (1+\pi_1 r) \\
              \end{array}
            \right)
$$
where we take $\pi_1=150$, and a volatility with an abrupt shift at $T/2$:

$$\Sigma(r)=\left(
              \begin{array}{cc}
                (1+f_1(r)) & 0 \\
                0 & (1+f_1(r)) \\
              \end{array}
            \right),
$$
where $f_1(r)=10\times\mathbf{1}_{(r\geq1/2)}(r)$. The lengths of the simulated series are $T=50,100,200$. The non correlation of the observed process is tested at the asymptotic nominal level 5\% taking $m=10$. From Figure
\ref{powhomouncorr} the $\widetilde{LB}_{OLS}$ test may appear more powerful than the $LB_{ALS}$ test when the samples are small. However this comes from the fact that the $\widetilde{LB}_{OLS}$ test
is strongly oversized in small samples. In accordance with the theoretical we see that the $LB_{ALS}$ test clearly outperform the $\widetilde{LB}_{OLS}$ and $LB_{OLS}$ as the samples become large.

%It
%appears that the GLS and ALS tests are slightly less powerful than
%the standard and OLS tests.  the errors are
%heteroscedastic
%
%
%and we found that the GLS test outperforms the other
%tests in this case. It also appears that the ALS test is slightly
%more powerful than the OLS tests. Finally the standard tests may
%appear more powerful than the other tests from Table \ref{tabpow2}.
%However this comparison is not fair since the standard tests are
%oversized in the heteroscedastic case.

\section{Illustrative examples}
\label{S7}

For our real data illustrations  we use two U.S. economic data sets. First we consider the
quarterly U.S. international finance data for the period from January 1, 1970 to October 1, 2009:
the balance on services and the balance on merchandise trade in billions of Dollars.  The length of the balance data series is $T=159$.
We also
consider monthly data on  the U.S.
consumer price indexes of energy and transportation for all urban consumers for the period from January 1, 1957 to February 1, 2011.
The length of the energy-transportation series is $T=648$.
The series are available seasonally adjusted from the
website of the research division of the Federal Reserve Bank of Saint
Louis.

\subsection{VAR modeling of the U.S. balance trade data}

\begin{figure}[h]\!\!\!\!\!\!\!\!\!\!
\vspace*{3.6cm} %\hspace*{8.5cm}$\tau_1$
%\hspace*{11.7cm}$\sigma_{21}$ \vspace*{-5.3 cm}

%\vspace*{4.3cm}\hspace*{2.15cm}$\tau_1$ \hspace*{2.3cm}$\sigma_{21}$ %\hspace*{3.8cm}$\tau_1$
%\vspace*{0.2 cm}

\protect \includegraphics{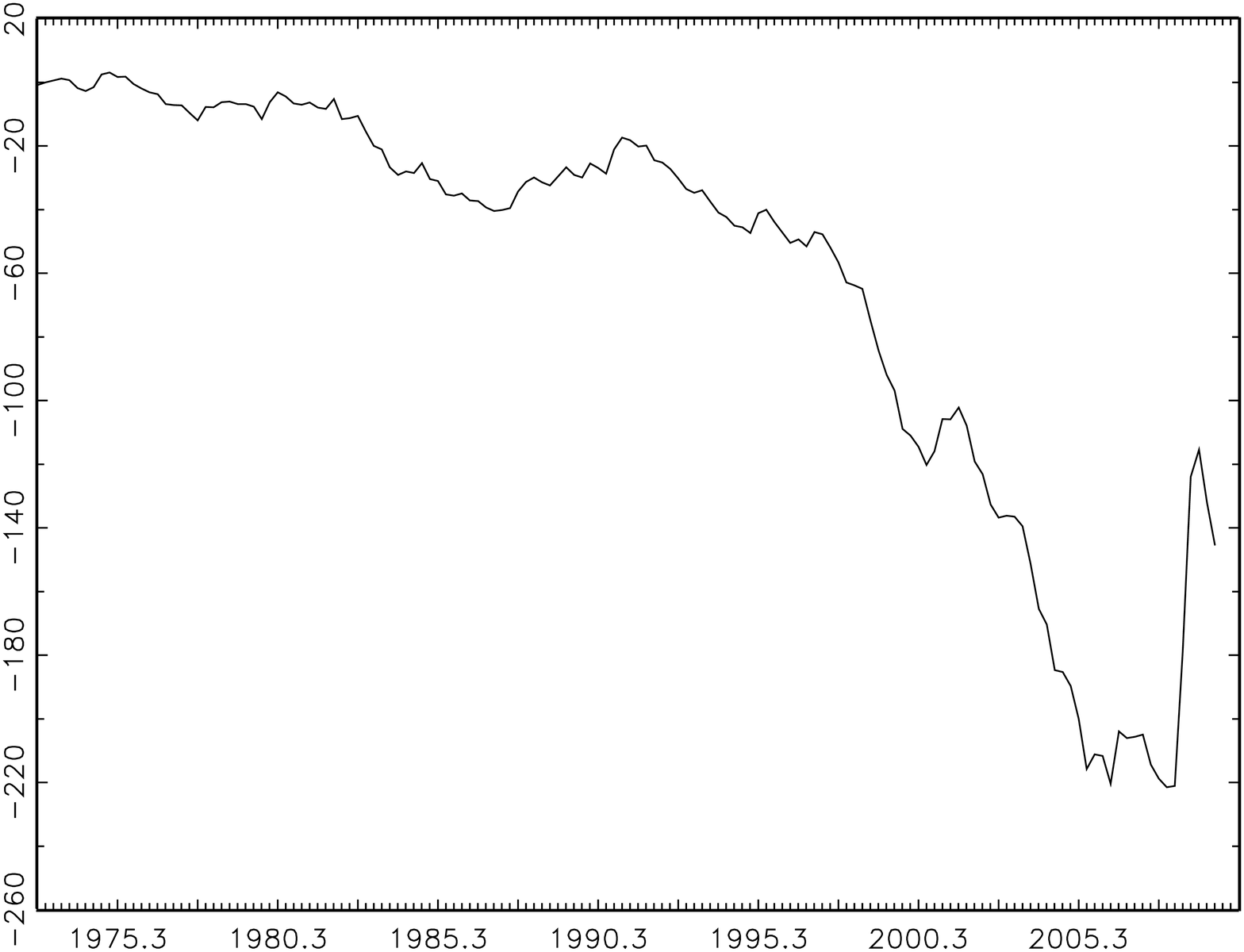} \protect \includegraphics{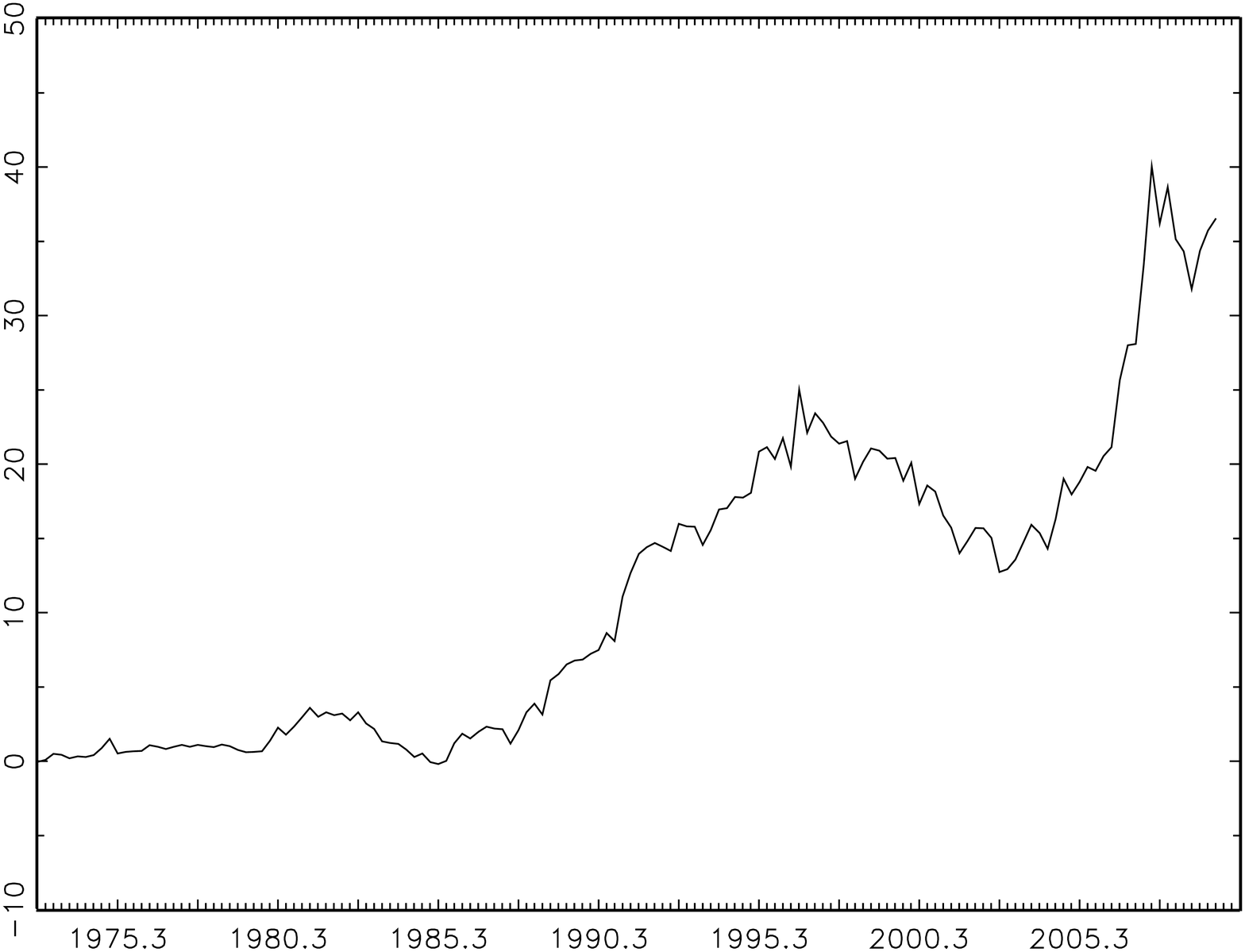} \caption{\label{data}
{\footnotesize The balance on merchandise trade for the U.S. on the left and the balance on services for the U.S. on the right in billions of dollars from 1/1/1970 to 10/1/2009, T=160. Data source: The research division of the federal reserve bank of Saint Louis, www.research.stlouis.org.}}
\end{figure}

\begin{figure}[h]\!\!\!\!\!\!\!\!\!\!
\vspace*{3.9cm} %\hspace*{8.5cm}$\tau_1$
%\hspace*{11.7cm}$\sigma_{21}$ \vspace*{-5.3 cm}

%\vspace*{4.3cm}\hspace*{2.15cm}$\tau_1$ \hspace*{2.3cm}$\sigma_{21}$ %\hspace*{3.8cm}$\tau_1$
%\vspace*{0.2 cm}

\protect \includegraphics{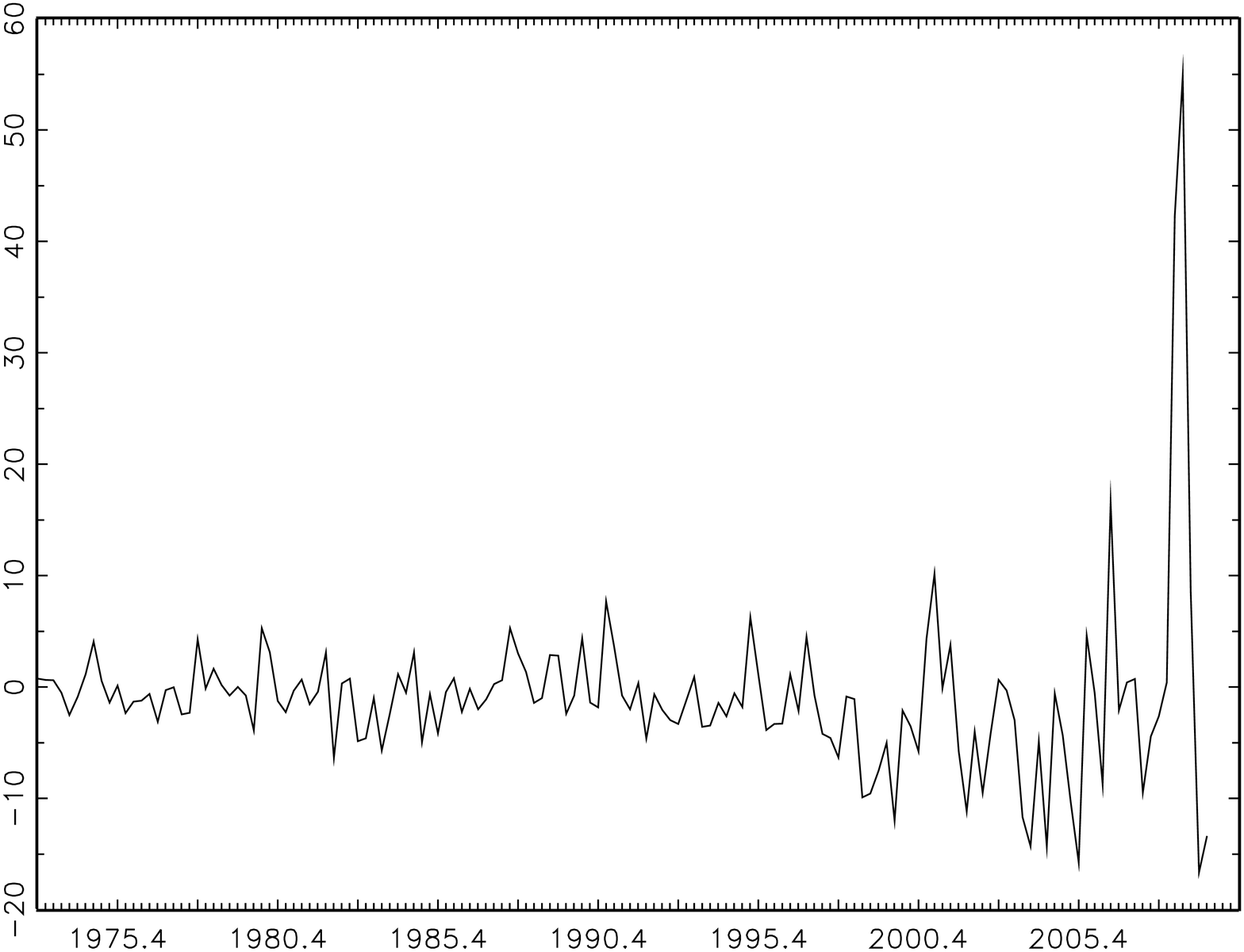} \protect \includegraphics{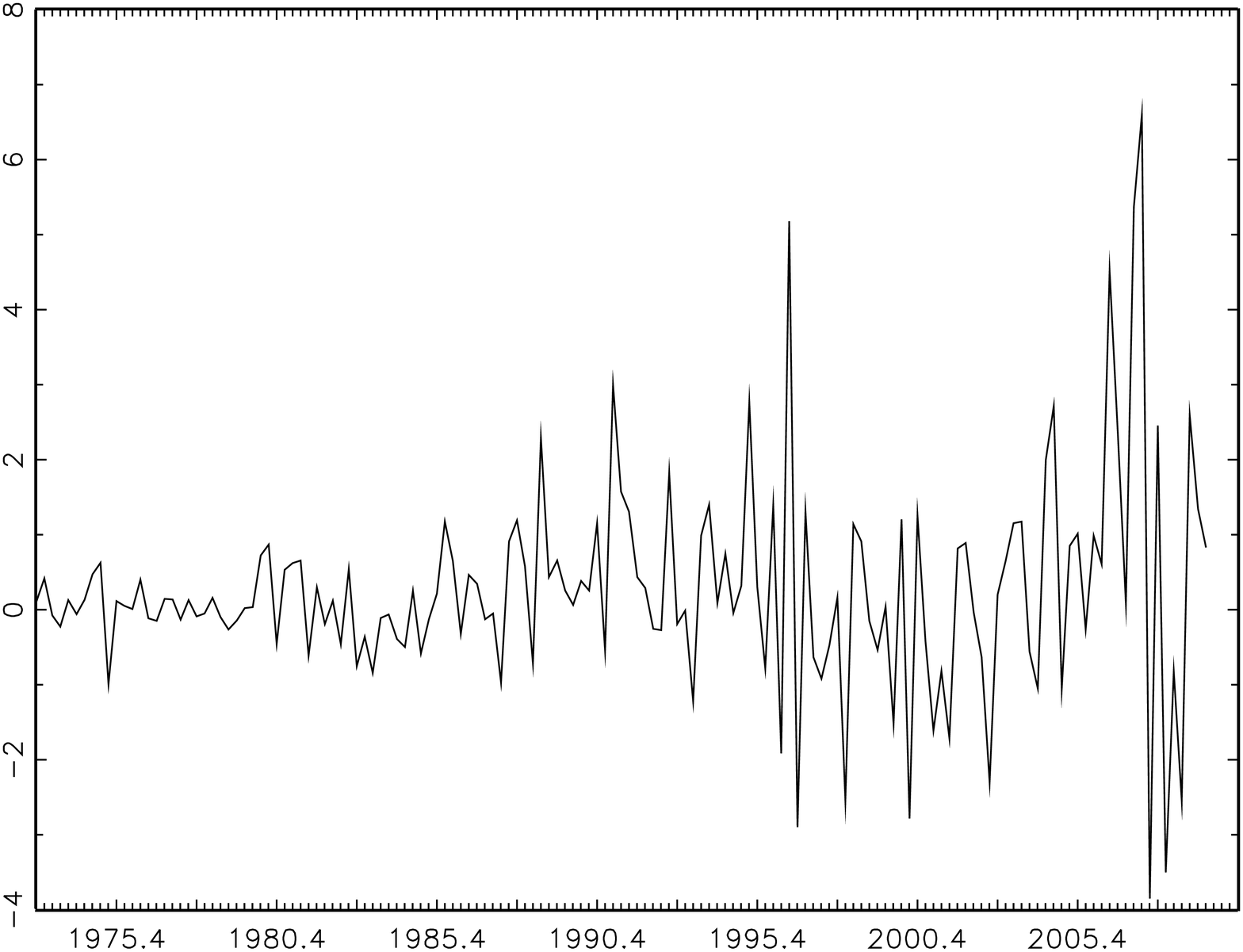} \caption{\label{datadiff}
{\footnotesize The differences of the balance on merchandise trade (on the left) and of the balance on services for the U.S. (on the right).}}
\end{figure}

\begin{figure}[h]\!\!\!\!\!\!\!\!\!\!
\vspace*{4.8cm} %\hspace*{8.5cm}$\tau_1$
%\hspace*{11.7cm}$\sigma_{21}$ \vspace*{-5.3 cm}

%\vspace*{4.3cm}\hspace*{2.15cm}$\tau_1$ \hspace*{2.3cm}$\sigma_{21}$ %\hspace*{3.8cm}$\tau_1$
%\vspace*{0.2 cm}

\protect \includegraphics{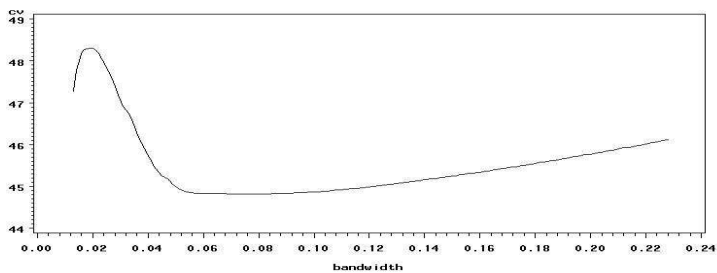}  \caption{\label{crossval}
{\footnotesize The cross validation score (CV) for the ALS estimation of the VAR(1) model for the differences of the balance on merchandise trade and on services in the U.S..}}
\end{figure}

%\footnote{www.research.stlouisfed.org.}
In our VAR system the first component corresponds to the balance on merchandise trade and the second corresponds to the balance on services trade.
From Figure \ref{data} it seems that the series have a random walk behavior. We applied the approach
of unit root testing proposed by Beare (2008) in presence of non constant volatility using the Augmented
Dickey-Fuller (ADF) test for each series. The ADF statistic is $0.72$ for the merchandise trade balance data
and is $-0.15$ for the services balance data. These statistics are greater than the 5\% critical value $-1.94$
of the ADF test, so that the stability hypothesis have to be rejected for the studied series. Furthermore we also
applied the Kolmogorov-Smirnov (KS) type test for homoscedasticity considered by Cavaliere and Taylor (2008, Theorem 3) for each series.
The KS statistic is $3.05$ for the merchandise trade balance data and is $7.62$ for the services balance data. These statistics are greater than the 5\% critical value $1.36$, so that the homoscedasticity hypothesis has to be rejected for our series. Therefore we consider the first differences of the series to get stable processes, so that the evolutions of the U.S. balance data are studied in the sequel. From Figure \ref{datadiff} we see that the first differences of the series are stable but have a non constant volatility. We adjusted a VAR(1) model to capture the linear dynamics of the series. The ALS and OLS estimators are given in Table \ref{estimates}. The standard deviations into brackets are computed using the results (\ref{res1}) and (\ref{res2}). In accordance with Patilea and Raïssi (2010), we find that the ALS estimation method seems better estimate the autoregressive parameters than the OLS estimation method, in the sense that the standard deviations of the ALS estimators are smaller than those of the OLS estimators. The bandwidth we use for the ALS estimation, $b=7.67\times10^{-2}$, is selected by cross-validation in a given range and using 200 grid points (Figure \ref{crossval}).

\begin{figure}[h]\!\!\!\!\!\!\!\!\!\!
\vspace*{4.8cm} %\hspace*{8.5cm}$\tau_1$
%\hspace*{11.7cm}$\sigma_{21}$ \vspace*{-5.3 cm}

%\vspace*{4.3cm}\hspace*{2.15cm}$\tau_1$ \hspace*{2.3cm}$\sigma_{21}$ %\hspace*{3.8cm}$\tau_1$
%\vspace*{0.2 cm}

\protect \includegraphics{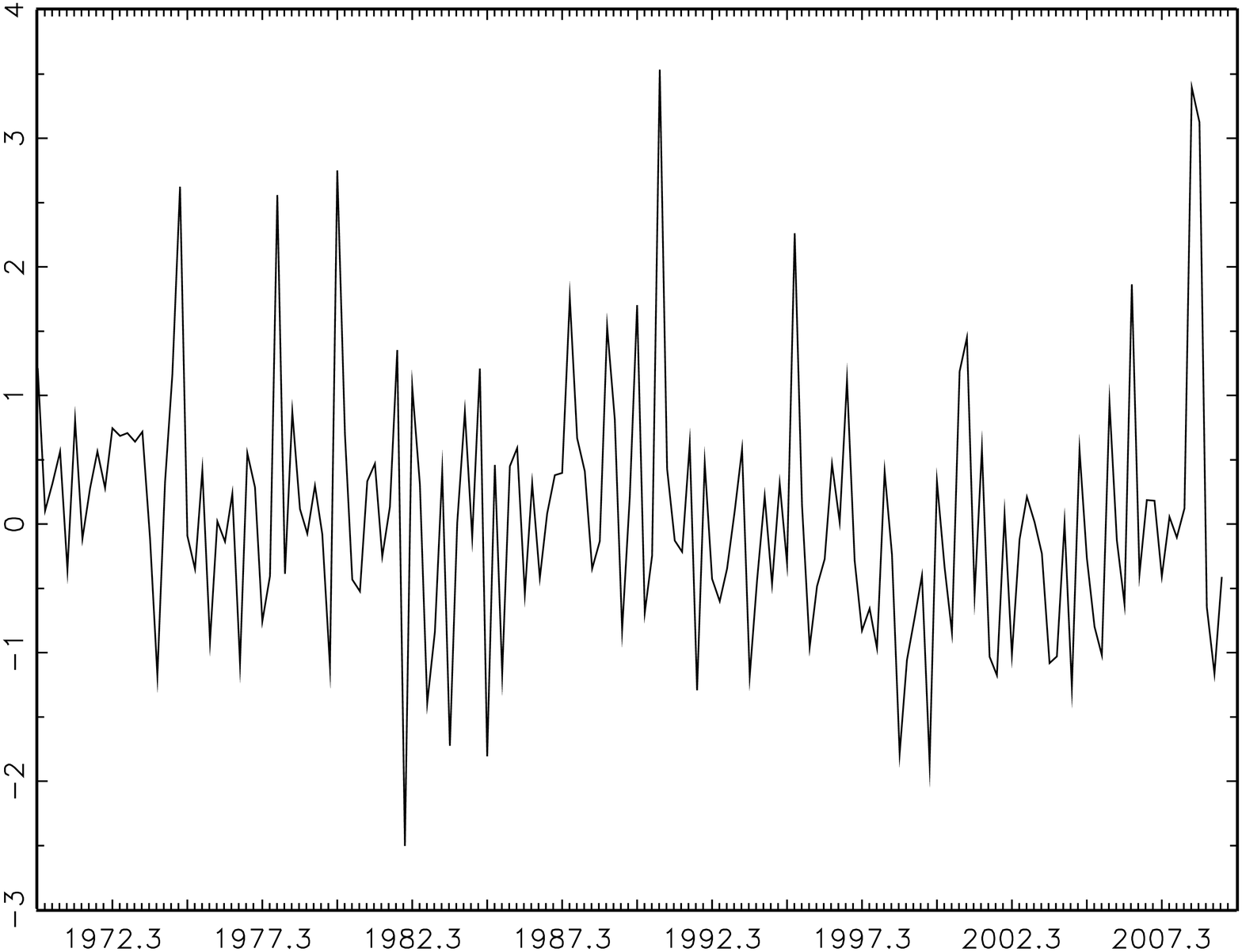} \protect \includegraphics{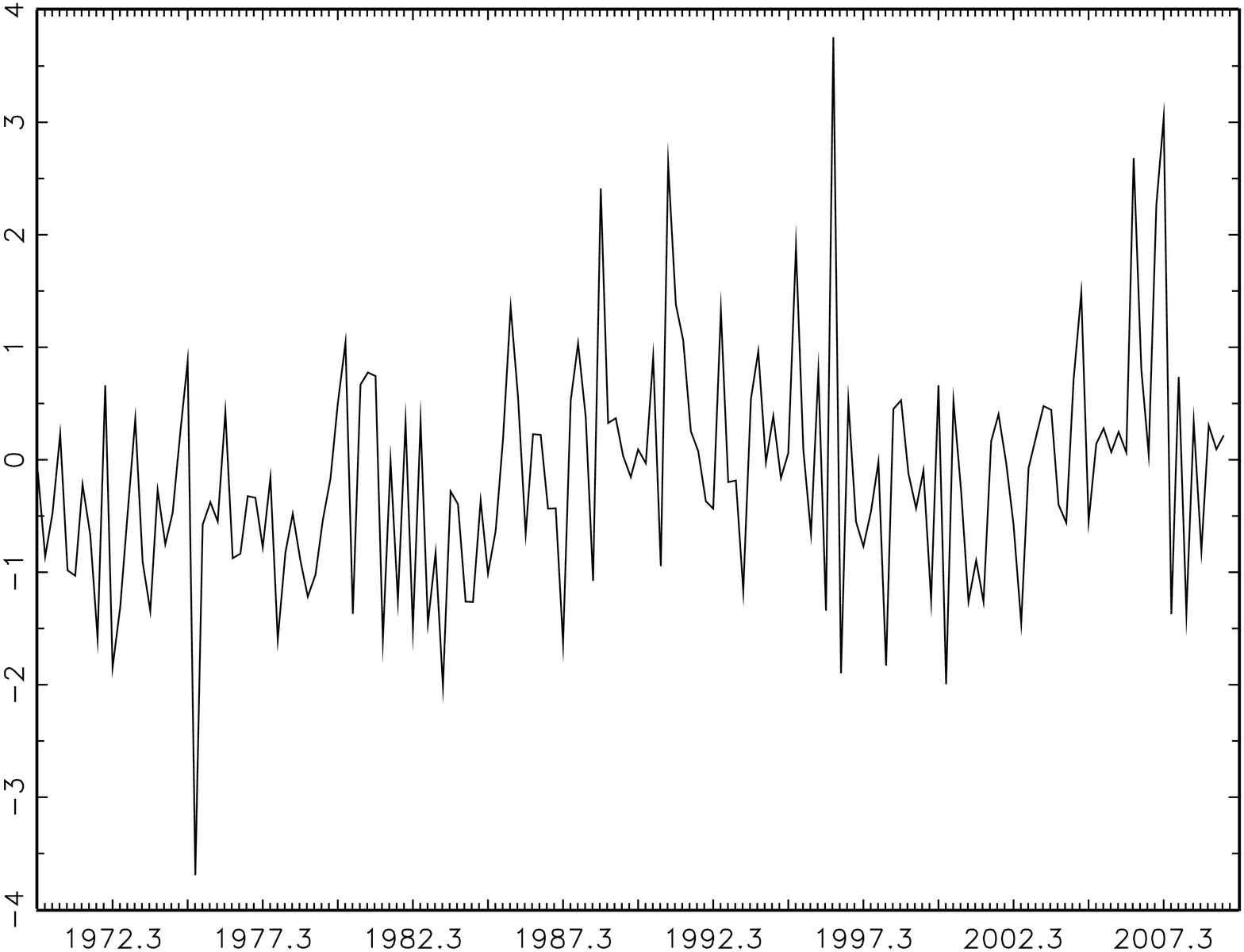} \caption{\label{residals}
{\footnotesize The ALS residuals of a VAR(1) for the differences of the balance on merchandise trade and on services in the U.S.. The first component of the ALS residuals is on the left and the second is on the right.}}
\end{figure}

\begin{figure}[h]\!\!\!\!\!\!\!\!\!\!
\vspace*{4.8cm} %\hspace*{8.5cm}$\tau_1$
%\hspace*{11.7cm}$\sigma_{21}$ \vspace*{-5.3 cm}

%\vspace*{4.3cm}\hspace*{2.15cm}$\tau_1$ \hspace*{2.3cm}$\sigma_{21}$ %\hspace*{3.8cm}$\tau_1$
%\vspace*{0.2 cm}

\protect \includegraphics{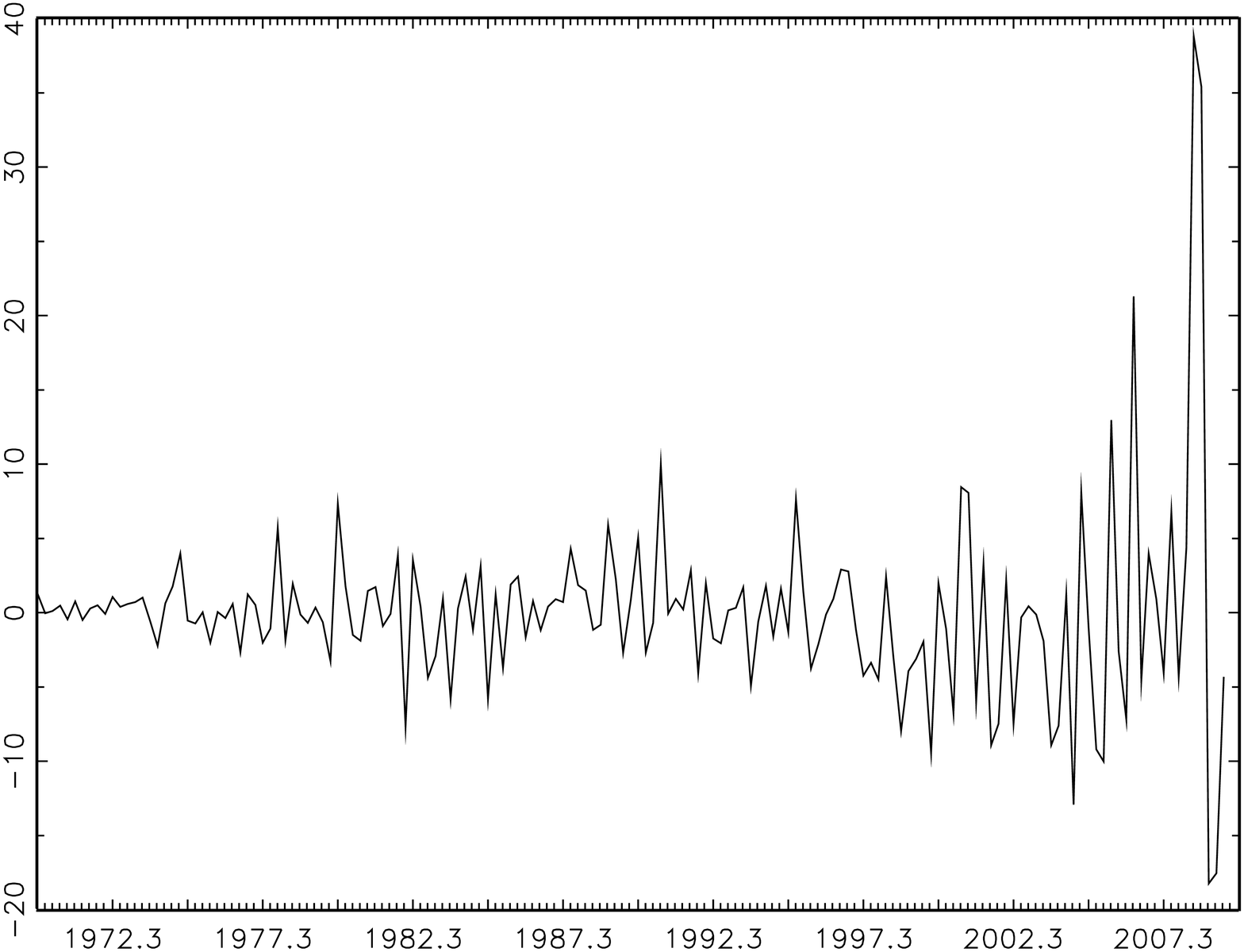} \protect \includegraphics{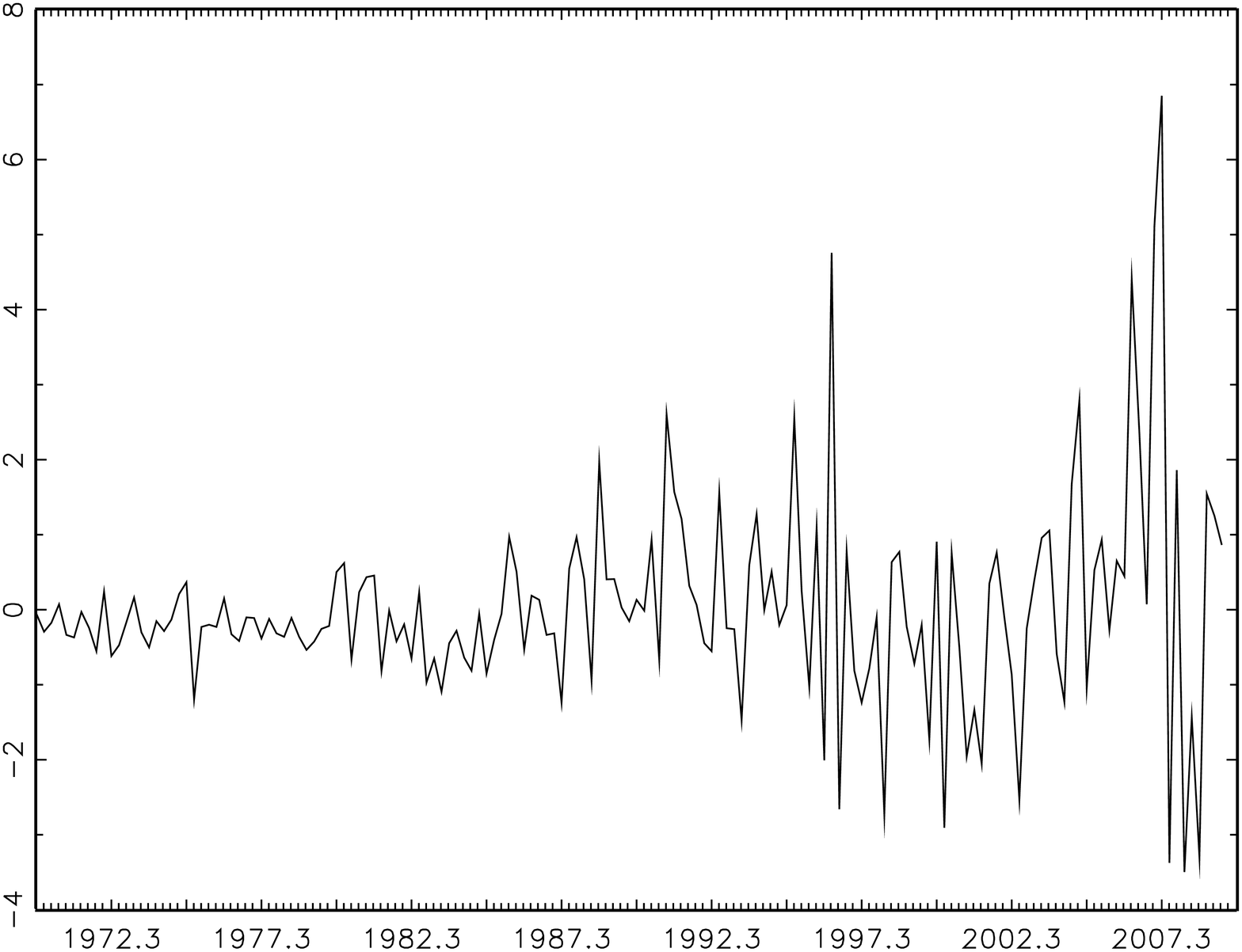} \caption{\label{residols}
{\footnotesize The same as in Figure \ref{residals} but for the OLS residuals.}}
\end{figure}

\begin{figure}[h]\!\!\!\!\!\!\!\!\!\!
%\vspace*{0.5cm} \hspace*{0.3cm}$\hat{R}_{OLS}^{21}(h)$
%\hspace*{5.2cm}$\hat{R}_{OLS}^{12}(h)$

\vspace*{2.1cm} \hspace*{6.3cm}$h$ \hspace*{6.0cm}$h$
\vspace*{1.4cm}

\protect \includegraphics{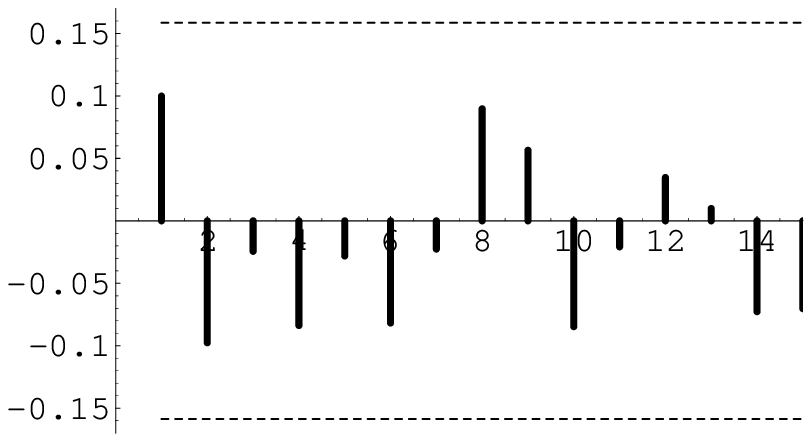} \protect \includegraphics{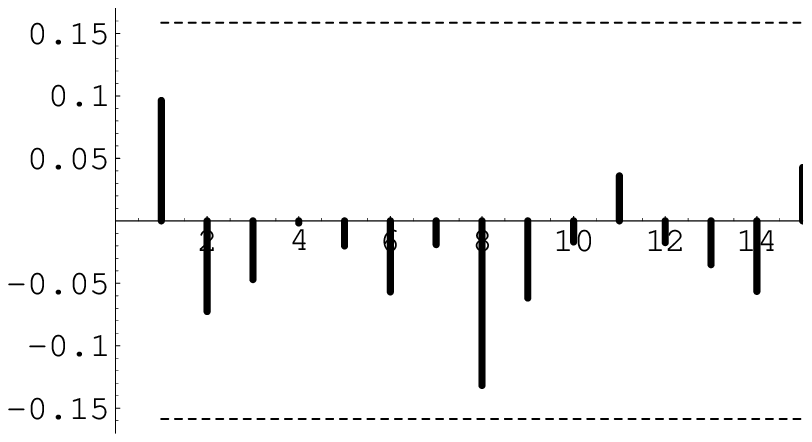} \caption{\label{autocovalscar}
{\footnotesize The balance data for the U.S.: the autocorrelations of the squares of the first component of the ALS residuals (on the left) and of the second component of the ALS residual (on the right).}}
\end{figure}

\begin{figure}[h]\!\!\!\!\!\!\!\!\!\!
\vspace*{5.6cm} %\hspace*{8.5cm}$\tau_1$
%\hspace*{11.7cm}$\sigma_{21}$ \vspace*{-5.3 cm}

%\vspace*{4.3cm}\hspace*{2.15cm}$\tau_1$ \hspace*{2.3cm}$\sigma_{21}$ %\hspace*{3.8cm}$\tau_1$
%\vspace*{0.2 cm}

\protect \includegraphics{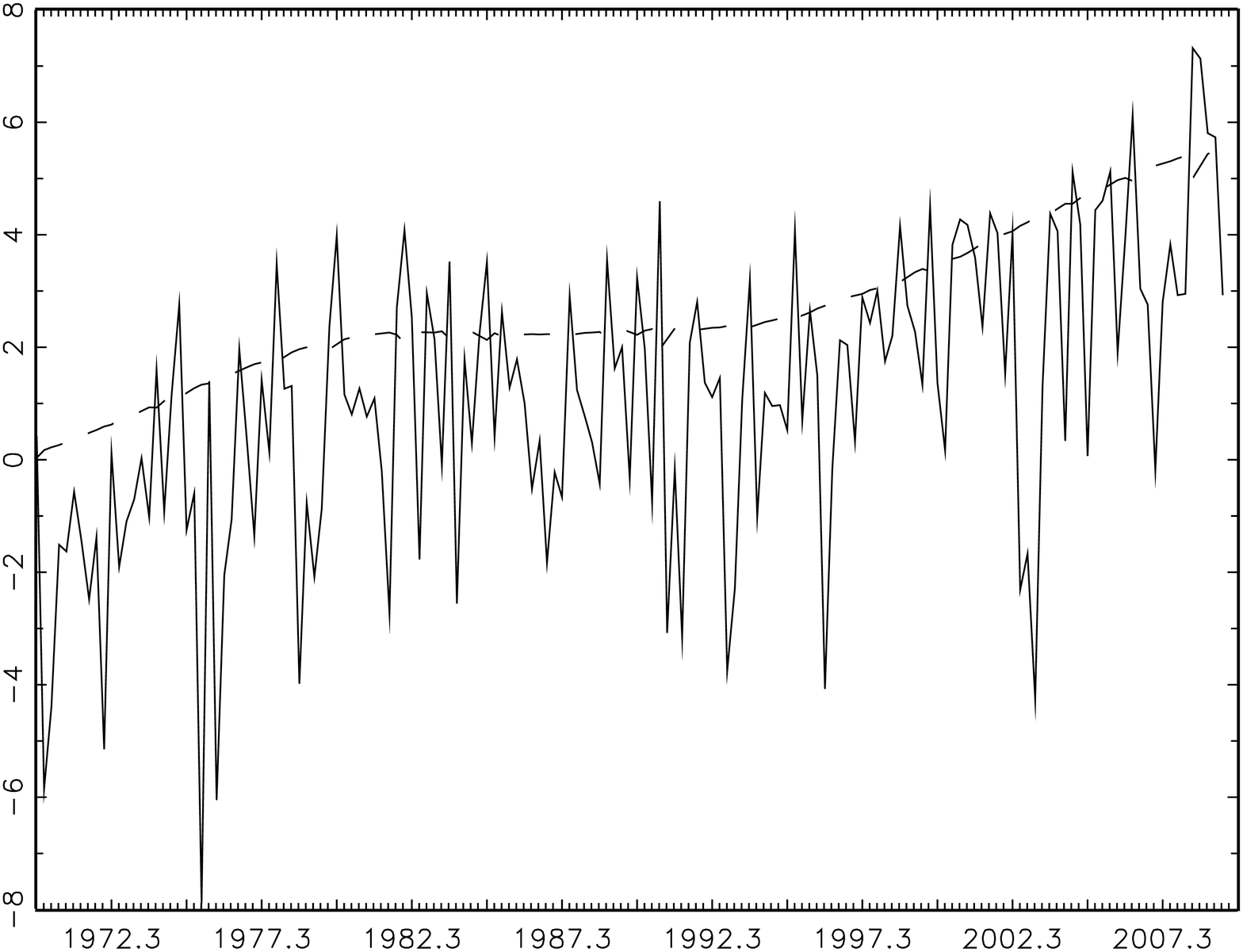} \protect \includegraphics{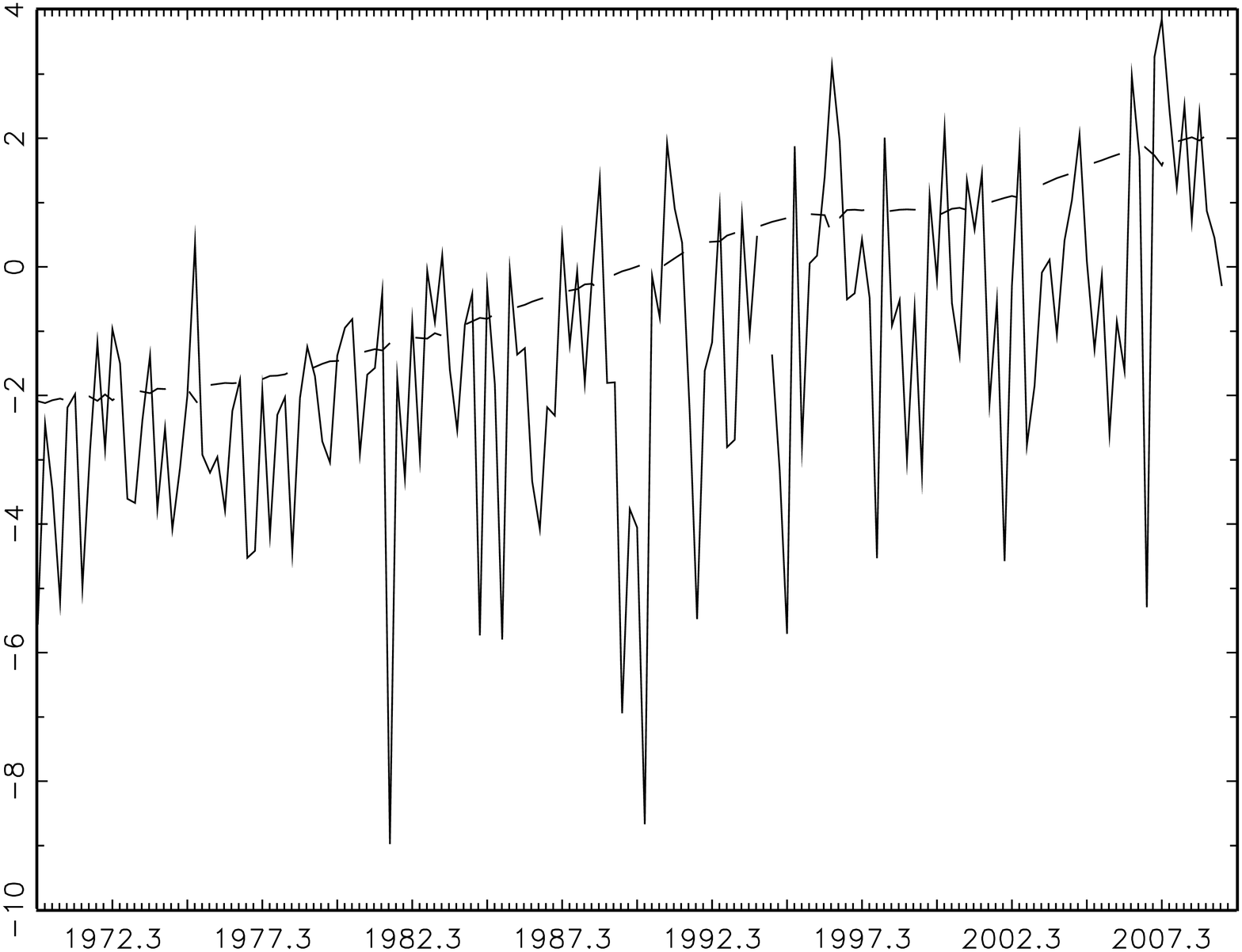} \caption{\label{varestim}
{\footnotesize The balance data for the U.S.: the logarithms of the $\hat{u}_{1t}^2$'s (full line) and the logarithms of the non parametric estimates of Var($u_{1t}$) (dotted line) on the left and the same for the $\hat{u}_{2t}^2$'s and Var($u_{2t}$) on the right.}}
\end{figure}

\begin{figure}[h]\!\!\!\!\!\!\!\!\!\!
\vspace*{4.4cm} %\hspace*{8.5cm}$\tau_1$
%\hspace*{11.7cm}$\sigma_{21}$ \vspace*{-5.3 cm}

%\vspace*{4.3cm}\hspace*{2.15cm}$\tau_1$ \hspace*{2.3cm}$\sigma_{21}$ %\hspace*{3.8cm}$\tau_1$
%\vspace*{0.2 cm}

\protect \includegraphics{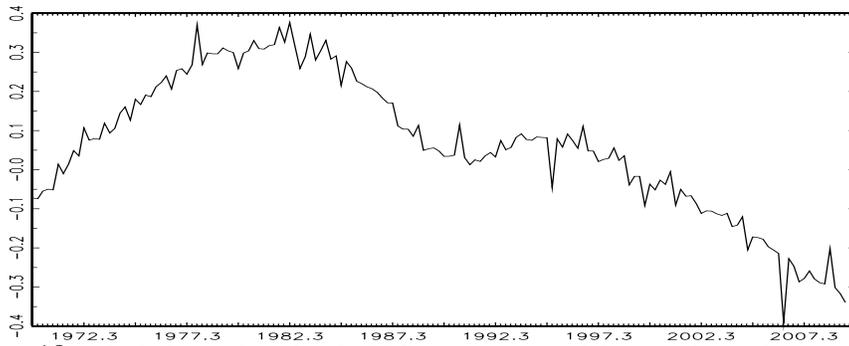}  \caption{\label{covestim}
{\footnotesize The balance data for the U.S.: estimation of the correlation between of the components of the error process.}}
\end{figure}

A quick inspection of Figures \ref{residals} and \ref{residols} suggests that the OLS residuals
have a varying volatility, while the stationarity of the ALS residuals is plausible.
Thus it seems that the volatility of the error process $(u_t)$ is satisfactorily estimated
by the adaptive method. Now we examine the possible presence conditional heteroscedasticity in the ALS residuals. From Figure \ref{autocovalscar}
we see that the autocorrelations of the squared ALS residuals components
are not significant.
 %and therefore the ALS residuals seem not conditionally heteroscedastic.
In addition we also considered the ARCH-LM test of Engle (1982) with different lags in Table \ref{archLM} for testing the presence of ARCH effects in the ALS residuals. It appears that the null hypothesis of conditional homoscedasticity cannot be rejected at the level 5\%. These diagnostics give some evidence that the conditional homoscedasticity assumption on $(\epsilon_t)$ is plausible in our case. To analyze the changes of the variance of the OLS error terms, we plotted the estimated variances and cross correlation of the components of the error process in Figures \ref{varestim} and \ref{covestim}. It appears from Figure \ref{varestim} that the variance of the first component of the residuals does not vary much until the end of the 90's and then increase. Similarly the volatility of the second component of the residuals does not vary much until the end of the 80's and then increase. From Figure \ref{covestim} we also note that the correlation between the components of the innovations seems to be positive until the beginning of the 90's and then become negative.

\begin{figure}[h]\!\!\!\!\!\!\!\!\!\!
\vspace*{0.5cm} \hspace*{0.3cm}$\hat{R}_{ALS}^{11}(h)$
\hspace*{5.2cm}$\hat{R}_{ALS}^{22}(h)$

\vspace*{1.5cm} \hspace*{6.3cm}$h$ \hspace*{6.0cm}$h$
\vspace*{1.4cm}

\protect \includegraphics{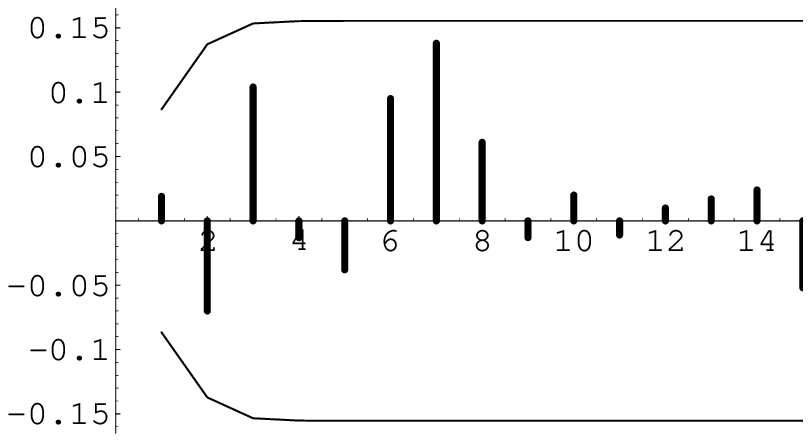} \protect \includegraphics{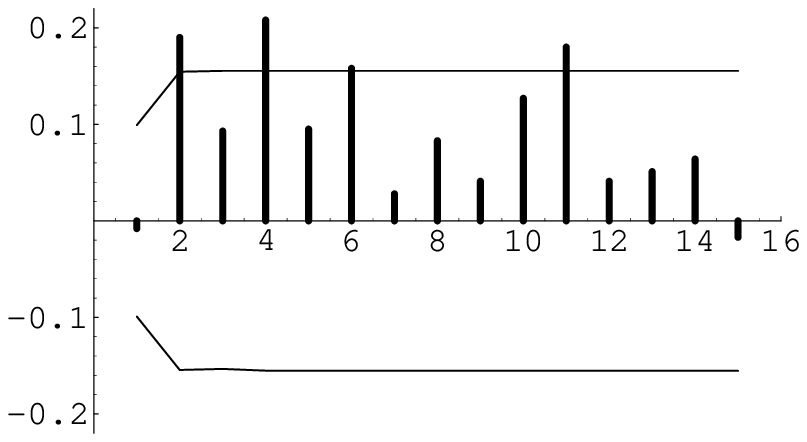} \caption{\label{autocorrals}
{\footnotesize The balance data for the U.S.: the ALS residual autocorrelations $\hat{R}_{ALS}^{11}(h)$ (on the left) and $\hat{R}_{ALS}^{22}(h)$ (on the right), with obvious notations. The 95\% confidence bounds are obtained using (\ref{gamgls}) and (\ref{equivalent}).}}
\end{figure}

\begin{figure}[h]\!\!\!\!\!\!\!\!\!\!
\vspace*{0.5cm} \hspace*{0.3cm}$\hat{R}_{ALS}^{21}(h)$
\hspace*{5.2cm}$\hat{R}_{ALS}^{12}(h)$

\vspace*{1.5cm} \hspace*{6.3cm}$h$ \hspace*{6.0cm}$h$
\vspace*{1.4cm}

\protect \includegraphics{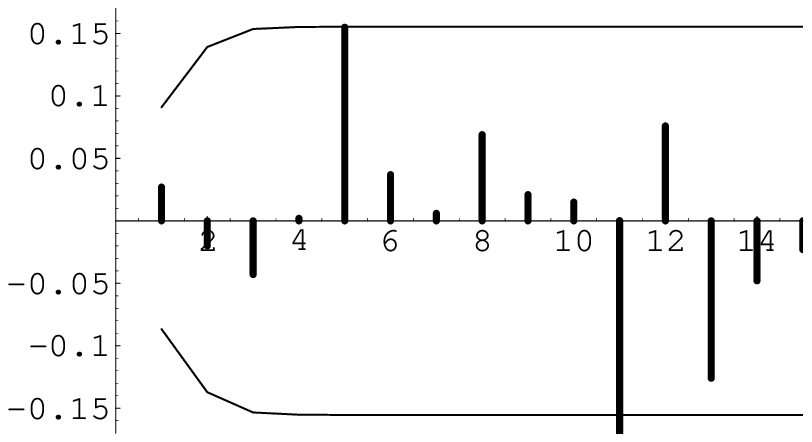} \protect \includegraphics{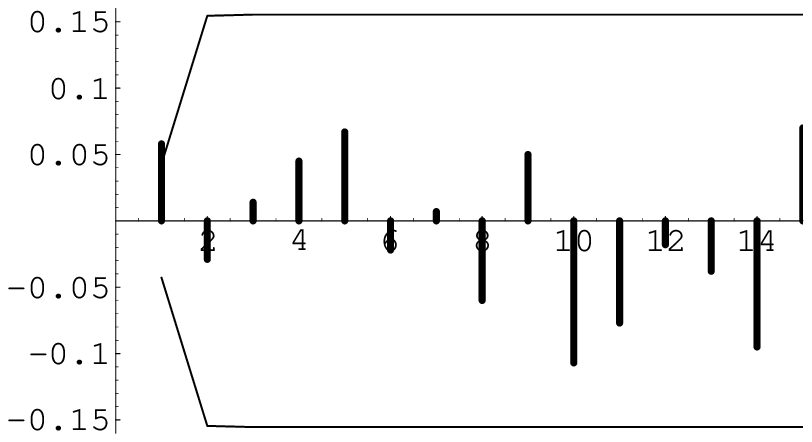} \caption{\label{autocovals}
{\footnotesize The same as in Figure \ref{autocorrals} but for $\hat{R}_{ALS}^{21}(h)$ (on the left) and $\hat{R}_{ALS}^{12}(h)$ (on the right).}}
\end{figure}

\begin{figure}[h]\!\!\!\!\!\!\!\!\!\!
\vspace*{0.5cm} \hspace*{0.3cm}$\hat{R}_{OLS}^{11}(h)$
\hspace*{5.2cm}$\hat{R}_{OLS}^{22}(h)$

\vspace*{1.5cm} \hspace*{6.3cm}$h$ \hspace*{6.0cm}$h$
\vspace*{1.4cm}

\protect \includegraphics{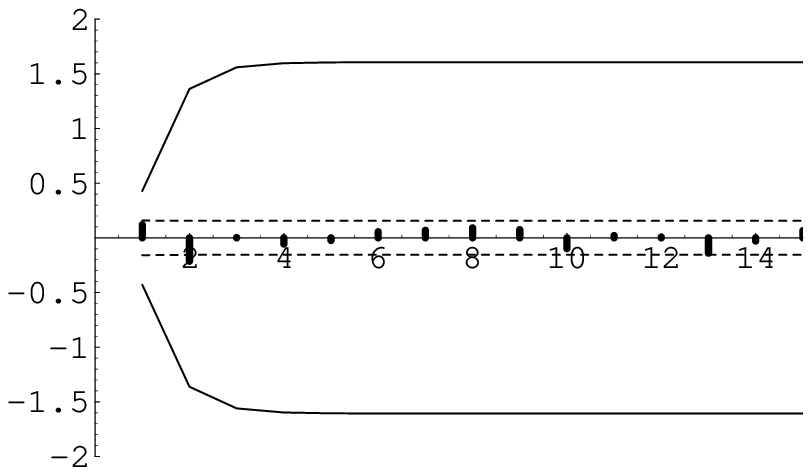} \protect \includegraphics{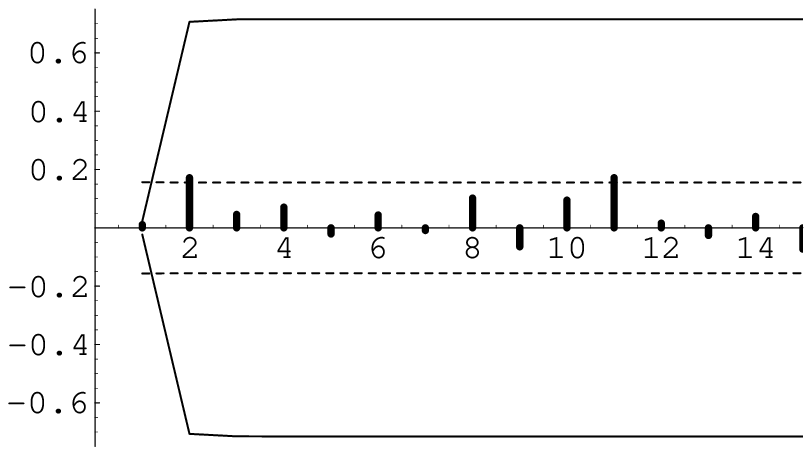} \caption{\label{autocorrols}
{\footnotesize The balance data for the U.S.: the OLS residual autocorrelations $\hat{R}_{OLS}^{11}(h)$ (on the left) and $\hat{R}_{OLS}^{22}(h)$ (on the right). The full lines 95\% confidence bounds are obtained using (\ref{rhools}). The dotted lines 95\% confidence bounds are obtained using the standard result (\ref{bruges}).}}
\end{figure}

\begin{figure}[h]\!\!\!\!\!\!\!\!\!\!
\vspace*{0.5cm} \hspace*{0.3cm}$\hat{R}_{OLS}^{21}(h)$
\hspace*{5.2cm}$\hat{R}_{OLS}^{12}(h)$

\vspace*{1.5cm} \hspace*{6.3cm}$h$ \hspace*{6.0cm}$h$
\vspace*{1.4cm}

\protect \includegraphics{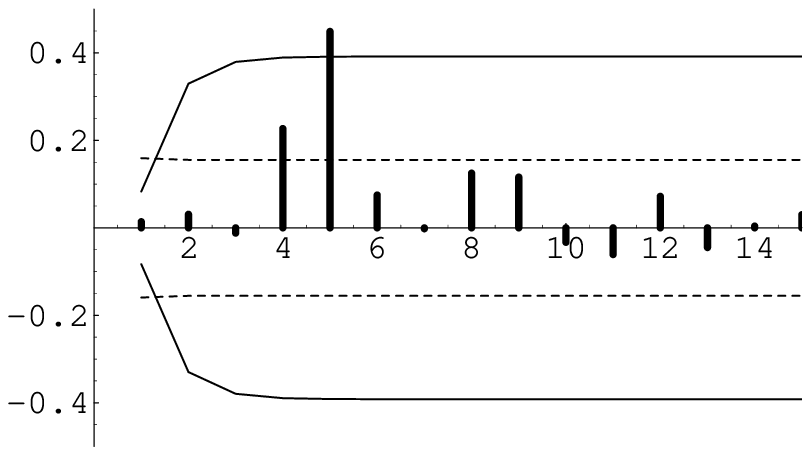} \protect \includegraphics{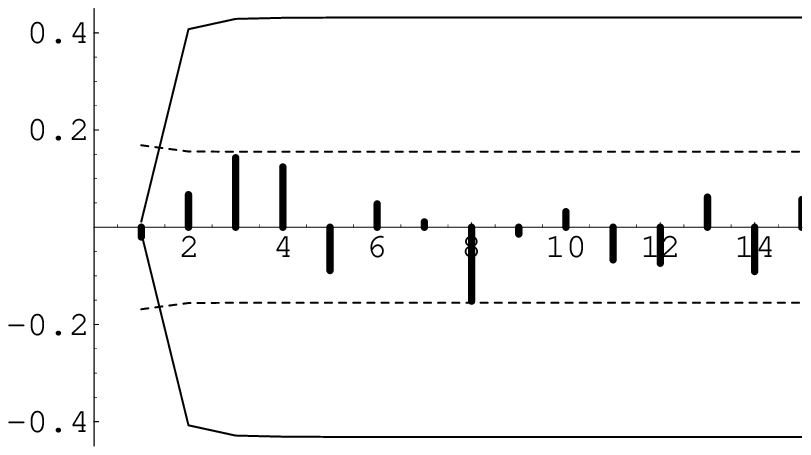} \caption{\label{autocovols}
{\footnotesize The same as in Figure \ref{autocorrals} but for $\hat{R}_{OLS}^{21}(h)$ (on the left) and $\hat{R}_{OLS}^{12}(h)$ (on the right).}}
\end{figure}

Now we turn to the check of the goodness-of-fit of the VAR(1) model adjusted to the first differences of the series. To illustrate the results of Proposition \ref{propostun} we plotted the ALS residual autocorrelations in Figures \ref{autocorrals} and \ref{autocovals}, and the OLS residual autocorrelations in Figures \ref{autocorrols} and \ref{autocovols}, where we denote

$$\hat{R}_{OLS}^{ij}(h)=\frac{T^{-1}\sum_{t=h+1}^T\hat{u}_{i\:t}\hat{u}_{j\:t-h}}
{\hat{\sigma}_u(i)\hat{\sigma}_u(j)}\quad\mbox{and}\quad \hat{R}_{ALS}^{ij}(h)=\frac{T^{-1}\sum_{t=h+1}^T
\check{\epsilon}_{i\:t}\check{\epsilon}_{j\:t-h}}{\check{\sigma}_{\epsilon}(i)\check{\sigma}_{\epsilon}(j)}.$$
The ALS 95\% confidence bounds obtained  using (\ref{rhogls}) and (\ref{equivalent})
are displayed in Figures \ref{autocorrals} and \ref{autocovals}. In Figures \ref{autocorrols} and
\ref{autocovols} the standard 95\% confidence bounds obtained using (\ref{bruges}) and the OLS 95 \% confidence
bounds obtained using (\ref{rhools}) are plotted. We can remark that the ALS residual autocorrelations are
inside or not much larger than the ALS significance limits. %in Figures \ref{autocorrals} and \ref{autocovals}.
A similar comment can be made for the OLS residual autocorrelations when compared to the OLS significance limits. %in Figures \ref{autocorrols} and \ref{autocovols}.
However we found that the OLS significance limits can be quite different from the standard significance
limits. This can be explained by the presence of unconditional volatility in the analyzed series.
In particular we note that the $\hat{R}_{OLS}^{21}(5)$ is far from the standard confidence bounds. We also
apply the different portmanteau tests considered in this paper for testing if the errors are uncorrelated. The test statistics and
$p$-values of the tests are displayed in Tables \ref{pvalues} and \ref{stats}. It appears that the
$p$-values of the standard tests are very small. %in Table \ref{pvalues}.
Therefore the standard tests
clearly reject the null hypothesis.
%of the adequacy of the VAR(1) model.
We also remark that the $p$-values
of the modified tests based on the OLS estimation and of the adaptive tests are far from zero. Thus in view of
the tests introduced in this paper the null hypothesis is not rejected. These contradictory results can be
explained by the fact that we found that the distribution in (\ref{truedist}) is very different from the
$\chi^2_{d^2(m-p)}$ standard distribution. For instance we obtained $\sup_{i\in\{1,\dots,d^2m\}}\left\{\hat{\delta}_i^{ols}\right\}=11.18$ for $m=15$ in our case. Our findings may be viewed as a consequence of the presence of unconditional heteroscedasticity in the data. Since the theoretical basis of the standard tests do not include the case of stable processes with non constant volatility, we can suspect that the results of the standard tests are not reliable. Therefore we can draw the conclusion that the practitioner is likely to select a too large autoregressive order in our case when using the standard tools for checking the adequacy of the VAR model. From Table \ref{stats} we see that the OLS and ALS statistics are quite different.  We also noted that the weights (not reported here) of the sums in (\ref{truedist}) and in (\ref{truedistadapt}) are quite different for our example.

\subsection{VAR modeling of the U.S. energy-transportation data}

\begin{figure}[h]\!\!\!\!\!\!\!\!\!\!
\vspace*{4.9cm} %\hspace*{8.5cm}$\tau_1$
%\hspace*{11.7cm}$\sigma_{21}$ \vspace*{-5.3 cm}

%\vspace*{4.3cm}\hspace*{2.15cm}$\tau_1$ \hspace*{2.3cm}$\sigma_{21}$ %\hspace*{3.8cm}$\tau_1$
%\vspace*{0.2 cm}

\protect \includegraphics{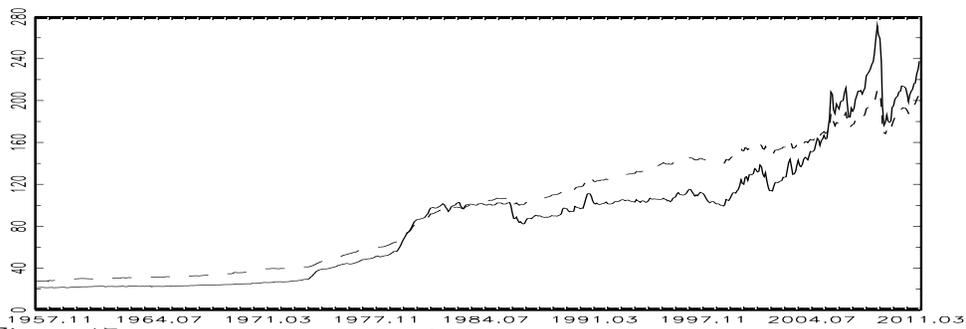} \caption{\label{datak}
{\footnotesize The energy price index (full line) and the transportation price index (dotted line) in the U.S. from 1/1/1957 to 2/1/2011, T=648. Data source: The research division of the federal reserve bank of Saint Louis, www.research.stlouis.org.}}
\end{figure}

\begin{figure}[h]\!\!\!\!\!\!\!\!\!\!
\vspace*{4.8cm} %\hspace*{8.5cm}$\tau_1$
%\hspace*{11.7cm}$\sigma_{21}$ \vspace*{-5.3 cm}

%\vspace*{4.3cm}\hspace*{2.15cm}$\tau_1$ \hspace*{2.3cm}$\sigma_{21}$ %\hspace*{3.8cm}$\tau_1$
%\vspace*{0.2 cm}

\protect \includegraphics{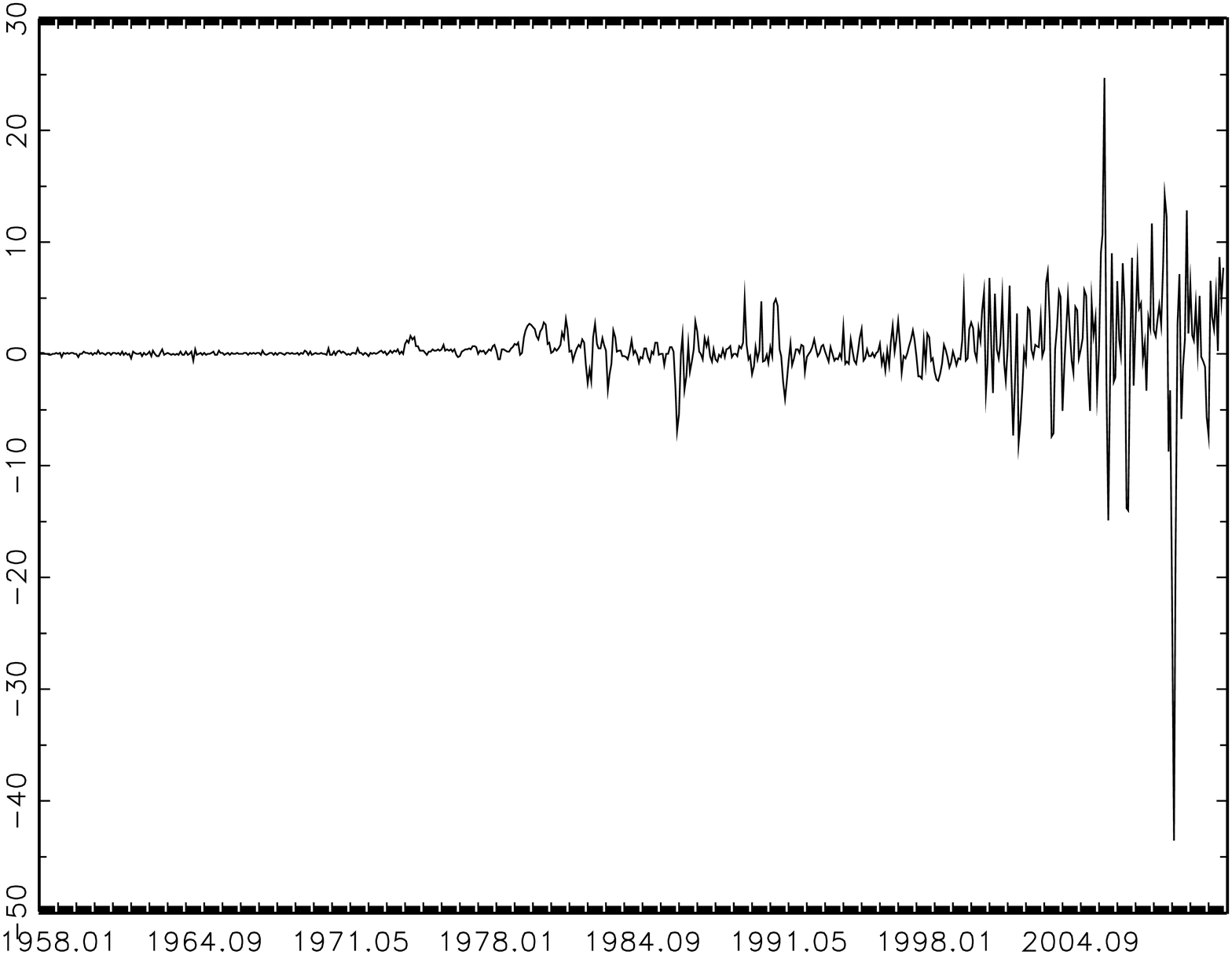} \protect \includegraphics{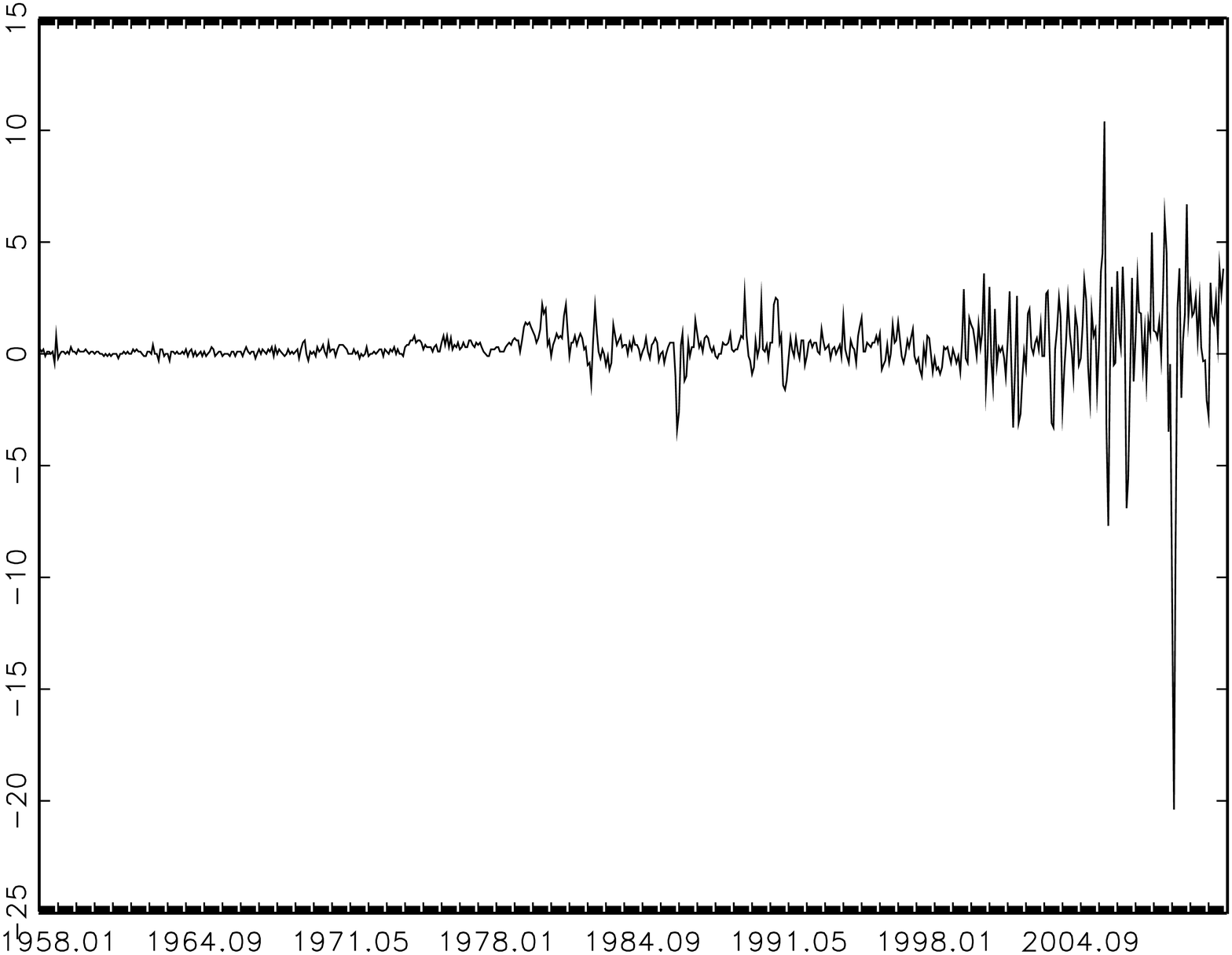} \caption{\label{datadiffk}
{\footnotesize The differences of the energy price (on the left) and of the transportation price indexes for the U.S. (on the right).}}
\end{figure}

\begin{figure}[h]\!\!\!\!\!\!\!\!\!\!
\vspace*{4.3cm} %\hspace*{8.5cm}$\tau_1$
%\hspace*{11.7cm}$\sigma_{21}$ \vspace*{-5.3 cm}

%\vspace*{4.3cm}\hspace*{2.15cm}$\tau_1$ \hspace*{2.3cm}$\sigma_{21}$ %\hspace*{3.8cm}$\tau_1$
%\vspace*{0.2 cm}

\protect \includegraphics{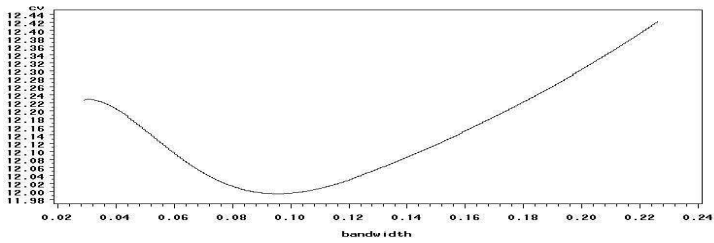}  \caption{\label{crossvalk}
{\footnotesize The cross validation score (CV) for the ALS estimation of the VAR(4) model for the differences of the energy-transportation price indexes in the U.S..}}
\end{figure}

\begin{figure}[h]\!\!\!\!\!\!\!\!\!\!
\vspace*{4.8cm} %\hspace*{8.5cm}$\tau_1$
%\hspace*{11.7cm}$\sigma_{21}$ \vspace*{-5.3 cm}

%\vspace*{4.3cm}\hspace*{2.15cm}$\tau_1$ \hspace*{2.3cm}$\sigma_{21}$ %\hspace*{3.8cm}$\tau_1$
%\vspace*{0.2 cm}

\protect \includegraphics{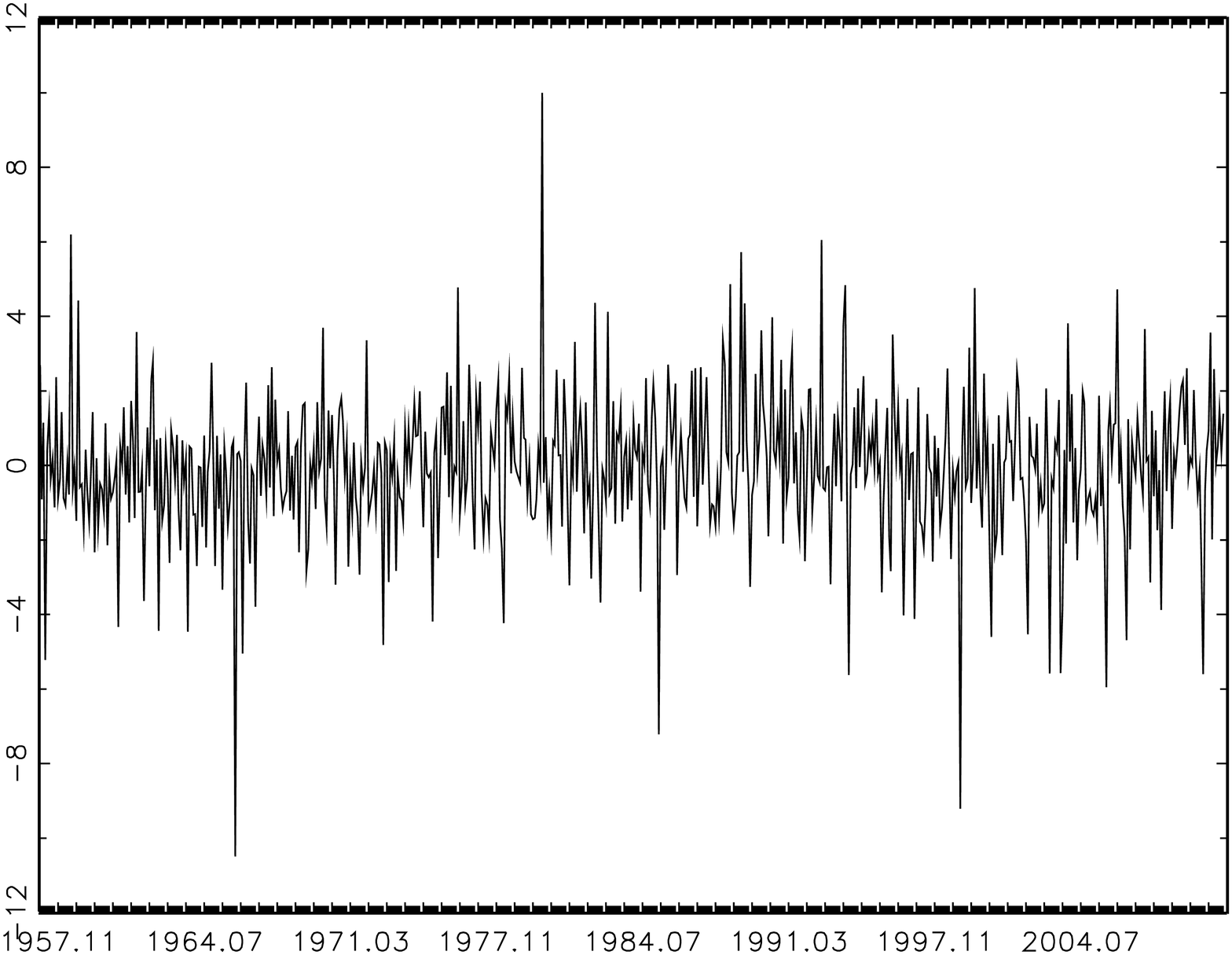} \protect \includegraphics{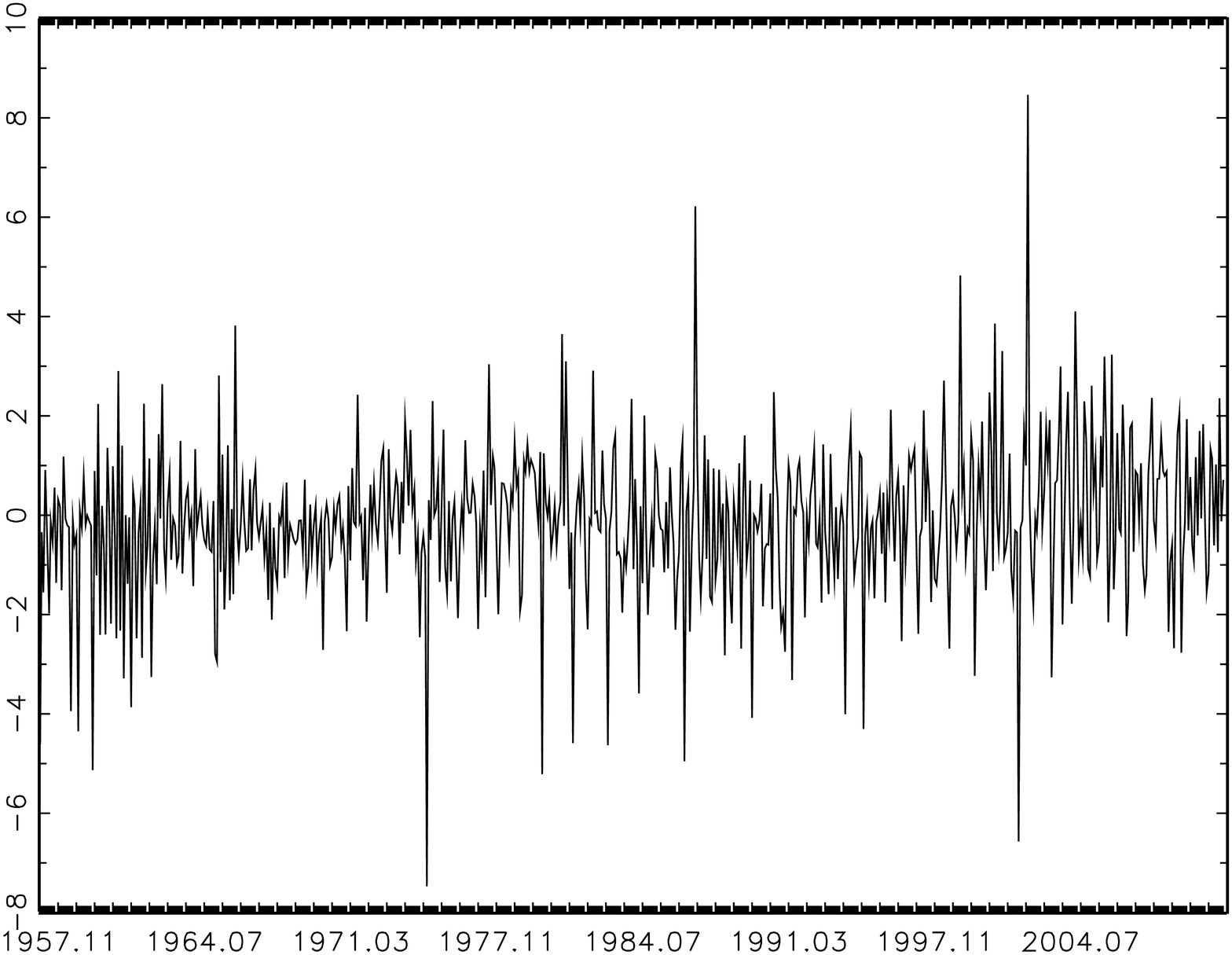} \caption{\label{residalsk}
{\footnotesize The ALS residuals of a VAR(4) model for the differences of the energy and transportation price indexes for the U.S.. The first component of the ALS residuals is on the left and the second is on the right.}}
\end{figure}

\begin{figure}[h]\!\!\!\!\!\!\!\!\!\!
\vspace*{4.8cm} %\hspace*{8.5cm}$\tau_1$
%\hspace*{11.7cm}$\sigma_{21}$ \vspace*{-5.3 cm}

%\vspace*{4.3cm}\hspace*{2.15cm}$\tau_1$ \hspace*{2.3cm}$\sigma_{21}$ %\hspace*{3.8cm}$\tau_1$
%\vspace*{0.2 cm}

\protect \includegraphics{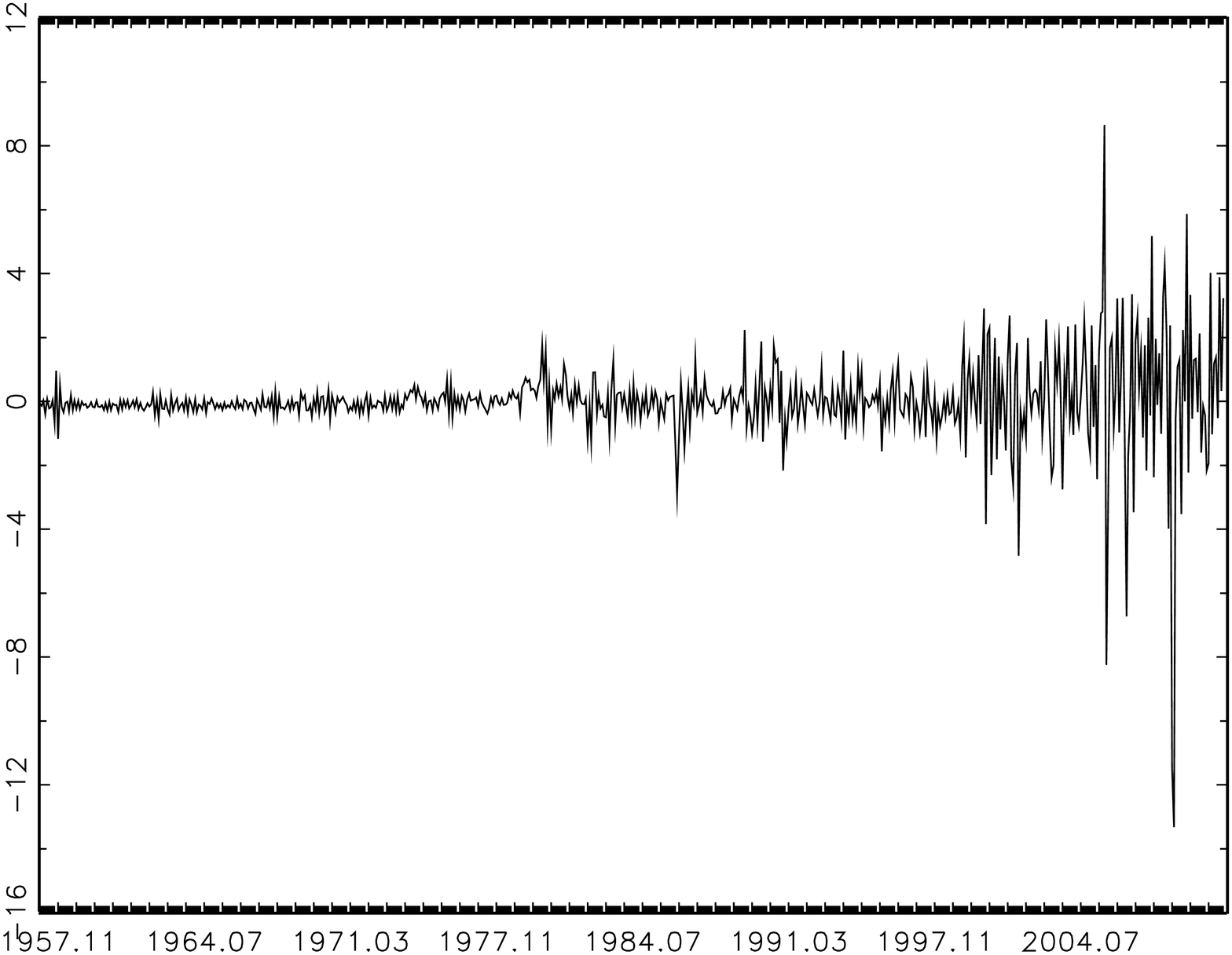} \protect \includegraphics{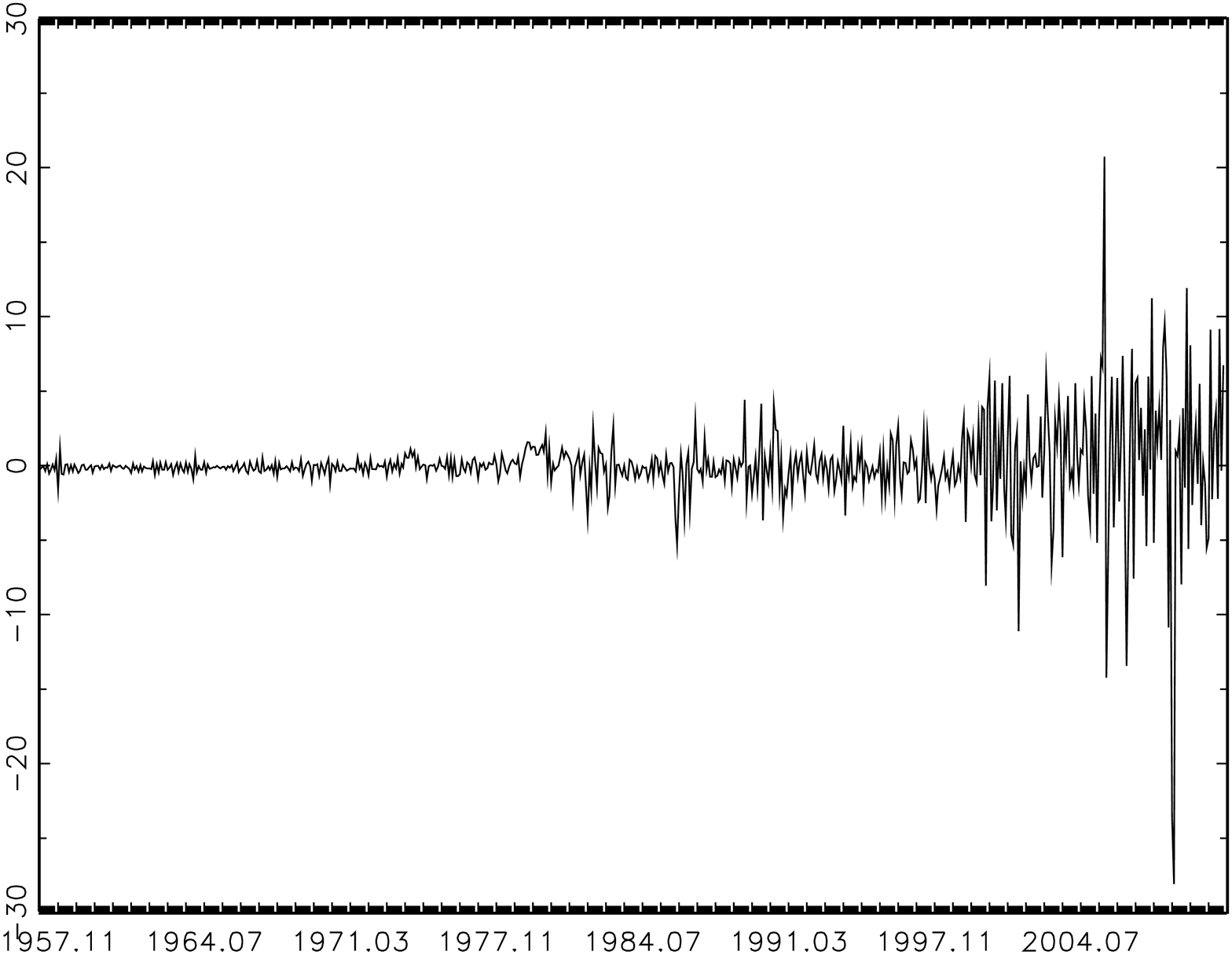} \caption{\label{residolsk}
{\footnotesize The same as in Figure \ref{residals} but for the OLS residuals.}}
\end{figure}

\begin{figure}[h]\!\!\!\!\!\!\!\!\!\!
\vspace*{4.8cm} %\hspace*{8.5cm}$\tau_1$
%\hspace*{11.7cm}$\sigma_{21}$ \vspace*{-5.3 cm}

%\vspace*{4.3cm}\hspace*{2.15cm}$\tau_1$ \hspace*{2.3cm}$\sigma_{21}$ %\hspace*{3.8cm}$\tau_1$
%\vspace*{0.2 cm}

\protect \includegraphics{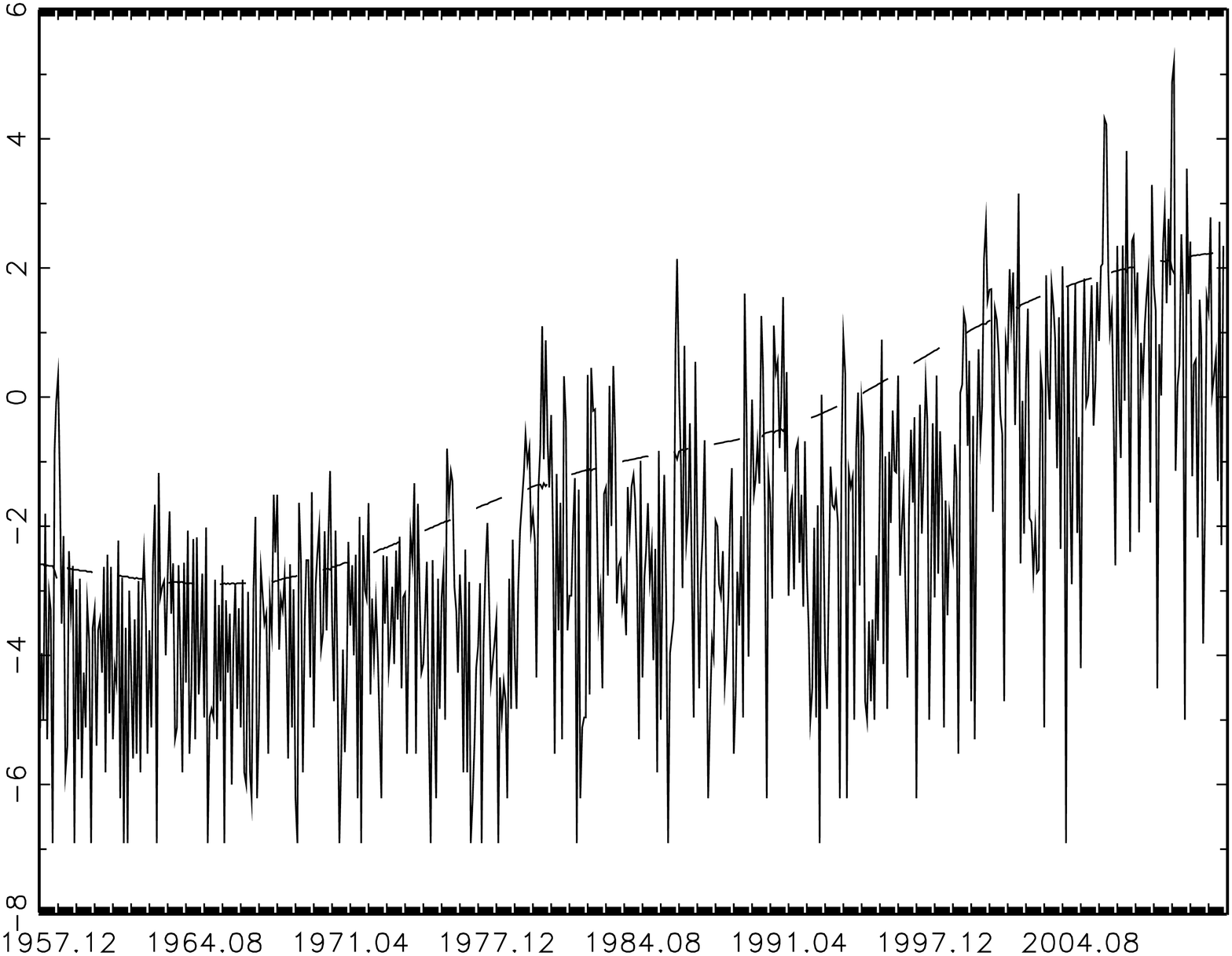} \protect \includegraphics{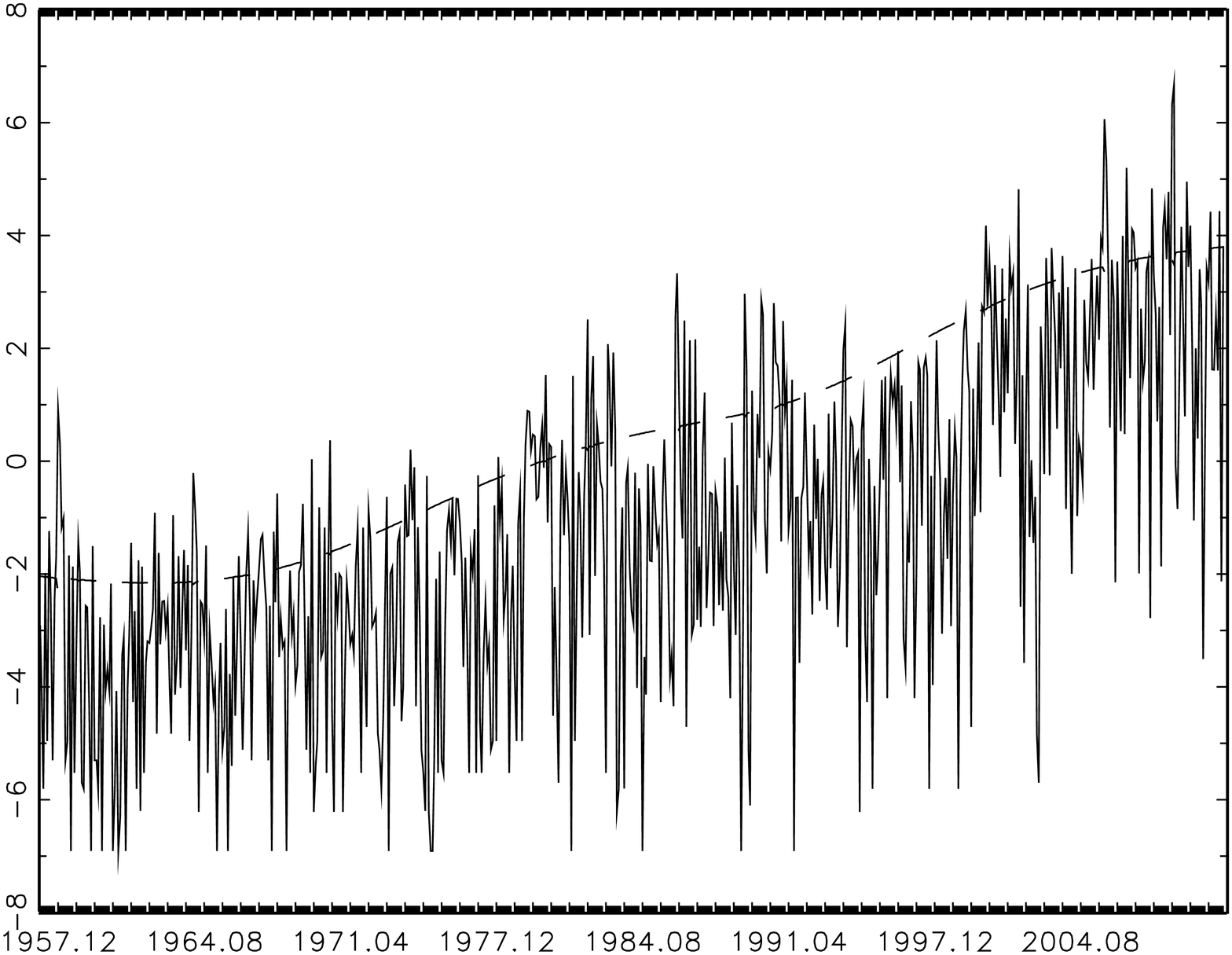} \caption{\label{varestimk}
{\footnotesize The energy-transportation data for the U.S.: the logarithms of the $\hat{u}_{1t}^2$'s (full line) and the logarithms of the non parametric estimation of Var($u_{1t}$) (dotted line) on the left and the same for the $\hat{u}_{2t}^2$'s and Var($u_{2t}$) on the right.}}
\end{figure}

In this example the first component of the VAR system corresponds to the transportation price index and the second corresponds to the energy price index.
We first briefly describe some features of the energy-transportation price indexes.
%since these features are similar to those of the balance data.
In Figure \ref{datak} we again see that the studied series seems to have a random walk behavior and then we again consider the first differences of the series. The KS type statistic is $7.05$ for the energy price index and is $6.81$ for the transportation price index, so that the homoscedasticity hypothesis has to be rejected for these series.
From Figure \ref{datadiffk} we see that the first differences of the series are stable but have a non constant volatility. We adjusted a VAR(4) model to capture the linear dynamics of the series. The results in Table \ref{estimatesk} indicate that the ALS approach is more precise than the OLS approach for the estimation of the autoregressive parameters. The bandwidth obtained by the cross-validation for the ALS estimation is $b=9.53\times10^{-2}$ (Figure \ref{crossvalk}).
From Figure \ref{residolsk} we see that the OLS residuals
seem to have a varying volatility. The stationarity of the ALS residuals is plausible (Figure \ref{residalsk}) and ARCH-LM tests (not reported here) show that the conditional homoscedasticity of the ALS residuals cannot be rejected.
From Figure \ref{varestimk} we see that the shape of the variance structure of the components of the OLS residuals are similar. More precisely it can be noted that the variance of the components of the OLS residuals is relatively low and seems constant until the beginning of the 80's. The OLS residual variance seems to switch to an other regime where the variance is increased but constant from the beginning of the 80's to the end of the 90's. The volatility of the OLS residual variance seems to increase from the end of the 90's. From Figure \ref{covestim} we also note that the components of the OLS residuals seem highly correlated.

Now we check the adequacy of the VAR(4) model. The ALS and OLS residual autocorrelations are given in Figures \ref{autocorralsk}, \ref{autocovalsk} and \ref{autocorrolsk}, \ref{autocovolsk}. The OLS residual autocorrelations with the standard bounds are given in Figures \ref{autocorrstgk} and \ref{autocovstgk}.
From Figures \ref{autocorralsk} and \ref{autocovalsk} it can be noted that the ALS residual autocorrelations are
inside or not much larger than the ALS significance limits.
The OLS confidence bounds seems not reliable since we remark that some confidence bounds (corresponding to the $\hat{R}_{OLS}^{ij}(2)$'s) are unexpectedly much larger than the OLS residual autocorrelations. From Figures \ref{autocorrstgk} and \ref{autocovstgk} it can also be noted that some of the OLS residual autocorrelations are far from the standard confidence bounds. However
considering the standard confidence bounds in presence of non constant variance can be misleading in view of our theoretical findings. In addition we also remark that some standard confidence bounds (corresponding to the $\hat{R}_{OLS}^{11}(2)$, $\hat{R}_{OLS}^{21}(2)$ and $\hat{R}_{OLS}^{22}(8)$, $\hat{R}_{OLS}^{12}(8)$) are unexpectedly far from the residual autocorrelations.
The non correlation of the residuals is tested using the different portmanteau tests considered in this paper.
The test statistics and
$p$-values of the tests are displayed in Tables \ref{pvaluesk} and \ref{statsk}.
The standard tests again lead to select complicated models since the adequacy of the VAR(4) model is rejected.
We also remark that the $p$-values
of the OLS tests are very large. In fact we found that the OLS tests do not reject the
hypothesis of the adequacy of a VAR(1) model for the studied series.
From Table \ref{weightsk} we see that some
of the estimated weights for the asymptotic distribution of the standard statistic take very large
values. It appears that the OLS method seems not able to estimate correctly the asymptotic distribution of the standard statistics and may be suspected to have a low power in this example.
Then the OLS portmanteau test seem
not reliable in this example on the contrary to the ALS portmanteau test. Indeed it can be noted that
the asymptotic distribution seems well estimated. Finally note that we found that the estimators of the $\Lambda_m^{\epsilon,\theta'}\Lambda_m^{\epsilon,\theta}$ and $\Phi_m'(\Lambda_m^{u,u})^{-1}\Phi_m$ are not invertible, so that the $\widetilde{LB}_{m}^{ALS}$ and $\widetilde{LB}_{m}^{OLS}$ are not feasible in this example. In view of the different outputs we presented, it seems that the $LB_m^{ALS}$ test is the only test which give reliable conclusions in this example. This may be explained by the fact that the variance strongly change in this second example when compared to the first example and hence make the tests based on inverted matrices or the tests which do not exploit the variance structure difficult to implement.

\section{Conclusion}
\label{S8}

In this paper the problem of specification of the linear autoregressive dynamics of multivariate series with deterministic but non constant volatility is studied. Considering such situations is important since it is well known that economic or financial series commonly exhibit non stationary volatility. The unreliability of the standard portmanteau tests for testing the adequacy of the autoregressive order of VAR processes is highlighted through theoretical results and empirical illustrations. From the statistical methodology point of view, the main contribution of the paper is two-fold. In the setup of a stable VAR with time-varying variance, (a) we show how to compute corrected critical values for the standard portmanteau statistics implemented in all specialized software; and (b) we propose new portmanteau statistics based on the model residuals filtered for the non constant variance. Moreover, we provide some theoretical and empirical power comparisons of the two approaches and we show that they are well-suited for replacing the usual test procedures even when the volatility is constant. The new portmanteau statistics require the estimation of the time-varying variance that is done by classical kernel smoothing of the outer product of the OLS residuals vectors. Then another contribution of the paper is represented by the fact that our asymptotic results are derived uniformly in the bandwidth used in the kernel. This makes the theory compatible with the practice where people usually use the data to determine the bandwidth. Our theoretical and empirical investigation could be extended to other setups, like for instance the co-integrated systems. We briefly mention this extension but a deeper investigation is left for future work.

\section{Appendix A: Proofs}
\label{S9}
For the sake of a simpler presentation, hereafter we stick to our identification condition for $H_t$ and hence we replace everywhere
$G(r)$ by $\Sigma(r)^{1/2}$.
Recall that
$$\tilde{X}_t=\sum_{i=0}^{\infty}\tilde{\psi}_iu_{t-i}^p=\sum_{i=0}^{\infty}
K^i\tilde{u}_{t-i},$$
(see pages \pageref{pg5} and \pageref{pg6}).
Let us introduce
$$\Upsilon_{t-1}^u=(u_{t-1}',\dots,u_{t-m}',\tilde{X}_{t-1}')'=(u_{t-1}^{m'},\tilde{X}_{t-1}')'$$
and
$$\Upsilon_{t-1}^{\epsilon}=
(\epsilon_{t-1}',\dots,\epsilon_{t-m}',\tilde{X}_{t-1}')'=(\epsilon_{t-1}^{m'},\tilde{X}_{t-1}')',$$
for a given $m>0$. To prove Propositions \ref{propostun} and \ref{propostun_b} we need several preliminary results that are gathered in Lemmas \ref{lem1} to \ref{lemtcllln} below.

\begin{lem}\label{lem1} Under Assumption {\bf A1} we have
\begin{equation}\label{Gamma}
\lim_{T\to\infty}E\left[\tilde{X}_{[Tr]-1}\tilde{X}_{[Tr]-1}'\right]=\sum_{i=0}^{\infty}
\tilde{\psi}_i\left\{\mathbf{1}_{p\times
p}\otimes\Sigma(r)\right\}\tilde{\psi}_i':=\Omega(r),
\end{equation}
\begin{equation}\label{Gamma1}
\lim_{T\to\infty}E\left[\Upsilon_{[Tr]-1}^u\Upsilon_{[Tr]-1}^{u'}\right]=\Omega^u(r),
\end{equation}
\begin{equation}\label{Gamma2}
\lim_{T\to\infty}E\left[\Upsilon_{[Tr]-1}^{\epsilon}\Upsilon_{[Tr]-1}^{\epsilon'}\right]
=\Omega^{\epsilon}(r),
\end{equation}
for values $r\in(0,1]$ where the functions $g_{ij}(\cdot)$ are
continuous. The matrices in (\ref{Gamma1}) and (\ref{Gamma2}) are
given by
$$\Omega^u(r)=\left(
             \begin{array}{cc}
               I_m\otimes\Sigma(r) & \Theta_m^{u}(r) \\
               \Theta_m^{u}(r)^\prime & \Omega(r) \\
             \end{array}
           \right),\quad\Theta_m^{u}(r)=\sum_{i=0}^{m-1}\left\{e_m(i+1)e_p(1)^\prime\otimes
\Sigma(r)\right\}K^{i\,\prime}$$ and
$$\Omega^{\epsilon}(r)=\left(
                      \begin{array}{cc}
                        I_{dm} & \Theta_m^{\epsilon}(r) \\
                        \Theta_m^{\epsilon } (r)^\prime& \Omega(r) \\
                      \end{array}
                    \right),\quad\Theta_m^{\epsilon}(r)=\sum_{i=0}^{m-1}\left\{e_m(i+1)e_p(1)^\prime\otimes
\Sigma(r)^{1/2}\right\}K^{i\,\prime}.
$$
\end{lem}

\noindent{\bf Proof of Lemma \ref{lem1}}\quad Statement \ref{Gamma} is a direct consequence
of Lemma 7.2 in Patilea and Raïssi (2010).
For the proof of
(\ref{Gamma1}) we write\footnote{Here we make a common abuse of notation because in Assumption \textbf{A1} the matrix-valued function $\Sigma(\cdot)$ is not defined for negative values. To remedy this problem it suffices to extend the function $\Sigma(\cdot)$ to the left of the origin, for instance by setting $\Sigma(r)$ equal to the identity matrix if $r\leq 0$.}
\begin{eqnarray*}
%E(u_{t-1}^m\otimes\tilde{X}_{t-1}')=
E(u_{t-1}^m\tilde{X}_{t-1}')
&=&\sum_{i=0}^{\infty}E(u_{t-1}^m\tilde{u}_{t-i-1}'K^{i\,\prime})
\\&=&\sum_{i=0}^{m-1}E(u_{t-1}^m\tilde{u}_{t-i-1}'K^{i\,\prime})
\\&=&\sum_{i=0}^{m-1}\{e_m(i+1)e_p(1)'\otimes
H_{t-i-1}H_{t-i-1}'\}K^{i\,\prime}\\&=&\sum_{i=0}^{m-1}\{e_m(i+1)e_p(1)'\otimes
\Sigma((t-i-1)/T)\}K^{i\,\prime}.
\end{eqnarray*}
Therefore
\begin{eqnarray*}
\lim_{T\to\infty}E(u_{[Tr]-1}^m\tilde{X}_{[Tr]-1}')&=&
\lim_{T\to\infty}\sum_{i=0}^{m-1}\{e_m(i+1)e_p(1)'\otimes
\Sigma(([Tr]-i-1)/T)\}K^{i\,\prime}\\&=&\sum_{i=0}^{m-1}\{e_m(i+1)e_p(1)'\otimes
\Sigma(r)\}K^{i\,\prime}.
\end{eqnarray*}
Similarly we have
\begin{equation}\label{resp7}
\lim_{T\to\infty}E(u_{[Tr]-1}^mu_{[Tr]-1}^{m'})=I_m\otimes\Sigma(r),
\end{equation}
so that using (\ref{Gamma}) we obtain the result (\ref{Gamma1}). The
proof of (\ref{Gamma2}) is similar.
$\quad\square$

\medskip

Let us define
$v_{t}^u=\mbox{vec}(\Upsilon_{t-1}^{u}\Upsilon_{t-1}^{u'}\otimes
u_tu_t')$ and
$v_{t}^{\epsilon}=\mbox{vec}(\Upsilon_{t-1}^{\epsilon}\Upsilon_{t-1}^{\epsilon'}
\otimes\epsilon_t\epsilon_t')$. The following lemma is similar to
Lemma 7.3 of Patilea and Raïssi (2010), and hence the proof is
omitted.

\begin{lem}\label{precedent} Under {\bf A1} we have
\begin{equation}\label{LLN1}
T^{-1}\sum_{t=i+1}^T\mbox{vec}(u_t\tilde{X}_{t-i}')\stackrel{P}{\longrightarrow}0,
\end{equation}
\begin{equation}\label{LLN12}
T^{-1}\sum_{t=i+1}^T\mbox{vec}(\epsilon_t\tilde{X}_{t-i}')\stackrel{P}{\longrightarrow}0,
\end{equation}
for $i>0$, and
\begin{equation}\label{LLN4}
T^{-1}\sum_{t=m+1}^T\mbox{vec}(u_{t-1}^m\tilde{X}_{t-1}')\stackrel{P}{\longrightarrow}
\lim_{T\to\infty}T^{-1}\sum_{t=m+1}^T\mbox{vec}\left\{E(u_{t-1}^m\tilde{X}_{t-1}')\right\},
\end{equation}
\begin{equation}\label{LLN42}
T^{-1}\sum_{t=m+1}^T\mbox{vec}(\epsilon_{t-1}^m\tilde{X}_{t-1}')\stackrel{P}{\longrightarrow}
\lim_{T\to\infty}T^{-1}\sum_{t=m+1}^T\mbox{vec}\left\{E(\epsilon_{t-1}^m\tilde{X}_{t-1}')\right\}.
\end{equation}
In addition we have
\begin{equation}\label{LLN2}
T^{-1}\sum_{t=1}^Tv_{t}^u\stackrel{P}{\longrightarrow}
\lim_{T\to\infty}T^{-1}\sum_{t=1}^TE(v_{t}^u)
=\lim_{T\to\infty}T^{-1}\sum_{t=1}^T
\mbox{vec}\left\{E(\Upsilon_{t-1}^{u}\Upsilon_{t-1}^{u'})\otimes\Sigma_t\right\},
\end{equation}
\begin{equation}\label{LLN3}
T^{-1}\sum_{t=1}^Tv_{t}^{\epsilon}
\stackrel{P}{\longrightarrow}\lim_{T\to\infty}T^{-1}\sum_{t=1}^TE(v_{t}^{\epsilon})
=\lim_{T\to\infty}T^{-1}\sum_{t=1}^T
\mbox{vec}\left\{E(\Upsilon_{t-1}^{\epsilon}\Upsilon_{t-1}^{\epsilon'})\otimes
I_d\right\}.
\end{equation}
\end{lem}

\quad

\begin{lem}\label{lemtcllln} Under {\bf A1}
we have
\begin{equation}\label{L2}
\hat{\Sigma}_{\tilde{\underline{X}}} %=
%T^{-1}\sum_{t=1}^T\tilde{X}_{t-1}\tilde{X}_{t-1}'\otimes\Sigma_t^{-1}
=\Lambda_1+o_p(1),
\end{equation}
\begin{equation}\label{L1}
\hat{\Sigma}_{\tilde{X}}
%=T^{-1}\sum_{t=1}^T\tilde{X}_{t-1}\tilde{X}_{t-1}'\otimes
%I_d
=\Lambda_3+o_p(1).
\end{equation}
In addition we also have
\begin{equation}\label{TCL2}
T^{-\frac{1}{2}}\sum_{t=1}^T\Upsilon_{t-1}^{u}\otimes u_t\Rightarrow
\mathcal{N}(0,\Xi_u),
\end{equation}
\begin{equation}\label{TCL1}
T^{-\frac{1}{2}}\sum_{t=1}^TJ_t^{-1}(\Upsilon_{t-1}^{\epsilon}\otimes\epsilon_t)\Rightarrow
\mathcal{N}(0,\Xi_{\epsilon}),
\end{equation}
where $$J_t=\left(
              \begin{array}{cc}
                I_{d^2m} & 0 \\
                0 & I_{dp}\otimes H_t \\
              \end{array}
            \right)$$
and
$$\Xi_u=\int_0^1\left(
                  \begin{array}{cc}
                    I_m\otimes\Sigma(r)^{\otimes2} & \Theta_m^u(r)\otimes\Sigma(r) \\
                    \Theta_m^{u\,\prime}(r)\otimes\Sigma(r) & \Omega(r)\otimes\Sigma(r) \\
                  \end{array}
                \right)dr = \left(
                  \begin{array}{cc}
                    \Lambda_m^{u,u} & \Lambda_m^{u,\theta}\\
                    \Lambda_m^{u,\theta\,\prime} & \Lambda_2 \\
                  \end{array}
                \right) ,$$
$$\Xi_{\epsilon}=\int_0^1\left(
                  \begin{array}{cc}
                    I_{d^2m} & \Theta_m^{\epsilon}(r)\otimes\Sigma(r)^{-\frac{1}{2}} \\
                    \Theta_m^{\epsilon'}(r)\otimes\Sigma(r)^{-\frac{1}{2}} & \Omega(r)\otimes \Sigma(r)^{-1} \\
                  \end{array}
                \right)dr= \left(
                  \begin{array}{cc}
                    \Lambda_m^{u,u} & \Lambda_m^{\epsilon,\theta}\\
                    \Lambda_m^{\epsilon,\theta\,\prime} & \Lambda_1 \\
                  \end{array}
                \right).$$
\end{lem}

\quad

\noindent{\bf Proof of Lemma \ref{lemtcllln}}\quad
Statements (\ref{L2}) and (\ref{L1}) lemma are
direct consequences of Lemma 7.4 of Patilea and Raïssi (2010). We
only give the proof of (\ref{TCL1}) and (\ref{TCL2}). To prove (\ref{TCL1}), using the well known identity $(B\otimes
C)(D\otimes F)=(BD)\otimes(CF)$ for matrices of appropriate
dimensions, we obtain
$$J_t^{-1}(\Upsilon_{t-1}^{\epsilon}\otimes\epsilon_t)
(\Upsilon_{t-1}^{\epsilon'}\otimes\epsilon_t')J_t^{-1}=
J_t^{-1}(\Upsilon_{t-1}^{\epsilon}\Upsilon_{t-1}^{\epsilon'}\otimes\epsilon_t\epsilon_t')J_t^{-1}.$$
From (\ref{LLN3}) we write
\begin{eqnarray*}
T^{-1}\sum_{t=1}^TJ_t^{-1}(\Upsilon_{t-1}^{\epsilon}
\Upsilon_{t-1}^{\epsilon'}\otimes\epsilon_t\epsilon_t')J_t^{-1}
&\stackrel{P}{\rightarrow}&\lim_{T\to\infty}T^{-1}\sum_{t=1}^TJ_t^{-1}
\left[E\left\{\Upsilon_{t-1}^{\epsilon}\Upsilon_{t-1}^{\epsilon'}\right\}\otimes
I_d\right]J_t^{-1}.
\end{eqnarray*}
Now let us denote the
discontinuous points of the functions $g_{ij}(.)$ by
$\xi_1,\xi_2,\dots,\xi_q$ where $q$ is a finite number independent
of $T$. We write
\begin{eqnarray*}
&&\lim_{T\to\infty}T^{-1}\sum_{t=1}^TJ_t^{-1}
\left[E(\Upsilon_{t-1}^{\epsilon}\Upsilon_{t-1}^{\epsilon'})\otimes
I_d\right]J_t^{-1}
\\&=&\lim_{T\to\infty}\sum_{t=1}^T\int_{t/T}^{(t+1)/T}
J_{[Tr]}^{-1}
\left[E(\Upsilon_{[Tr]-1}^{\epsilon}\Upsilon_{[Tr]-1}^{\epsilon'})
\otimes
I_d\right]J_{[Tr]}^{-1}dr+o_p(1)\\&=&\lim_{T\to\infty}\int_{1/T}^{\xi_1}
J_{[Tr]}^{-1}
\left[E(\Upsilon_{[Tr]-1}^{\epsilon}\Upsilon_{[Tr]-1}^{\epsilon'})
\otimes
I_d\right]J_{[Tr]}^{-1}dr+\dots\\&\dots+&\int_{\xi_q}^{(T+1)/T}
J_{[Tr]}^{-1}
\left[E(\Upsilon_{[Tr]-1}^{\epsilon}\Upsilon_{[Tr]-1}^{\epsilon'})
\otimes I_d\right]J_{[Tr]}^{-1}dr+o_p(1).
\end{eqnarray*}
Then from (\ref{Gamma2}) we obtain
\begin{eqnarray*}
&&T^{-1}\sum_{t=1}^TJ_t^{-1}(\Upsilon_{t-1}^{\epsilon}
\Upsilon_{t-1}^{\epsilon'}\otimes\epsilon_t\epsilon_t')J_t^{-1}\\&\stackrel{P}{\longrightarrow}&
\int_0^1J(r)^{-1}\left(
                   \begin{array}{cc}
                     I_{d^2m} & \Theta_m^{\epsilon}(r)\otimes I_d \\
                     \Theta_m^{\epsilon'}(r)\otimes I_d & \Omega(r)\otimes I_d \\
                   \end{array}
                 \right)J(r)^{-1}dr,
\end{eqnarray*}
where
$$J(r)=\left(
         \begin{array}{cc}
           I_{d^2m} & 0 \\
           0 & I_{dp}\otimes\Sigma(r)^{-\frac{1}{2}} \\
         \end{array}
       \right)$$
and $\Sigma(r)^{1/2}=G(r)=H_{[Tr]}$. Noting
that $J_t^{-1}(\Upsilon_{t-1}^{\epsilon}\otimes\epsilon_t)$ are
martingale differences, we obtain the result (\ref{TCL1}) using the
Lindeberg central limit theorem. Using relations (\ref{Gamma1}) and
(\ref{LLN2}),
the proof of (\ref{TCL2}) is similar. Finally, the equivalent compact expressions of $\Xi_{\epsilon}$ and $\Xi_u$ can be easily derived using
elementary properties of the Kronecker product. $\quad\square$

\medskip

\noindent{\bf Proof of Proposition \ref{propostun}}\quad First we
establish the result (\ref{gamols}). Let us define
$$\Gamma_u(h)=T^{-1}\sum_{t=h+1}^Tu_tu_{t-h}'\quad\mbox{and}
\quad
c_m^u=\mbox{vec}\left\{(\Gamma_u(1),\dots,\Gamma_u(m))\right\}.$$ Let us first show the asymptotic normality of
$T^{1/2}(c_m^{u\,\prime},(\hat{\theta}_{OLS}-\theta_0)')'$. Note
that
$$c_m^{u}=T^{-1}\sum_{t=1}^T\tilde{u}_{t-1}^m\otimes u_t
\quad\mbox{and}\quad\hat{\theta}_{OLS}-\theta_0=\hat{\Sigma}_{\tilde{X}}^{-1}\left\{T^{-1}\sum_{t=1}^T(\tilde{X}_{t-1}\otimes
u_t)\right\},$$
with $\tilde{u}_{t-1}^m=(\mathbf{1}_{(0,\infty)}(t-1)\times u_{t-1}',
\dots,\mathbf{1}_{(0,\infty)}(t-m)\times u_{t-m}')'$. From (\ref{L1}) we write
\begin{equation*}
T^{\frac{1}{2}}\left(
                 \begin{array}{c}
                   c_m^{u} \\
                   \hat{\theta}_{OLS}-\theta_0 \\
                 \end{array}
               \right)=\dot{\Lambda}_3^{-1}
               \left\{T^{-\frac{1}{2}}\sum_{t=1}^T\Upsilon_{t-1}^u\otimes u_t\right\}+o_p(1),
\end{equation*}
with $$\dot{\Lambda}_3=\left(
                          \begin{array}{cc}
                            I_{d^2m} & 0 \\
                            0 & \Lambda_3 \\
                          \end{array}
                        \right).$$
Then we can obtain from (\ref{TCL2})
\begin{equation}\label{normal}
T^{\frac{1}{2}}\left(
\begin{array}{c}
c_m^{u} \\
\hat{\theta}_{OLS}-\theta_0 \\
\end{array}
\right)\Rightarrow
\mathcal{N}(0,\dot{\Lambda}_3^{-1}\Xi_u\dot{\Lambda}_3^{-1}),
\end{equation}
with $\Xi_u$ defined in Lemma \ref{lemtcllln}. Now, define $u_t(\theta)=X_t-(\tilde{X}_{t-1}'\otimes I_d)\theta$
with $\theta\in \mathbb{R}^{d^2p}$. Considering
$\hat{\gamma}_m^{u,OLS}$ and $c_m^u$ as values of the same function at
the points $\theta_0$ and $\hat{\theta}_{OLS}$, by the Mean Value Theorem
$$\hat{\gamma}_m^{u,OLS}=c_m^u+T^{-1}\sum_{t=1}^T\left\{\tilde{u}_{t-1}^m(\theta)
\otimes\frac{\partial u_t(\theta)}{\partial\theta'}
+\frac{\partial \tilde{u}_{t-1}^m(\theta)}{\partial\theta'}\otimes
u_t(\theta)\right\}_{\theta=\theta^*}(\hat{\theta}_{OLS}-\theta_0),$$
with $\theta^*$ between $\hat{\theta}_{OLS}$ and $\theta_0$.\footnote{The value $\theta^*$ between $\hat{\theta}_{OLS}$ and $\theta_0$ may be different for different components of $\hat{\gamma}_m^{OLS}$ and $c_m^u$.}
Using $T^{1/2}(\hat{\theta}_{OLS}-\theta_0)=O_p(1)$ and
since $\partial
u_{t-i}(\theta)\!/\partial\theta'\! =-(\tilde{X}_{t-i-1}'\otimes I_d)$,
it follows from (\ref{LLN1}) and (\ref{LLN4}) that
\begin{eqnarray}\label{petitop1}
\hat{\gamma}_m^{u,OLS}&=&c_m^u+\lim_{T\to\infty}T^{-1}\sum_{t=1}^TE\left\{\tilde{u}_{t-1}^m\otimes\frac{\partial
u_t}{\partial\theta'}
\right\}(\hat{\theta}_{OLS}-\theta_0)+o_p(T^{-\frac{1}{2}})\nonumber\\&=&
c_m^u+\lim_{T\to\infty}T^{-1}\sum_{t=1}^T-E\left\{\tilde{u}_{t-1}^m\otimes\tilde{X}_{t-1}'\otimes
I_d\right\}(\hat{\theta}_{OLS}-\theta_0)\\&+&o_p(T^{-\frac{1}{2}})\nonumber.
\end{eqnarray}
From (\ref{Gamma1}) and using arguments like in the proof
of (\ref{TCL1}), it is easy to see that
$$\lim_{T\to\infty}T^{-1}\sum_{t=1}^T-E\left\{\tilde{u}_{t-1}^m\otimes\tilde{X}_{t-1}'\otimes
I_d\right\}=-\int_0^1\Theta_m^u(r) dr\otimes
I_d+o_p(1)=-\Phi_m^u+o_p(1).$$ Finally from (\ref{petitop1}) we have
\begin{eqnarray}\label{petitop}
\hat{\gamma}_m^{u,OLS}&=&
c_m^u-\Phi_m^u(\hat{\theta}_{OLS}-\theta_0)+o_p(T^{-\frac{1}{2}}),
\end{eqnarray}
so that it follows from (\ref{normal}) that
$T^{1/2}\hat{\gamma}_m^{OLS}$ is asymptotically normal with covariance
matrix
\begin{equation}\label{jaifaim}\Sigma^{u,OLS}=\Lambda^{u,u}_m
+\Phi^u_m\Lambda_3^{-1}\Lambda_2\Lambda_3^{-1}\Phi^{u\,\prime}_{m}-
\Lambda^{u,\theta}_m\Lambda_3^{-1}\Phi^{u\,\prime}_{m}-\Phi^u_m\Lambda_3^{-1}\Lambda^{u,\theta\,\prime}_{m}.
\end{equation}
The proof of  (\ref{gamgls}) is very similar, here we only present a
sketch. Let us define
$$\Gamma_{\epsilon}(h)=T^{-1}\sum_{t=h+1}^T\epsilon_t\epsilon_{t-h}'\quad\mbox{and}
\quad
c_m^{\epsilon}=\mbox{vec}\left\{(\Gamma_{\epsilon}(1),\dots,\Gamma_{\epsilon}(m))\right\}.$$
Using (\ref{L2}) and (\ref{TCL1}) it can be shown
that
\begin{equation}\label{normal2}
T^{\frac{1}{2}}\left(
\begin{array}{c}
c_m^{\epsilon} \\
\hat{\theta}_{GLS}-\theta_0 \\
\end{array}
\right)\Rightarrow
\mathcal{N}(0,\dot{\Lambda}_1^{-1}\Xi_{\epsilon}\dot{\Lambda}_1^{-1}),
\end{equation}
with $\Xi_\epsilon$ defined in Lemma \ref{lemtcllln} and
$$\dot{\Lambda}_1=\left(
                          \begin{array}{cc}
                            I_{d^2m} & 0 \\
                            0 & \Lambda_1 \\
                          \end{array}
                        \right).$$
From (\ref{LLN12}), (\ref{LLN42}) and since $\partial
\epsilon_{t-i}(\theta)/\partial\theta'=-(\tilde{X}_{t-i-1}'\otimes
H_t^{-1})$ we write
\begin{eqnarray*}
\hat{\gamma}_m^{\epsilon, GLS}&=&
c_m^{\epsilon}+\lim_{T\to\infty}T^{-1}\sum_{t=m+1}^T-E\left\{\epsilon_{t-1}^m\otimes\tilde{X}_{t-1}'\otimes
H_t^{-1}\right\}(\hat{\theta}_{GLS}-\theta_0)+o_p(T^{-\frac{1}{2}})
\\&=&c_m^{\epsilon}-\Lambda^{\epsilon,\theta}_{m}(\hat{\theta}_{GLS}-\theta_0)+o_p(T^{-\frac{1}{2}}).
\end{eqnarray*}
By (\ref{normal2}), $T^{1/2}\hat{\gamma}_m^{\epsilon,GLS}$ is
asymptotically normal with covariance matrix $
\Sigma^{\epsilon,GLS}=I_{d^2m}
-\Lambda^{\epsilon,\theta}_m\Lambda_1^{-1}\Lambda^{\epsilon,\theta\,\prime}_{m}.
$ The particular case where the order of the VAR model is $p=0$ is an easy consequence of the arguments above in this proof. $\quad\square$

\quad

\noindent{\bf Proof of Proposition \ref{propostun_b}}
For the proof of (\ref{rhools}), we write
$$u_{it}=\sum_{j=1}^dh_{ij,t}\epsilon_{jt}\quad\mbox{and}\quad E(u_{it}^2)=\sum_{j=1}^dh_{ij,t}^2=\sigma_{ii,t}^2,\quad\mbox{say.}$$\\
It is clear from (\ref{resp7}) that
$$\lim_{T\to\infty}E(u_{i[Tr]}^2)=\sigma_{ii}^2(r),$$
where $\sigma_{ii}^2(r)$ is the $i$th diagonal
element of $\Sigma(r)$. Following similar arguments used in Phillips
and Xu (2005 p 303) for the proof of Lemma 1 (iii), we write
$$T^{-1}\sum_{t=1}^Tu_{it}^2=\int_0^1\sigma_{ii}^2(r)dr+o_p(1).$$
Let us define $\Sigma^{u}=T^{-1}\sum_{t=1}^Tu_tu_t'$ and
$\hat{\Sigma}^{u}=T^{-1}\sum_{t=1}^T\hat{u}_t\hat{u}_t'$. We have
using again the Mean Value Theorem
$$\mbox{vec}(\hat{\Sigma}^{u})=\mbox{vec}(\Sigma^{u})
+T^{-1}\sum_{t=1}^T\left\{u_{t}(\theta)\otimes\frac{\partial
u_t(\theta)}{\partial\theta'} +\frac{\partial
u_{t}(\theta)}{\partial\theta'}\otimes
u_t(\theta)\right\}_{\theta=\theta^*}(\hat{\theta}_{OLS}-\theta_0),$$
with $\theta^*$ is between $\hat{\theta}_{OLS}$ and $\theta_0$.\footnote{The value $\theta^*$  may be different for different components of $\mbox{vec}(\hat\Sigma^{u})$ and $\mbox{vec}(\Sigma^{u})$.}
Therefore using $\partial
u_{t}(\theta)/\partial\theta'=-(\tilde{X}_{t-1}'\otimes I_d)$ and from the consistency of $\hat{\theta}_{OLS}$, we write
$$T^{\frac{1}{2}}\mbox{vec}\hat{\Sigma}^{u}=T^{\frac{1}{2}}\mbox{vec}\Sigma^{u}+o_p(1),$$
and
$$T^{-1}\sum_{t=1}^T\hat{u}_{it}^2=\int_0^1\sigma_{ii}^2(r)dr+o_p(1),$$
so that the result follows from the Slutsky lemma. We obtain the expression (\ref{rhools}) noting that
$$\hat{\rho}_m^{OLS}=\{I_m\otimes(\hat{S}_u\otimes\hat{S}_u)^{-1}\}\hat{\gamma}_m^{OLS}.$$
The proof of (\ref{rhogls}) is similar to that of
(\ref{rhools}) and hence is omitted. $\quad\square$

\medskip

\noindent{\bf Proof of Proposition \ref{lemALS}}
In the following, $c$, $C$, ... denote constants with possibly different values from line to line. To simplify notation, let $b$ denote the $d(d+1)/2$ vector of bandwidths $b_{kl}$, $1\leq k\leq l\leq d$. Below we will simply write \emph{uniformly w.r.t. $b$} instead of \emph{uniformly w.r.t. $b_{kl}$, $1\leq k\leq l\leq d$}, and  $\sup_b$ instead of $\sup_{b_{kl}\in\mathcal{B}_T, 1\leq k\leq l\leq d}$. Here the norm $\| \cdot\|$ is the Frobenius norm which in particular is a sub-multiplicative norm, that is $\| AB\| \leq \| A\| \| B\|$, and for a positive definite matrix $A$, $\|A\| \leq C[\lambda_{min} (A) ]^{-1}$ with $C$ a constant depending only on the dimension of $A$. Moreover, $\|A \otimes B \| = \| A\| \|B \|$.

To obtain the asymptotic equivalences in equation (\ref{equivalent}) it suffices to notice that for all $1\leq i \leq d$, $\hat\sigma^2_\epsilon (i) - 1 = o_p(1) $, and to prove
\begin{equation}\label{sig_1}
\sup_{1\leq i \leq d} \sup_b \left|\check\sigma^2_\epsilon (i) - \hat\sigma^2_\epsilon (i)\right| = o_p(1)
\end{equation}
and
\begin{equation}\label{gam_1}
\sup_b \left| T^{\frac{1}{2}} \left\{  \Gamma_{ALS}(h) -\Gamma_{GLS}(h)\right\} \right| = o_p(1),
\end{equation}
for any fixed $h\geq 1$. Let us write
\begin{eqnarray*}
\check \epsilon_t - \hat\epsilon_t &=& (\check \Sigma_t^{-\frac{1}{2}} - \Sigma_t^{-\frac{1}{2}}) u_t + \check \Sigma_t^{-\frac{1}{2}}(\tilde X_{t-1}^\prime \otimes I_d) (\hat \theta_{GLS}
- \hat \theta_{ALS} ) \\ &&+ (\Sigma_t^{-\frac{1}{2}} - \check \Sigma_t^{-\frac{1}{2}})(\tilde X_{t-1}^\prime \otimes I_d) (\hat \theta_{GLS}
- \theta_{0} )\\
&=:& (\check \Sigma_t^{-\frac{1}{2}} - \Sigma_t^{-\frac{1}{2}}) u_t + \delta^\epsilon_t
\end{eqnarray*}
where $\| \delta^\epsilon_t \| \leq \|\tilde X_{t-1}\| \check R_T
(b)$ with
\[
\check R_T (b) = d \! \left\{ \| \hat \theta_{GLS} \!-\! \hat
\theta_{ALS} \| \sup_{1\leq t \leq T} \left\| \check
\Sigma_t^{-\frac{1}{2}} \right\| +  \| \hat \theta_{GLS} -
\theta_{0} \| \sup_{1\leq t \leq T} \left\| \check
\Sigma_t^{-\frac{1}{2}} -  \Sigma_t^{-\frac{1}{2}} \right\| \right\}
\]
By Lemma \ref{lem_Ht}-(a,b) and given that $\hat \theta_{GLS}
- \theta_{0} = O_p(T^{-1/2})$ and $\sup_b \|\hat \theta_{GLS} - \hat \theta_{ALS}  \| = o_p(T^{-1/2})$,
we obtain that
\begin{equation}\label{small_d_t}
\sup_b \check R_T (b) = o_p(T^{-1/2}).
\end{equation}
From this and the moment conditions on $(X_t)$ induced by Assumption
{\bf A1}, deduce that  (\ref{sig_1}) holds true. On the other hand,
\begin{eqnarray*}
\Gamma_{ALS}(h) -\Gamma_{GLS}(h) & =& \frac{1}{T} \sum_{t=h+1}^T \hat\epsilon_t (\check \epsilon_{t-h} - \hat\epsilon_{t-h} )^\prime
+ \frac{1}{T} \sum_{t=h+1}^T  (\check \epsilon_t - \hat\epsilon_t )\hat\epsilon_{t-h}^\prime\\
&& + \frac{1}{T} \sum_{t=h+1}^T (\check \epsilon_t - \hat\epsilon_t )(\check \epsilon_{t-h} - \hat\epsilon_{t-h} )^\prime\\
&=:& R_{1T}(h) +  R_{2T} (h) +  R_{3T} (h).
\end{eqnarray*}
The terms $R_{1T}(h)$ and $R_{2T} (h)$ could be handled in a similar manner, hence we will only analyze $R_{2T}(h)$. Let us write
\begin{eqnarray*}
R_{2T} (h) \!\!\! &=& \!\!\!\frac{1}{T} \sum_{t=h+1}^T \!\!\left[
(\check \Sigma_t^{-\frac{1}{2}} \!-\! \Sigma_t^{-\frac{1}{2}}) u_t +
\delta^\epsilon_t \right]\!\! \left[ \Sigma_{t-h}^{-\frac{1}{2}}
u_{t-h} \!-\! \Sigma_{t-h}^{-\frac{1}{2}} (\tilde X_{t-h-1}^\prime
\!\otimes \! I_d) (\hat \theta_{GLS}\!
- \!\theta_{0} ) \right]^\prime\\
\!\!\! \!\!\!&=:& \!\!\!\! \frac{1}{T} \sum_{t=h+1}^T (\check
\Sigma_t^{-\frac{1}{2}} \!- \! \Sigma_t^{-\frac{1}{2}}) u_t
u_{t-h}^\prime \Sigma_{t-h}^{-\frac{1}{2}\, \prime} + R_{22T} (h;b)
=: R_{21T} (h;b) + R_{22T} (h;b).
\end{eqnarray*}
By (\ref{small_d_t}) and the moment conditions on the innovation
process $(\epsilon_t)$, and the rate of convergence of
$\hat\theta_{GLS}$ and \ref{lem_Ht}-(b), it is clear that $\sup_b \|
R_{22T} (h;b)\| = o_p(T^{-1/2})$. Next let us write
\begin{eqnarray*}
R_{21T} (h) &=& \frac{1}{T} \sum_{t=h+1}^T  \left(\check \Sigma_t^{-\frac{1}{2}} - (\check \Sigma_t^{0})^{-\frac{1}{2}}\right) u_t \epsilon_{t-h}^\prime +
\frac{1}{T} \sum_{t=h+1}^T   \left((\check \Sigma_t^{0})^{-\frac{1}{2}} - \stackrel{\circ}{\Sigma}\;\!\!\! _t^{-\frac{1}{2}}\right) u_t \epsilon_{t-h}^\prime \\
&&+ \frac{1}{T}\sum_{t=h+1}^T  \left(\stackrel{\circ}{\Sigma}\;\!\!\! _t^{-\frac{1}{2}} - \bar \Sigma_t^{-\frac{1}{2}}\right) u_t \epsilon_{t-h}^\prime +
\frac{1}{T} \sum_{t=h+1}^T   \left(\bar \Sigma_t^{-\frac{1}{2}} - \Sigma_t^{-\frac{1}{2}}\right) u_t \epsilon_{t-h}^\prime\\
&=:& R_{211T} (h;b)+R_{212T} (h;b)+R_{213T} (h;b)+R_{214T} (h;b),
\end{eqnarray*}
where, like in Patilea and Ra\"{i}ssi (2010),
\begin{equation*}\label{sigma_rond_p}
\stackrel{\circ}{\Sigma}_t = \stackrel{\circ}{\Sigma}_t (b) =
\sum_{i=1}^T w_{ti}\odot {u}_i{u}_i'\quad \text{and} \quad \bar
\Sigma_t =  \bar \Sigma_t (b) = \sum_{i=1}^T w_{ti} \odot \Sigma_i.
\end{equation*}
From classical matrix norm inequalities (see for instance Horn and
Johnson, 1994), we have that for any $d\times d-$positive definite
matrices $A$ and $B$, for $a=1$ or $a=-1$,
\begin{equation}\label{sqrt_eq_m}
\| A^{-\frac{a}{2}} - B^{-\frac{a}{2}}\| \leq c_a \left(\max\{ \|A^a\|, \|B^a\| \} \right)^{\frac{1}{2}}  \left\|A^{-\frac{1+a}{2}}\right\| \left\|B^{-\frac{1+a}{2}}\right\| \| A-B\|,
\end{equation}
where $c_a$ is a constant that depends only on $d$ (by definition $A^0=B^0=I_d$). Applying this inequality twice we deduce
\begin{eqnarray*}
\left\|\check \Sigma_t^{-\frac{1}{2}} - (\check \Sigma_t^{0})^{-\frac{1}{2}}\right\| &\leq & \nu_T c_1 c_{-1}\left\|\check \Sigma_t^{-1}\right\| \left\| (\check \Sigma_t^{0})^{-1}\right\|\\
&& \!\!\!\! \times \left(\max\{ \|\check \Sigma_t\|, \|\check
\Sigma_t^{0}\| \} \right)^{\frac{1}{2}}\left(\max\{ \|[(\check
\Sigma_t^{0})^{2}+\nu_T I_d]^{-1}\|, \|(\check \Sigma_t^{0})^{-2}\|
\} \right)^{\frac{1}{2}}\!.
\end{eqnarray*}
Take the norm of $R_{211T}$, use the inequality in the last display,
Lemma \ref{lem_Ht}-(a) below, the moment conditions on the
innovation process and the condition $T\nu_T^2 \rightarrow \infty$
to deduce that $\sup_b \| R_{211T} (h;b)\| = o_p(T^{-1/2})$. Next,
using similar matrix norm inequalities, Lemma \ref{lem_Ht}-(a) and
the Cauchy-Schwarz inequality,
\begin{eqnarray*}
\sup_b\| R_{212T} (h;b)\| \!\!&\leq & \!\!O_p(1) \sup_b\left\{
\frac{1}{T} \sum_{t=h+1}^T   \left\|(\check \Sigma_t^{0}) -
\stackrel{\circ}{\Sigma}\;\!\!\! _t\right\| \|u_t
\epsilon_{t-h}^\prime\|\right\} \\ &\leq &\!\!\!\! O_p(1) \left(
\sup_b\left\{\!\! \frac{1}{T} \sum_{t=h+1}^T \left\|(\check
\Sigma_t^{0}) - \stackrel{\circ}{\Sigma}\;\!\!\! _t\right\|^2
\right\} \right)^{\!\! 1/2}\!\!
\left( \frac{1}{T} \!\sum_{t=h+1}^T \!\! \|u_t \epsilon_{t-h}^\prime\|^2 \right)^{\!\!1/2}\\
&=& O_p(1) O_p(T^{-1}b_T^{-1}) \left( \frac{1}{T} \sum_{t=h+1}^T
\|u_t \epsilon_{t-h}^\prime\|^2 \right)^{1/2},
\end{eqnarray*}
where for the equality we used Lemma 7.6-(i) in Patilea and Raïssi
(2010). Deduce that $\sup_b \| R_{212T} (h;b)\| = o_p(T^{-1/2})$.
The uniform rate of convergence for $R_{213T} (h;b)$ is obtained
after replacing $\stackrel{\circ}{\Sigma}_t^{-1/2}
- \bar \Sigma_t^{-1/2}$ by a Taylor expansion of the power
$-1/2$ function for positive definite matrix, a key and apparently
new ingredient we provide in section \ref{treyu} below. The reminder
term of the Taylor expansion could be controlled taking expectation,
using Cauchy-Schwarz inequality and Lemma \ref{lem_Ht}-(d). The term
under the integral that represents  the first order term of this
Taylor expansion could be treated similarly to the term
$\bar{\Sigma}_t^{-1} [ \Sigma_i - u_i u_i^\prime ]
\bar{\Sigma}_t^{-1}u_t\tilde{X}_{t-1}'$ in the proof of the
Proposition 4.1 of Patilea and Ra\"{i}ssi (2010). That means we use
the CLT for m.d. sequences indexed by classes of functions, see Bae,
Jun and Levental (2010), see also Bae and Choi (1999). Here the
uniformity to be considered is also with respect to the integration
variable $v$, but this can be handled with very little additional
effort, like in Patilea and Ra\"{i}ssi (2010). The details are
omitted. Finally, to derive the uniform order $R_{214T} (h;b)$, let
us write it as
\[
R_{214T} (h;b) = R_{214T} (h;b) - R_{214T} (h;b_T) + R_{214T} (h;b_T) = : r_{214T} (b) + R_{214T} (h;b_T).
\]
The term $R_{214T} (h;b_T)$ is centered and  the variance of each
element of this matrix decreases to zero at the rate $o(1/T)$ (use
Lemma \ref{lem_Ht}-(d) and Assumption {\bf A1'} to derive the rate
of the variance). Deduce that $R_{214T} (h;b_T)=o_p(T^{-1/2})$. Next
consider the $d^2$ stochastic processes corresponding to the
elements of $r_{214T} (b) $ and indexed by $\vartheta\in[c_{min},
c_{max}]$ where $b=\vartheta b_T$. For each such process apply
Theorem 1 of Bae, Jun and Levental (2010) to deduce that $\sup_b \|
r_{214T} (b)\| = o_p(T^{-1/2})$. Finally, deduce that $\sup_b \|
R_{214T} (h;b)\| = o_p(T^{-1/2})$

To handle the term $R_{3T}(h)$, let us write
\begin{eqnarray*}
R_{3T} (h) &=& \!\frac{1}{T} \sum_{t=h+1}^T \left[ (\check \Sigma_t^{-\frac{1}{2}} - \Sigma_t^{-\frac{1}{2}}) u_t + \delta^\epsilon_t \right]\! \left[  (\check \Sigma_{t-h}^{-\frac{1}{2}} - \Sigma_{t-h}^{-\frac{1}{2}}) u_{t-h} + \delta^\epsilon_{t-h}  \right]^\prime\\
&=:& \frac{1}{T} \sum_{t=h+1}^T (\check \Sigma_t^{-\frac{1}{2}} -
\Sigma_t^{-\frac{1}{2}}) u_t    u_{t-h}^\prime (\check
\Sigma_{t-h}^{-\frac{1}{2}} - \Sigma_{t-h}^{-\frac{1}{2}})^\prime +
R_{32T} (h)\\& =:& R_{31T} (h) + R_{32T} (h).
\end{eqnarray*}
The term $R_{32T} (h)$ could be easily handled taking the norm,
using the bound on $\delta^\epsilon_t$ and Lemma \ref{lem_Ht}-(b)
below. For $R_{31T} (h)$, we could decompose $\check \Sigma_t^{-1/2}
- \Sigma_t^{-1/2}$ in four term exactly as we did for $R_{21T} (h)$
and apply the same techniques. The details are omitted and are
available from the authors upon request. $\quad\square$

\medskip

\begin{lem}\label{lem_Ht} Let $\|\cdot\|$ denote the Frobenius norm. Under the Assumptions of Proposition \ref{lemALS}
we have:

(a)
 As $T\rightarrow\infty$, for $a=1$ or $a=-1$, we have
\begin{equation*}\label{qrtr_b}
\sup_{1\leq t \leq T} \sup_{b\in\mathcal{B}_T}\left\{ \left\| \check \Sigma_t^{-\frac{1}{2}} \right\| +\left\| \check \Sigma_t^{a} \right\|  + \left\| (\check \Sigma_t^{0})^{a} \right\| + \left\|\stackrel{\circ}{\Sigma}\;\!\!\! _t^{a} \right\|  + \left\|\bar \Sigma_t^{a} \right\|\right\}= O_p (1).
\end{equation*}

(b)
 As $T\rightarrow\infty$,
\begin{equation*}\label{qrtr_c}
\sup_{1\leq t \leq T} \sup_{b\in\mathcal{B}_T} \left\| \check \Sigma_t^{-\frac{1}{2}} - \Sigma_t^{-\frac{1}{2}}\right\| = o_p (1).
\end{equation*}

(c)  As $T\rightarrow\infty$,
\[
\sup_{b\in\mathcal{B}_T} \frac{1}{T}\sum_{t=1}^{T} \left\|\bar \Sigma_t - \Sigma_t \right\|^2 = o(1).
\]

(d) As $T\rightarrow\infty$,
\begin{equation}\label{rtrt_bb}
\max_{1\leq t\leq T} E\left( \sup_{b\in\mathcal{B}_T}\|\stackrel{\circ}{\Sigma}_t - \bar \Sigma_t \|^4\right) = O(\left(1/(Tb_T)^2  \right).
\end{equation}

\end{lem}

The proof of Lemma \ref{lem_Ht} is a direct consequence of Lemmas
7.5 and 7.6 of Patilea and Ra\"{i}ssi (2010) and Lemma A of Xu and
Phillips (2008) applied elementwise, and hence will be omitted.

\quad

\noindent{\bf Proof of Proposition \ref{bahadur_comp}} The notation
in this proof are those of section \ref{S5b}. First let us notice
that  $q_A (x) = x/2 \{1+o(1)\}$ for large values of $x$, provided
the asymptotic law of a test statistic $Q_A$ under the null
hypothesis is $\chi_m^2$ with some $m\geq 1$. In the case where a
test statistic $Q_A$ has the asymptotic distribution of
$U(\delta_m^{OLS})$ defined in equation (\ref{truedist}),
\begin{eqnarray*}
q_A(x)&=& - \log P(U(\delta_m^{OLS}) >x ) \leq  -\log P\left( \max_i\{ \delta_{i}^{OLS}\}    U^2  >x \right)
\\&=&  \frac{x}{2 \max_i\{ \delta_{i}^{OLS} \} }\{1+o(1) \},
\end{eqnarray*}
and
\begin{eqnarray*}
q_A(x)&=& - \log P(U(\delta_m^{OLS}) >x ) \geq  -\log P\left(
\max_i\{ \delta_{i}^{OLS}\}  \Sigma_{j=1}^{d^2 m}  U_j^2    >x
\right)
\\&=&  \frac{x}{2 \max_i\{ \delta_{i}^{OLS} \} }\{1+o(1) \},
\end{eqnarray*}
with $U$ and $U_j$ independent $\mathcal{N}(0,1)$  variables. Thus
to prove (i) it suffices to show that
\begin{equation}\label{rezp}
\int_0^1\Sigma dr \otimes\left(\int_0^1\Sigma dr\right)^{\!\!-1}\!\!
\ll \max_i\{ \delta_{i}^{OLS}\} \left(\! \int_0^1 \!\!\! \Sigma dr
\!\otimes\! I_d \! \right)  \Sigma_{G^{\otimes 2}}^{-1}
\left(\!\int_0^1 \!\!\! \Sigma dr \!\otimes\! I_d \! \right).
\end{equation}
Herein, for any $A$ and $B$ symmetric matrices, $A\ll B$ means that
$B-A$ is positive semidefinite. Now, in the last display, multiply
both sides of the order relationship  on the left and on the right
by $\left(\int_0^1\Sigma dr\right)^{-1/2}
\otimes\left(\int_0^1\Sigma dr\right)^{1/2}$ and deduce that it
suffices to prove
\begin{eqnarray*}
I_d \otimes I_d &\ll& \!\!\max_i\{ \delta_{i}^{OLS}\} \left[\left(\!
\int_0^1 \!\!\! \Sigma dr \!\right)^{\!\!1/2}\!\otimes \! \left(\!
\int_0^1 \!\!\! \Sigma dr \!\right)^{\!\!1/2} \right]\!
\Sigma_{G^{\otimes 2}}^{-1} \left[\left(\! \int_0^1 \!\!\! \Sigma dr
\! \right)^{\!\!1/2}\! \!\otimes \! \left(\! \int_0^1 \!\!\! \Sigma
dr
\! \right)^{\!\!1/2} \right]\\
&=:& \!\!\max_i\{ \delta_{i}^{OLS}\} \tilde\Delta_m^{-1}.
\end{eqnarray*}
To obtain (\ref{rezp}) it remains to notice that $\Delta_m^{OLS} =
I_m \otimes \tilde\Delta_m$ and that $\delta_{i}^{OLS}$, $1\leq i
\leq d^2 m$ are the eigenvalues of $\Delta_m^{OLS}$.

In (ii) we suppose $\Sigma(\cdot) = \sigma^2(\cdot) I_d$ and in this
case it suffices to notice that
$$
\left(\! \int_0^1 \!\!\! \Sigma dr \!\otimes\! I_d \! \right)
\Sigma_{G^{\otimes 2}}^{-1} \left(\!\int_0^1 \!\!\! \Sigma dr
\!\otimes\! I_d \! \right) = \frac{\left( \int_0^1 \sigma^2
(r)dr\right)^2 }{\int_0^1 \sigma^4 (r) dr} I_d \otimes I_d \ll  I_d
\otimes I_d,
$$
where for the order relationship we use Cauchy-Schwarz inequality,
while
$$
\left(\int_0^1G(r)'\otimes G(r)^{-1}dr\right)^2 = I_d \otimes I_d.
$$
$\quad\square$

\medskip

\subsection{A Taylor expansion of the matrix function $f(A) = A^{-1/2}$}
\label{treyu}
 Recall that the differential of a function $F$ that
maps a $r\times r$ matrix $X$ into a $r\times r$ matrix $F(X)$ is
defined by the equation
\begin{equation*}
vec(dF)=d\mathbf{f}
\end{equation*}%
where $\mathbf{f}$ is a $r^{2}\times 1$ vector function such that $\mathbf{f}%
(vec(X))=vec(F(X))$. In other words, the (first-order) differential of $F$
at $X$ in the $r\times r$ matrix obtained by unstacking the differential of $%
d\mathbf{f}$ at $vec(X)$. See also Schott (2005), section 9.3. Basic
properties of vector differentials implies
\begin{equation*}
0=d(X^{-1/2}XX^{-1/2})=d(X^{-1/2})X^{1/2}+X^{-1/2}d(X)X^{-1/2}+X^{1/2}d%
\left( X^{-1/2}\right) .
\end{equation*}%
Now, recall that for any $A$ positive definite matrix, the Lyapunov equation
$AY+YA=B$ has a unique solution that can be represented as
\begin{equation*}
Y=\int_{0}^{\infty }\exp (-vA)B\exp (-vA)dv.
\end{equation*}%
See Horn and Johnson (1991), section 6.5. All these facts brings us to the
following technical result.

\begin{lem}
\label{technical3} Let $A$ and $\widehat A$ be two positive definite $%
r\times r-$matrices such that $0< c_1\leq \lambda_{min} (B) <\infty$ for
some constant $c_1$ and $B = A$ and $B = \widehat A$, where $\lambda_{min}
(B)$ is the smallest eigenvalue of the symmetric matrix $B$. Moreover,
suppose that $\| \widehat A - A \| \leq c_2 $ for some small constant $c_2$.
Then
\begin{equation*}
\widehat A^{-1/2} - A^{-1/2} = - \int_0^\infty \exp(-vA) A^{-1/2} \{
\widehat A - A\} A^{-1/2} \exp(-vA) dv + R_n
\end{equation*}
where $R_n$ is a $r\times r-$symmetric matrix with $\|R_n\| \leq C
\|\widehat A - A\|^2 $ and $C$ is a constant depending only on $c_1$, $c_2$.
\end{lem}

\noindent{\bf Proof of Lemma \ref{technical3}}
Let $\Delta $ be some arbitrary matrix. By Taylor expansion, for
sufficiently small values of $\varepsilon $ and for some matrices $G_{i}$, $%
i=1,2,...$
\begin{equation}
(A+\varepsilon \Delta )^{-1/2}=A^{-1/2}+\varepsilon G_{1}+\varepsilon
^{2}G_{2}+...=A^{-1/2}+\varepsilon G_{1}+R_{1}  \label{eq_ly1}
\end{equation}%
with $\Vert R_{1}\Vert \leq C_{1}\varepsilon ^{2}$ and $C_{1}$ a constant
depending on $c_{1}$ and the norm of $\Delta $. This kind of representation
could be derived from a Taylor formula for the vector function $\mathbf{f}$
defined by the equation $\mathbf{f}(vec(X))=vec(X^{-1/2})$ considered in a
neighborhood of the $vec(A)$ for a positive definite matrix $A$%
. See, for instance, Schott (2005), section 9.6. On the other hand, recall
that for $B$ a square matrix with $\Vert B\Vert <1$, $%
(I-B)^{-1}=I+B+B^{2}+...=I+B+R_{2}$ where $R_{2}$ is the reminder of the
expansion with $\Vert R_{2}\Vert \leq \Vert B\Vert ^{2}(1-\Vert B\Vert
)^{-1} $. Thus for sufficiently small values of we can write
\begin{eqnarray}
(A+\varepsilon \Delta )^{-1} &=&A^{-1/2}(I+\varepsilon A^{-1/2}\Delta
A^{-1/2})^{-1}A^{-1/2}  \label{eq_ly2} \\
&=&A^{-1/2}(I-\varepsilon A^{-1/2}\Delta A^{-1/2}+\varepsilon
^{2}A^{-1/2}\Delta A^{-1}\Delta A^{-1/2}+...)A^{-1/2}  \notag
\end{eqnarray}%
%
%
%with $\|R_2\|\leq C_2 \varepsilon^2$ and $C_2$ a constant depending
%on $c_1$ and the norm of  $\Delta$.
Taking the square on both sides of the first equality in (\ref{eq_ly1}) and
identifying the coefficients of the power of $\varepsilon $ in equation (\ref%
{eq_ly2}) deduce that $G_{1}$ is the solution of the Lyapunov equation
\begin{equation*}
A^{-1/2}Y+YA^{-1/2}=-A^{-1}\Delta A^{-1}.
\end{equation*}%
Finally, the result follows by taking $\Delta =(\widehat{A}-A)/\Vert
\widehat{A}-A\Vert $ and $\varepsilon =\Vert \widehat{A}-A\Vert $. $\quad\square$

\qquad

\section*{References}
\begin{description}
\item[]{\sc Ahn, S.K.} (1988) Distribution for residual autocovariances in multivariate autoregressive models with structured parameterization. \textit{Biometrika} 75, 590-593.
\item[]{\sc Anderson, T.W.} (1951) Estimating linear restrictions on
regression coefficients for multivariate normal distributions.
\textit{Annals of Mathematical Statistics} 22, 327-351.
\item[]{\sc Bae, J., and Choi, M.J.} (1999) The uniform CLT for martingale difference of function-indexed process
under uniform integrable entropy.
\textit{Communications Korean Mathematical Society} 14, 581-595.
\item[] {\sc Bae, J., Jun, D., and Levental, S. (2010)} The uniform CLT for martingale differences arrays under the uniformly integrable entropy.
\textit{Bulletin of the Korean Mathematical  Society} 47, 39-51.
\item[]{\sc Baek, E., and Brock, W.} (1992) A general test for nonlinear Granger causality: bivariate model. Working paper, University of Wisconsin-Madison.

\item[] {\sc Bahadur, R.R. (1960)} Stochastic comparison of tests. \textit{Annals of Mathematical Statistics} 31, 276-295.

\item[]{\sc Beare B.K.} (2008) Unit root testing with unstable volatility. Working paper, Nuffield College, University of Oxford.
\item[]{\sc Boswijk, H.P.} (2010) Nuisance parameter free inference on cointegration
parameters in the presence of a variance shift. \textit{Economics Letters} 107, 190-193.
\item[]{\sc Boswijk, H.P., and Franses, H.F.} (1992) Dynamic specification and coitengration.
\textit{Oxford Bulletin of Economics and Statistics} 54, 369-381.
\item[]{\sc Boubacar Mainassara, Y.} (2010) Multivariate portmanteau test for structural VARMA models
with uncorrelated but non-independent error terms. Working paper. EQUIPPE Université Lille 3.
\item[]{\sc Box, G.E.P.,  and Pierce, D.A.}  (1970) Distribution of residual
autocorrelations in autoregressive-integrated moving average time
series models. \textit{Journal of the American Statistical Association}
65, 1509-1526.
\item[]{\sc Brockwell, P.J., and Davis, R.A.} (1991) \textit{Time Series: Theory and methods}. Springer, New York.
\item[]{\sc Br\"{u}ggemann R., L\"{u}tkepohl H., and Saikkonen P.} (2006) Residual autocorrelation testing for vector error correction models. \textit{Journal of Econometrics} 134, 579-604.
\item[]{\sc Cavaliere, G.} (2004) Unit root tests under time-varying
variance shifts. \textit{Econome\-tric Reviews} 23, 259-292.
\item[]{\sc Cavaliere, G., Rahbek, A. and Taylor, A.M.R.} (2010)
Testing for co-integration in vector autoregressions with
non-stationary volatility. \textit{Journal of Econometrics} 158, 7-24.
\item[]{\sc Cavaliere, G., and Taylor, A.M.R.} (2007) Testing for unit roots in time
series models with non-stationary volatility. \textit{Journal of Econometrics} 140, 919-947.
\item[]{\sc Cavaliere, G., and Taylor, A.M.R.} (2008) Time-transformed unit root tests for models with non-stationary volatility. \textit{Journal of Time Series Analysis} 29, 300-330.
\item[]{\sc
Chitturi, R. V.} (1974) Distribution of residual autocorrelations
in multiple autoregressive schemes. \textit{Journal of the American
Statistical Association} 69, 928-934.
\item[]{\sc Duchesne, P.} (2005) Testing for serial correlation of unknown form in cointegrated time series models. \textit{Annals of the Institute of Statistical Mathematics} 57, 575-595.
\item[]{\sc Edgerton, D., and Shukur, G.} (1999) Testing autocorrelation in a system perspective. \textit{Econometric Reviews} 18, 343-386.
\item[]{\sc Engle, R.F.} (1982) Autoregressive conditional heteroscedasticity with estimates of the variance of UK inflation. \textit{Econometrica} 50, 987-1008.
\item[]{\sc Francq, C., and Raïssi, H.} (2007) Multivariate portmanteau test for autoregressive models with uncorrelated but nonindependent
errors. \textit{Journal of Time Series Analysis} 28, 454-470.
\item[]{\sc Francq, C., Roy, R., and Zako\"{\i}an, J.-M.} (2005) Diagnostic checking in ARMA models with uncorrelated errors.
\textit{Journal of the American Statistical Association} 13, 532-544.

\item[]{\sc Horn, R.A., and Johnson, C.R.} (1994) \emph{Topics in Matrix Analysis}. Cambridge University Press, Cambridge.

\item[]{\sc Hosking, J. R. M.} (1980) The
multivariate portmanteau statistic. \textit{Journal of the American
Statistical Association} 75, 343-386.
\item[]{\sc Imhof, J. P.} (1961) Computing the distribution of quadratic forms in normal variables.
\textit{Biometrika} 48, 419-426.
\item[]{\sc Johansen, S.} (1995) {\em Likelihood-Based Inference in Cointegrated Vector Autoregressive
Models}. Oxford University Press, New York.
\item[]{\sc Jones,
J.D.} (1989) A comparison of lag-length selection techniques in
tests of Granger causality between money growth and inflation:
evidence for the US, 1959-86. \textit{Applied Economics} 21,
809-822.
\item[]{\sc Katayama, N.} (2008) An improvement of the portmanteau statistic. \textit{Journal of Time Series Analysis} 29, 359-370.
\item[]{\sc Kim, T.-H., Leybourne, S., and Newbold, P.} (2002) Unit root tests with a break in innovation variance. \textit{Journal of Econometrics} 103, 365-387.
\item[]{\sc Kuonen, D.} (1999)
Saddlepoint approximations for distributions of quadratic forms in
normal variables. \textit{Biometrika} 86, 929-935.
\item[]{\sc Ljung, G.M. and Box, G.E.P. } (1978)
On measure of lack of fit in time series models. \textit{Biometrika}
65, 297-303.
\item[]{\sc Lobato, I., Nankervis, J.C., and Savin, N.E.} (2002) Testing for zero autocorrelation in the presence of statistical dependence. \textit{Econometric Theory} 18, 730-743.
\item[]{\sc Lütkepohl, H.} (2005) {\em New Introduction to Multiple Time Series Analysis}. Springer, Berlin.
\item[]{\sc Patilea, V., and Raïssi, H.} (2010) Adaptive estimation of vector autoregressive
models with time-varying variance: application to testing linear causality in mean. Working
document IRMAR-INSA;   \href{http://arxiv.org/abs/1007.1193}{arXiv:1007.1193v2}
\item[]{\sc Phillips, P.C.B., and Xu, K.L.} (2005)
Inference in autoregression under heteroskedasticity.
\textit{Journal of Time Series Analysis} 27, 289-308.
\item[]{\sc Ramey, V.A., and Vine, D.J.} (2006) Declining volatility in the U.S. automobile industry. \textit{The American Economic Review} 96, 1876-1889.
\item[]{\sc Raïssi, H.} (2010) Autocorrelation based tests for vector error correction models with uncorrelated but nonindependent errors. \textit{Test} 19, 304-324.
%\item[]{\sc Reinsel, G.C.} (1993) \textit{Elements of Multivariate Time Series Analysis}. Springer, New-York.
\item[]{\sc Sensier, M., and van Dijk, D.} (2004) Testing for volatility changes in U.S. macroeconomic time series. \textit{Review of Economics and Statistics} 86, 833-839.
\item[]{\sc Schott, J.R.} (2005) \textit{Matrix analysis for statistics} (2nd ed.).Wiley series in probability and statistics, Wiley,  Hoboken, N.J.
\item[]{\sc
Stock, J.H., and Watson, M.W.} (1989) Interpreting the evidence on
money-income causality. \emph{Journal of Econometrics} 40, 161-181.

\item[]{\sc van der Vaart, A.W.} (1998) \!\textit{Asymptotic Statistics}. Cambridge University Press, Cambridge.

\item[]{\sc Thornton, D.L. and Batten, D.S.} (1985) Lag-length selection and tests of Granger causality between money and income. \emph{Journal of Money, Credit, and Banking} 17, 164-178.

\item[]{\sc
Vilasuso, J.} (2001) Causality tests and conditional
heteroskedasticity: Monte Carlo evidence. \emph{Journal of
Econometrics} 101, 25-35.

\item[]{\sc Watson, M.W.} (1999) Explaining the increased variability in long-term interest rates. \textit{Federal Reserve Bank of Richmond Economic Quarterly} 85/4, 71-96.

\item[]{\sc
White, H.} (1980) A heteroskedasticity consistent covariance matrix
estimator and a direct test for heteroskedasticity.
\emph{Econometrica} 48, 817-838.

\item[]{\sc Xu, K.L., and Phillips, P.C.B.} (2008)
Adaptive estimation of autoregressive models with time-varying
variances. \textit{Journal of Econometrics} 142, 265-280.
\end{description}

\newpage
\section{Appendix B: Tables and Figures}

\begin{table}[hh]\!\!\!\!\!\!\!\!\!\!
\begin{center}
\caption{\small{Empirical size (in \%) of the portmanteau tests with iid standard Gaussian errors.}}
\begin{tabular}{|c|c|c|c||c|c|c|}
\hline
Case & \multicolumn{3}{|c||}{$m=5$}&\multicolumn{3}{|c|}{$m=15$}\\
\hline\hline
  T & 50 & 100 & 200  &50 & 100 & 200 \\
  \hline\hline
  %$BP_{m}^{S}$ & {\bf 1.0} &  3.7 &  5.1& {\bf 0.5}&{\bf 2.0} & {\bf 2.7} \\
  $LB_{m}^{S}$ &{\bf 2.6}& 4.6 & 5.5& 4.4& 4.1 & 4.6\\
  \hline\hline
  %$BP_{m}^{OLS}$ & {\bf 2.6} &4.5 & 4.4&{\bf 1.9} &{\bf 3.2}& 4.5\\
  $LB_{m}^{OLS}$ & 4.2 & 4.9 & 5.2 &{\bf 11.5} &{\bf 8.1} & {\bf  6.9} \\
  \hline%\hline
  %$BP_{m}^{ALS}$ &{\bf 0.3} &{\bf 3.2} & 5.0 &{\bf 0.2} &{\bf 1.8} &  {\bf 2.5}\\
  $LB_{m}^{ALS}$ &{\bf 2.2} & 4.1 &  5.1& 4.1 & 3.7&4.4 \\
  \hline%
  %$BP_{m}^{GLS}$ & {\bf 0.6} & {\bf 3.3} & 5.1 &{\bf 0.3} & {\bf 1.6}&{\bf  2.4}\\
  $LB_{m}^{GLS}$ &{\bf 2.0} & 3.9 & 5.1 & 3.7 & 3.8& 4.3\\
  \hline\hline
    %$BP_{m}^{OLS}$ & {\bf 2.6} &4.5 & 4.4&{\bf 1.9} &{\bf 3.2}& 4.5\\
  $\widetilde{LB}_{m}^{OLS}$ & {\bf 14.2} & {\bf 9.0} & {\bf 6.5} &{\bf 30.9}&{\bf 15.4}&{\bf  10.5} \\
  \hline%\hline
  %$BP_{m}^{ALS}$ &{\bf 0.3} &{\bf 3.2} & 5.0 &{\bf 0.2} &{\bf 1.8} &  {\bf 2.5}\\
  $\widetilde{LB}_{m}^{ALS}$ &6.3 & 5.9 &  5.6&{\bf 15.8} & {\bf 7.4}&{\bf 6.8} \\
  \hline%\hline
  %$BP_{m}^{GLS}$ & {\bf 0.6} & {\bf 3.3} & 5.1 &{\bf 0.3} & {\bf 1.6}&{\bf  2.4}\\
  $\widetilde{LB}_{m}^{GLS}$ &4.6 & 4.7 & 4.8 &{\bf 8.6} & {\bf 8.1}& {\bf 8.1}\\
  \hline
\end{tabular}
\label{tab1}
\end{center}
\end{table}

\begin{table}[hh]\!\!\!\!\!\!\!\!\!\!
\begin{center}
\caption{\small{Empirical size (in \%) of the portmanteau tests. The innovations are
heteroscedastic with an abrupt break at $T/2$.}}
\begin{tabular}{|c|c|c|c||c|c|c|}
\hline
Case & \multicolumn{3}{|c||}{$m=5$}&\multicolumn{3}{|c|}{$m=15$}\\
\hline\hline
  T & 50 & 100 & 200  &50 & 100 & 200 \\
  \hline\hline
  %$BP_{m}^{S}$ & 5.8 & {\bf 8.1} &  {\bf 12.6}& {\bf 1.9}&{\bf 8.1} &{\bf  17.8} \\
  $LB_{m}^{S}$ & {\bf 27.9} & {\bf 35.3} & {\bf 40.1}& {\bf 35.7}&{\bf 63.0} & {\bf 76.7}\\
  \hline\hline
  %$BP_{m}^{OLS}$ & {\bf 2.9} & 3.8 & 4.2&{\bf 2.2} &{\bf 3.4}& {\bf 3.6}\\
  $LB_{m}^{OLS}$ & 4.5 & {\bf 3.3} & 4.8 &5.4 & 6.1 & 6.0 \\
  \hline%\hline
  %$BP_{m}^{ALS}$ &{\bf 1.5} & {\bf 3.0} & 4.6 &{\bf 0.3} &{\bf 2.1} & {\bf 3.1}\\
  $LB_{m}^{ALS}$ &{\bf 3.2} &  3.7 &  5.0&3.8 & 3.8&3.9 \\
  \hline%\hline
  %$BP_{m}^{GLS}$ & {\bf 0.9} & {\bf 3.1} &  4.8 &{\bf 0.4} & {\bf 1.6}&{\bf 2.2}\\
  $LB_{m}^{GLS}$ &{\bf 2.6} &  4.2 & 5.7 & 3.7 & 4.2& 4.7\\
  \hline\hline
    %$BP_{m}^{OLS}$ & {\bf 2.6} &4.5 & 4.4&{\bf 1.9} &{\bf 3.2}& 4.5\\
  $\widetilde{LB}_{m}^{OLS}$ & {\bf 28.9} & {\bf 13.8} & {\bf 9.7} &{\bf 30.9}&{\bf 21.9}&{\bf  15.2} \\
  \hline%\hline
  %$BP_{m}^{ALS}$ &{\bf 0.3} &{\bf 3.2} & 5.0 &{\bf 0.2} &{\bf 1.8} &  {\bf 2.5}\\
  $\widetilde{LB}_{m}^{ALS}$ &{\bf 18.6} & {\bf 8.7} &  {\bf 7.1}&{\bf 35.0} & {\bf 15.9}&{\bf 9.0} \\
  \hline%\hline
  %$BP_{m}^{GLS}$ & {\bf 0.6} & {\bf 3.3} & 5.1 &{\bf 0.3} & {\bf 1.6}&{\bf  2.4}\\
  $\widetilde{LB}_{m}^{GLS}$ &{\bf 6.4} & 5.3 & 6.3 &{\bf 12.5} & {\bf 10.0}& {\bf 9.7}\\
  \hline
\end{tabular}
\label{tab2}
\end{center}
\end{table}

\begin{table}[hh]\!\!\!\!\!\!\!\!\!\!
\begin{center}
\caption{\small{Empirical size (in \%) of the portmanteau tests. The innovations are
heteroscedastic with trending behaviour.}}
\begin{tabular}{|c|c|c|c||c|c|c|}
\hline
Case & \multicolumn{3}{|c||}{$m=5$}&\multicolumn{3}{|c|}{$m=15$}\\
\hline\hline
  T & 50 & 100 & 200  &50 & 100 & 200 \\
  \hline\hline
  %$BP_{m}^{S}$ & 5.8 & {\bf 8.1} &  {\bf 12.6}& {\bf 1.9}&{\bf 8.1} &{\bf  17.8} \\
  $LB_{m}^{S}$ & {\bf 12.8} & {\bf 15.1} & {\bf 19.2}& {\bf 18.4}&{\bf 27.5} & {\bf 36.8}\\
  \hline\hline
  %$BP_{m}^{OLS}$ & {\bf 2.9} & 3.8 & 4.2&{\bf 2.2} &{\bf 3.4}& {\bf 3.6}\\
  $LB_{m}^{OLS}$ & 4.9 & 4.5 & 4.8 &{\bf 9.2} & {\bf 7.3} & 6.3 \\
  \hline%\hline
  %$BP_{m}^{ALS}$ &{\bf 1.5} & {\bf 3.0} & 4.6 &{\bf 0.3} &{\bf 2.1} & {\bf 3.1}\\
  $LB_{m}^{ALS}$ &4.4 &  5.0 &  5.0&{\bf 8.0} & 6.0&6.0 \\
  \hline%\hline
  %$BP_{m}^{GLS}$ & {\bf 0.9} & {\bf 3.1} &  4.8 &{\bf 0.4} & {\bf 1.6}&{\bf 2.2}\\
  $LB_{m}^{GLS}$ &{\bf 2.3} &  3.7 & 5.2 & {\bf 2.8} & 4.0& 4.0\\
  \hline\hline
    %$BP_{m}^{OLS}$ & {\bf 2.6} &4.5 & 4.4&{\bf 1.9} &{\bf 3.2}& 4.5\\
  $\widetilde{LB}_{m}^{OLS}$ & {\bf 32.4} & {\bf 22.6} & {\bf 15.3} &{\bf 38.3}&{\bf 25.1}&{\bf  16.5} \\
  \hline%\hline
  %$BP_{m}^{ALS}$ &{\bf 0.3} &{\bf 3.2} & 5.0 &{\bf 0.2} &{\bf 1.8} &  {\bf 2.5}\\
  $\widetilde{LB}_{m}^{ALS}$ &{\bf 10.0} & 4.2 &  {\bf 3.3}&{\bf 22.7} & {\bf 6.7}&5.2 \\
  \hline%\hline
  %$BP_{m}^{GLS}$ & {\bf 0.6} & {\bf 3.3} & 5.1 &{\bf 0.3} & {\bf 1.6}&{\bf  2.4}\\
  $\widetilde{LB}_{m}^{GLS}$ &{\bf 5.7} & 5.3 & 6.3 &{\bf 10.8} & {\bf 9.7}& {\bf 9.2}\\
  \hline
\end{tabular}
\label{tab3}
\end{center}
\end{table}

\begin{table}[hh]\!\!\!\!\!\!\!\!\!\!
\begin{center}
\caption{The empirical means and standard deviations of the weights in the sums (\ref{truedist}), (\ref{truedistadapt}) and their GLS counterparts over the $N=1000$ iterations. The innovations are
heteroscedastic with trending behaviour.}
\begin{tabular}{|c|c|c|c|c|c|}
\hline
  $i$ & 1 & 2 & 3  &4 & 5  \\
  \hline\hline
  $\hat{\delta}_i^{ols}$&$0.02_{[0.02]}$& $0.06_{[0.04]}$& $0.11_{[0.06]}$& $0.2_{[0.09]}$& $0.89_{[0.11]}$\\
  \hline%\hline
  $\hat{\delta}_i^{als}$&$0.03_{[0.02]}$& $0.05_{[0.02]}$& $0.09_{[0.03]}$& $0.11_{[0.03]}$& $1.00_{[0.00]}$\\
  \hline
  $\hat{\delta}_i^{gls}$ &$0.05_{[0.04]}$& $0.06_{[0.04]}$& $0.15_{[0.08]}$& $0.17_{[0.08]}$& $1.00_{[0.00]}$ \\

  \hline\hline
   $i$ & 6 & 7 & 8 &9 & 10 \\
   \hline\hline

     $\hat{\delta}_i^{ols}$& $0.89_{[0.11]}$& $0.89_{[0.11]}$& $0.91_{[0.11]}$& $1.11_{[0.13]}$& $1.11_{[0.13]}$\\
  \hline%\hline
  $\hat{\delta}_i^{als}$& $1.00_{[0.00]}$& $1.00_{[0.00]}$& $1.00_{[0.00]}$& $1.00_{[0.00]}$& $1.00_{[0.00]}$\\
  \hline
  $\hat{\delta}_i^{gls}$ &  $1.00_{[0.00]}$& $1.00_{[0.00]}$& $1.00_{[0.00]}$& $1.00_{[0.00]}$& $1.00_{[0.00]}$ \\

  \hline\hline
    $i$ & 11 & 12 & 13  &14 & 15 \\
  \hline\hline

    $\hat{\delta}_i^{ols}$&$1.11_{[0.13]}$& $1.12_{[0.13]}$& $1.36_{[0.17]}$& $1.36_{[0.17]}$& $1.36_{[0.17]}$\\
  \hline%\hline
  $\hat{\delta}_i^{als}$&$1.00_{[0.00]}$& $1.00_{[0.00]}$& $1.00_{[0.00]}$& $1.00_{[0.00]}$& $1.00_{[0.00]}$\\
  \hline
  $\hat{\delta}_i^{gls}$ &$1.00_{[0.00]}$& $1.00_{[0.00]}$& $1.00_{[0.00]}$& $1.00_{[0.00]}$& $1.00_{[0.00]}$\\

  \hline\hline
    $i$ & 16 & 17 & 18 &19 & 20 \\
  \hline\hline

      $\hat{\delta}_i^{ols}$& $1.36_{[0.17]}$& $1.69_{[0.27]}$& $1.71_{[0.27]}$& $1.71_{[0.27]}$& $1.71_{[0.27]}$\\
  \hline%\hline
  $\hat{\delta}_i^{als}$&$1.00_{[0.00]}$& $1.00_{[0.00]}$& $1.00_{[0.00]}$& $1.00_{[0.00]}$& $1.00_{[0.00]}$\\
  \hline
  $\hat{\delta}_i^{gls}$ &$1.00_{[0.00]}$& $1.00_{[0.00]}$& $1.00_{[0.00]}$& $1.00_{[0.00]}$& $1.00_{[0.00]}$\\
  \hline
\end{tabular}
\label{weights}
\end{center}
\end{table}

\begin{table}[hh]\!\!\!\!\!\!\!\!\!\!
\begin{center}
\caption{\small{The estimators of the autoregressive parameters of the VAR(1) model for the balance data for the U.S..}}
\begin{tabular}{|c|c|c|c|c|}
\hline
    Parameter & $\theta_1$  & $\theta_2$ & $\theta_3$ & $\theta_4$ \\
\hline\hline
   ALS estimate & $0.33_{[0.08]}$ & $0.02_{[0.02]}$&$-0.35_{[0.30]}$ & $-0.07_{[0.08]}$\\
   OLS estimate &$0.45_{[0.23]}$ & $0.00_{[0.02]}$& $-1.02_{[0.60]}$& $0.1_{[0.17]}$\\
  \hline
\end{tabular}
\label{estimates}
\end{center}
\end{table}

\begin{table}[hh]\!\!\!\!\!\!\!\!\!\!
\begin{center}
\caption{\small{The balance data for the U.S.: the $p$-values of the ARCH-LM tests (in \%) for the components of the ALS-residuals of a VAR(1).}}
\begin{tabular}{|c|c|c|c|}
\hline
    lags & 2  & 5 & 10\\
\hline\hline
   $\check{\epsilon}_{1t}$ & 22.26 & 45.05& 36.44\\
   \hline
   $\check{\epsilon}_{2t}$&25.32 & 73.32& 77.18\\
  \hline
\end{tabular}
\label{archLM}
\end{center}
\end{table}

\begin{table}[hh]\!\!\!\!\!\!\!\!\!\!
\begin{center}
\caption{\small{The $p$-values of the portmanteau tests (in \%) for the checking of the adequacy of the VAR(1) model for the U.S. trade balance data.}}
\begin{tabular}{|c|c|c|}
\hline
    $m$ & 5  & 15 \\
\hline\hline
   %$BP_{m}^{S}$ & 0.00 & 0.02\\
   $LB_{m}^{S}$&0.00 & 0.01\\
  \hline\hline
   %$BP_{m}^{OLS}$ & 52.96 & 99.96\\
   $LB_{m}^{OLS}$& 50.80 & 99.94\\
  \hline%\hline
   %$BP_{m}^{ALS}$ & 7.33 & 23.44\\
   $LB_{m}^{ALS}$ & 6.36 & 15.95\\
  \hline\hline
   %$BP_{m}^{OLS}$ & 52.96 & 99.96\\
   $\widetilde{LB}_{m}^{OLS}$& 0.00 & 7.87\\
  \hline%\hline
   %$BP_{m}^{ALS}$ & 7.33 & 23.44\\
   $\widetilde{LB}_{m}^{ALS}$ & 5.61 & 15.50\\
  \hline
\end{tabular}
\label{pvalues}
\end{center}
\end{table}

\begin{table}[hh]\!\!\!\!\!\!\!\!\!\!
\begin{center}
\caption{\small{The balance data for the U.S.: the test statistics of the portmanteau tests used for checking the adequacy of the VAR(1) model. The $\underline{\tilde{Q}}_{m}^{OLS}$ and $\underline{\tilde{Q}}_{m}^{ALS}$ correspond to the statistics of the LB version of the Katayama portmanteau tests with standard asymptotic distribution.}}
\begin{tabular}{|c|c|c|}
\hline
    $m$ & 5  & 15 \\
\hline\hline
   %$Q_{m}^{OLS}$ & 6.67 & 102.22\\
   $\tilde{Q}_{m}^{OLS}$&6.84 & 106.34\\
  \hline\hline
   %$Q_{m}^{ALS}$ & 25.17 & 63.65\\
   $\tilde{Q}_{m}^{ALS}$&25.73 & 66.83\\
  \hline
     %$Q_{m}^{OLS}$ & 6.67 & 102.22\\
   $\underline{\tilde{Q}}_{m}^{OLS}$&6.84 & 106.34\\
  \hline\hline
   %$Q_{m}^{ALS}$ & 25.17 & 63.65\\
   $\underline{\tilde{Q}}_{m}^{ALS}$&48.73 & 66.83\\
  \hline
\end{tabular}
\label{stats}
\end{center}
\end{table}

\begin{table}[hh]\!\!\!\!\!\!\!\!\!\!
\begin{center}
\caption{\small{The estimators of the autoregressive parameters of the VAR(4) model for the U.S. energy-transportation price indexes.}}
\begin{tabular}{|c|c|c|c|c|}
\hline
    Parameter & $\theta_1$  & $\theta_2$ & $\theta_3$ & $\theta_4$ \\
\hline\hline
   ALS estimate & $0.36_{[0.08]}$ & $0.37_{[0.14]}$&$0.10_{[0.04]}$ & $0.43_{[0.08]}$\\
   OLS estimate &$0.74_{[0.32]}$ & $1.08_{[0.67]}$& $-0.08_{[0.13]}$& $0.10_{[0.28]}$\\
  \hline\hline
    Parameter & $\theta_5$  & $\theta_6$ & $\theta_7$ & $\theta_8$ \\
\hline\hline
   ALS estimate & $0.06_{[0.08]}$ & $-0.02_{[0.14]}$&$-0.09_{[0.04]}$ & $-0.13_{[0.08]}$\\
   OLS estimate &$-0.53_{[0.35]}$ & $-1.26_{[0.73]}$& $0.10_{[0.14]}$& $0.27_{[0.30]}$\\
  \hline\hline
    Parameter &  $\theta_{9}$  & $\theta_{10}$ & $\theta_{11}$ & $\theta_{12}$ \\
\hline\hline
   ALS estimate & $0.18_{[0.08]}$ & $0.13_{[0.14]}$&$-0.05_{[0.04]}$ & $0.01_{[0.08]}$\\
   OLS estimate &$0.21_{[0.24]}$ & $0.13_{[0.52]}$& $-0.05_{[0.14]}$& $0.03_{[0.31]}$\\
  \hline\hline
    Parameter & $\theta_{13}$  & $\theta_{14}$ & $\theta_{15}$ & $\theta_{16}$ \\
\hline\hline
   ALS estimate & $0.17_{[0.08]}$ & $0.15_{[0.14]}$&$-0.07_{[0.04]}$ & $0.03_{[0.08]}$\\
   OLS estimate &$0.32_{[0.26]}$ & $0.64_{[0.57]}$& $-0.17_{[0.15]}$& $0.31_{[0.32]}$\\
  \hline
\end{tabular}
\label{estimatesk}
\end{center}
\end{table}

%\begin{table}[hh]\!\!\!\!\!\!\!\!\!\!
%\begin{center}
%\caption{\small{The balance data for the U.S.: the p-values of the ARCH-LM tests (in \%) for the components of the ALS-residuals of a VAR(1).}}
%\begin{tabular}{|c|c|c|c|}
%\hline
%    lags & 2  & 5 & 10\\
%\hline\hline
%   $\check{\epsilon}_{1t}$ & 22.26 & 45.05& 36.44\\
%   \hline
%   $\check{\epsilon}_{2t}$&25.32 & 73.32& 77.18\\
%  \hline
%\end{tabular}
%\label{archLM}
%\end{center}
%\end{table}

\begin{table}[hh]\!\!\!\!\!\!\!\!\!\!
\begin{center}
\caption{\small{The $p$-values of the portmanteau tests (in \%) for the checking of the adequacy of the VAR(4) model for the U.S. energy-transportation price indexes (n.a.: not available).}}
\begin{tabular}{|c|c|c|c|}
\hline
    $m$ & 3  & 6 & 12 \\
\hline\hline
   %$BP_{m}^{S}$ & 0.00 & 0.02\\
   $LB_{m}^{S}$& n.a. & 2.08 & 0.00\\
  \hline%\hline
   %$BP_{m}^{OLS}$ & 52.96 & 99.96\\
   $LB_{m}^{OLS}$& 100.00 & 100.00 & 100.00\\
  \hline%\hline
   %$BP_{m}^{ALS}$ & 7.33 & 23.44\\
   $LB_{m}^{ALS}$ & 85.44 & 98.72 & 10.07\\
  \hline
  $\widetilde{LB}_{m}^{OLS}$ & n.a.& n.a.& n.a.\\
    \hline
  $\widetilde{LB}_{m}^{ALS}$ & n.a.& n.a.& n.a.\\
  \hline
\end{tabular}\\
\label{pvaluesk}
\end{center}
\end{table}

\begin{table}[hh]\!\!\!\!\!\!\!\!\!\!
\begin{center}
\caption{\small{VAR modeling of the energy-transportation price indexes: the test statistics of the portmanteau tests used for checking the adequacy of the VAR(4) model.}}
\begin{tabular}{|c|c|c|c|}
\hline
    $m$ & 3  & 6 & 12 \\
\hline\hline
   %$Q_{m}^{OLS}$ & 6.67 & 102.22\\
   $\tilde{Q}_{m}^{OLS}$&1.87 & 18.06& 117.51\\
  \hline\hline
   %$Q_{m}^{ALS}$ & 25.17 & 63.65\\
   $\tilde{Q}_{m}^{ALS}$&5.03 & 9.50&57.69\\
  \hline
     $\tilde{\underline{Q}}_{m}^{OLS}$&n.a. & n.a.&n.a.\\
  \hline
     $\tilde{\underline{Q}}_{m}^{ALS}$&n.a. & n.a.&n.a.\\
  \hline
\end{tabular}
\label{statsk}
\end{center}
\end{table}

\begin{table}[hh]\!\!\!\!\!\!\!\!\!\!
\begin{center}
\caption{\small{VAR modeling of the energy-transportation price indexes: the weights of the non standard distributions of the portmanteau tests used for checking the adequacy of the VAR(4) model with $m=3$.}}
{\footnotesize\begin{tabular}{|c|c|c|c|c|c|c|c|c|c|c|c|c|}
\hline
    $i$ & 1  & 2 & 3 & 4& 5& 6& 7& 8& 9& 10&11&12\\
\hline\hline
   $\hat{\delta}_{i}^{ols}$& 0.05& 0.48& 1.94& 2.05& 2.66& 2.85& 4.97& 6.58& 10.51& 16.06& 64.14& 312.18\\
  \hline\hline
   $\hat{\delta}_{i}^{als}$& 0.01& 0.09& 0.24& 0.89& 1.00& 1.00& 1.00& 1.00& 1.00& 1.00& 1.00&1.00\\
  \hline
\end{tabular}}
\label{weightsk}
\end{center}
\end{table}

\clearpage

%\begin{figure}[h]\!\!\!\!\!\!\!\!\!\!
%\vspace*{0.5cm} \hspace*{0.3cm}
%\hspace*{5.2cm}
%
%\vspace*{1.5cm} \hspace*{6.3cm} \hspace*{6.0cm}
%\vspace*{1.4cm}
%
%\protect \special{psfile=power_homo.eps hoffset=-55 voffset=-12
%hscale=115 vscale=85} \caption{\label{powhomo}
%{\footnotesize Empirical power (in \%) of the portmanteau tests with $m=10$. The innovations are
%homoscedastic. The relative rejection frequencies of the $LB_m^{S}$ are in black, the $LB_m^{OLS}$ in green, the $LB_m^{ALS}$ in blue,
%the $LB_m^{GLS}$ in cyan, the $\widetilde{LB}_m^{OLS}$ in red, the $\widetilde{LB}_m^{ALS}$ in orange, the $\widetilde{LB}_m^{GLS}$ in pink.}}
%\end{figure}
%
%
%
%\begin{figure}[h]\!\!\!\!\!\!\!\!\!\!
%\vspace*{0.5cm} \hspace*{0.3cm}
%\hspace*{5.2cm}
%
%\vspace*{2.4cm} \hspace*{6.3cm} \hspace*{6.0cm}
%\vspace*{1.4cm}
%
%\protect \special{psfile=power_trend.eps hoffset=-55 voffset=-12
%hscale=115 vscale=85} \caption{\label{powtrend}
%{\footnotesize The same as in Figure \ref{powhomo} but for
%heteroscedastic innovation variance with trending behaviour.}}
%\end{figure}
%
%\begin{figure}[h]\!\!\!\!\!\!\!\!\!\!
%\vspace*{0.5cm} \hspace*{0.3cm}
%\hspace*{5.2cm}
%
%\vspace*{1.8cm} \hspace*{6.3cm} \hspace*{6.0cm}
%\vspace*{1.4cm}
%
%\protect \special{psfile=power_break.eps hoffset=-55 voffset=-12
%hscale=115 vscale=85} \caption{\label{powbreak}
%{\footnotesize The same as in Figure \ref{powhomo} but for
%heteroscedastic innovation variance which exhibits an abrupt break at $T/2$.}}
%\end{figure}

\clearpage

\clearpage

%\begin{figure}[h]\!\!\!\!\!\!\!\!\!\!
%%\vspace*{0.5cm} \hspace*{0.3cm}$\hat{R}_{OLS}^{21}(h)$
%%\hspace*{5.2cm}$\hat{R}_{OLS}^{12}(h)$
%
%\vspace*{2.1cm} \hspace*{6.3cm}$h$ \hspace*{6.0cm}$h$
%\vspace*{1.4cm}
%
%\protect \special{psfile=alsresidcarre1.eps hoffset=-65 voffset=-12
%hscale=75 vscale=73} \protect \special{psfile=alsresidcarre2.eps
%hoffset=+112 voffset=-12 hscale=75 vscale=73} \caption{\label{autocovalscar}
%{\footnotesize The energy-transportation data for the U.S.: the autocorrelations of the squares of the first component of the ALS residuals (on the left) and of the second component of the ALS residual (on the right).}}
%\end{figure}

\begin{figure}[h]\!\!\!\!\!\!\!\!\!\!
\vspace*{4.1cm} %\hspace*{8.5cm}$\tau_1$
%\hspace*{11.7cm}$\sigma_{21}$ \vspace*{-5.3 cm}

%\vspace*{4.3cm}\hspace*{2.15cm}$\tau_1$ \hspace*{2.3cm}$\sigma_{21}$ %\hspace*{3.8cm}$\tau_1$
%\vspace*{0.2 cm}

\protect \includegraphics{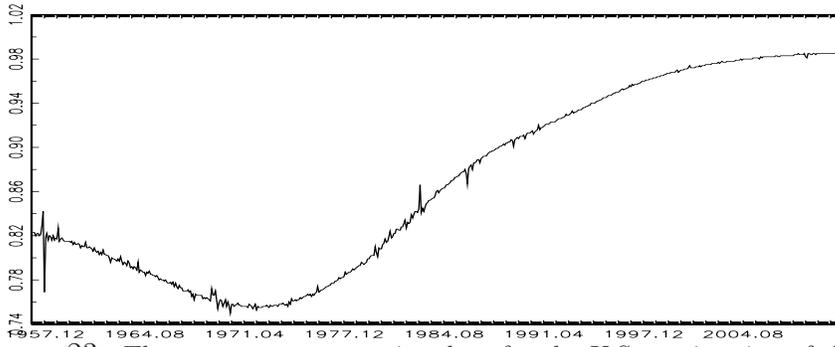}  \caption{\label{covestimk}
{\footnotesize The energy-transportation data for the U.S.: estimation of the correlation between the components of the error process.}}
\end{figure}
\clearpage
\begin{figure}[h]\!\!\!\!\!\!\!\!\!\!
\vspace*{0.5cm} \hspace*{0.3cm}$\hat{R}_{ALS}^{11}(h)$
\hspace*{5.2cm}$\hat{R}_{ALS}^{22}(h)$

\vspace*{1.5cm} \hspace*{6.3cm}$h$ \hspace*{6.0cm}$h$
\vspace*{1.4cm}

\protect \includegraphics{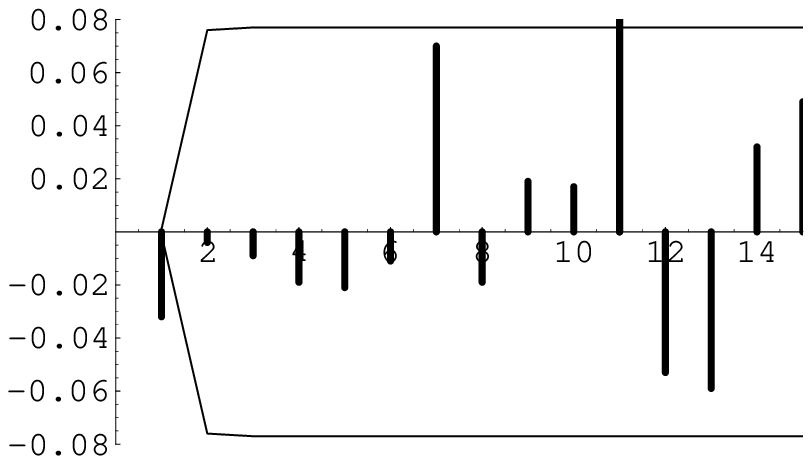} \protect \includegraphics{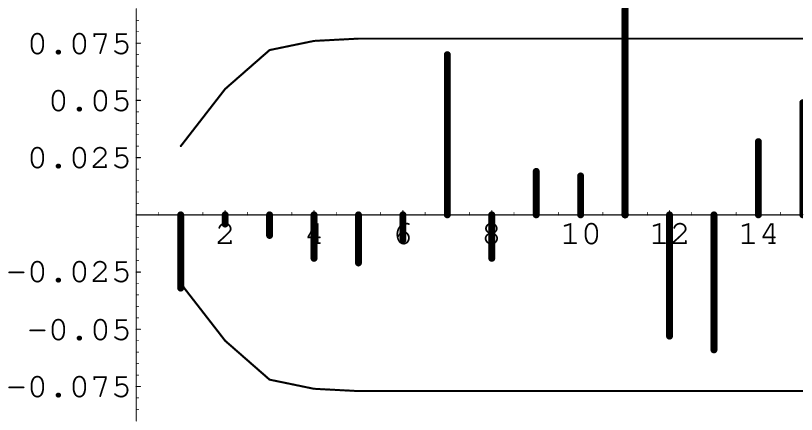} \caption{\label{autocorralsk}
{\footnotesize The energy-transportation data for the U.S.: the ALS residual autocorrelations $\hat{R}_{ALS}^{11}(h)$ (on the left) and $\hat{R}_{ALS}^{22}(h)$ (on the right), with obvious notations. The 95\% confidence bounds are obtained using (\ref{gamgls}) and (\ref{equivalent}).}}
\end{figure}

\begin{figure}[h]\!\!\!\!\!\!\!\!\!\!
\vspace*{0.5cm} \hspace*{0.3cm}$\hat{R}_{ALS}^{21}(h)$
\hspace*{5.2cm}$\hat{R}_{ALS}^{12}(h)$

\vspace*{1.5cm} \hspace*{6.3cm}$h$ \hspace*{6.0cm}$h$
\vspace*{1.4cm}

\protect \includegraphics{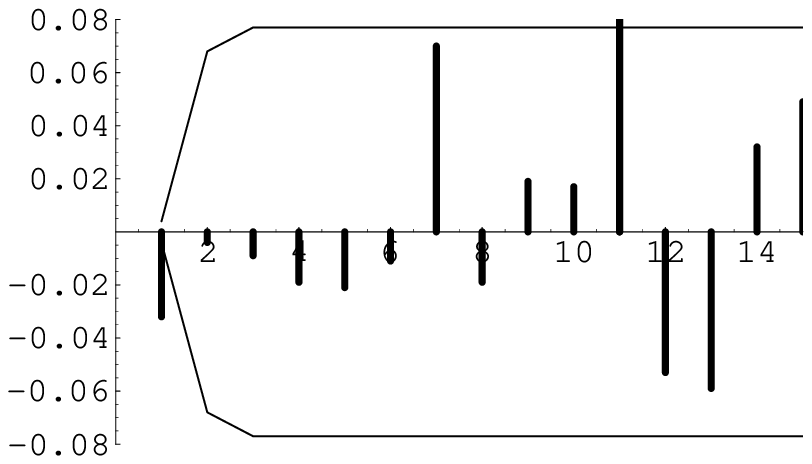} \protect \includegraphics{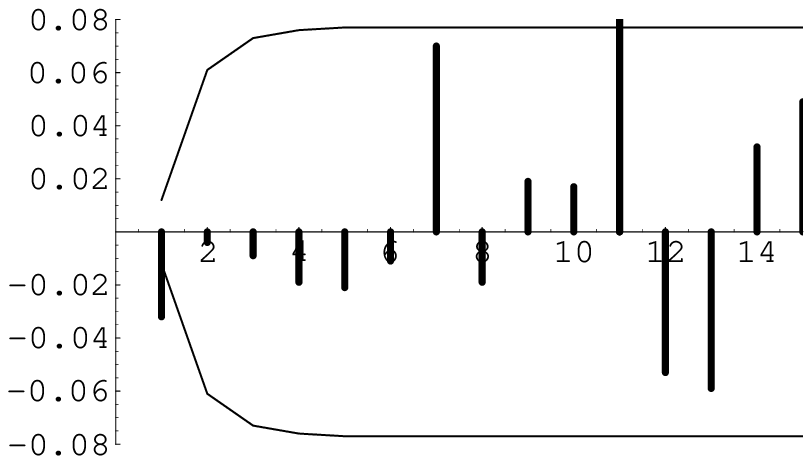} \caption{\label{autocovalsk}
{\footnotesize The same as in Figure \ref{autocorralsk} but for $\hat{R}_{ALS}^{21}(h)$ (on the left) and $\hat{R}_{ALS}^{12}(h)$ (on the right).}}
\end{figure}

\begin{figure}[h]\!\!\!\!\!\!\!\!\!\!
\vspace*{0.5cm} \hspace*{0.3cm}$\hat{R}_{OLS}^{11}(h)$
\hspace*{5.2cm}$\hat{R}_{OLS}^{22}(h)$

\vspace*{1.5cm} \hspace*{6.3cm}$h$ \hspace*{6.0cm}$h$
\vspace*{1.4cm}

\protect \includegraphics{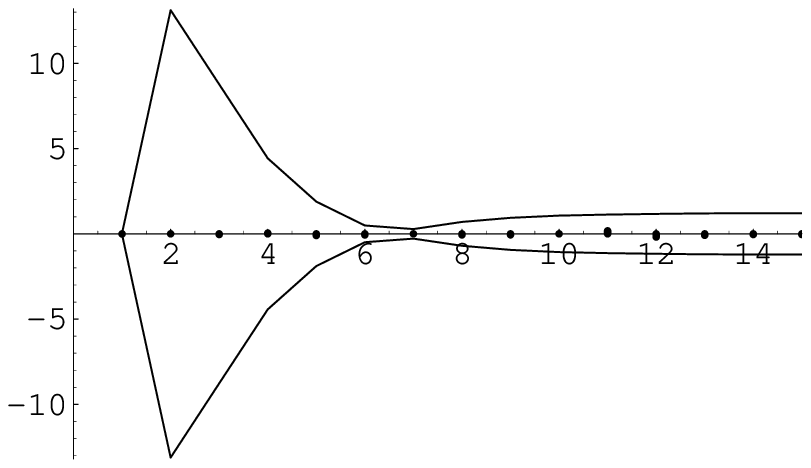} \protect \includegraphics{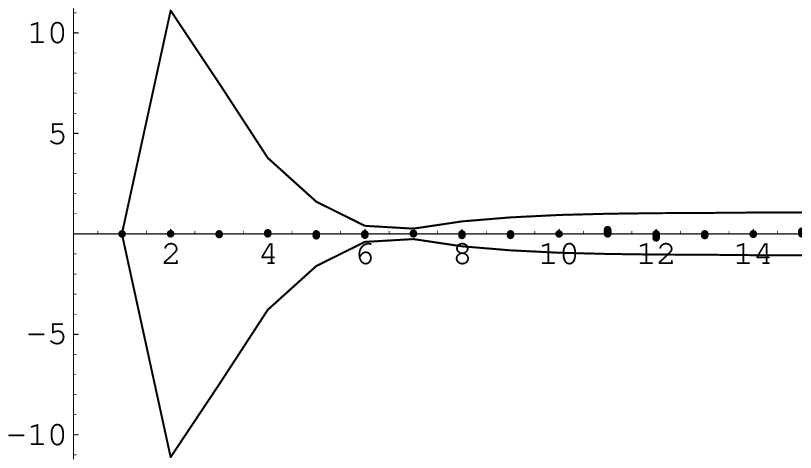} \caption{\label{autocorrolsk}
{\footnotesize The energy-transportation data for the U.S.: the OLS residual autocorrelations $\hat{R}_{OLS}^{11}(h)$ (on the left) and $\hat{R}_{OLS}^{22}(h)$ (on the right). The full line 95\% confidence bounds are obtained using (\ref{rhools}). The dotted lines 95\% confidence bounds are obtained using the standard result (\ref{bruges}).}}
\end{figure}

\begin{figure}[h]\!\!\!\!\!\!\!\!\!\!
\vspace*{0.5cm} \hspace*{0.3cm}$\hat{R}_{OLS}^{21}(h)$
\hspace*{5.2cm}$\hat{R}_{OLS}^{12}(h)$

\vspace*{1.5cm} \hspace*{6.3cm}$h$ \hspace*{6.0cm}$h$
\vspace*{1.4cm}

\protect \includegraphics{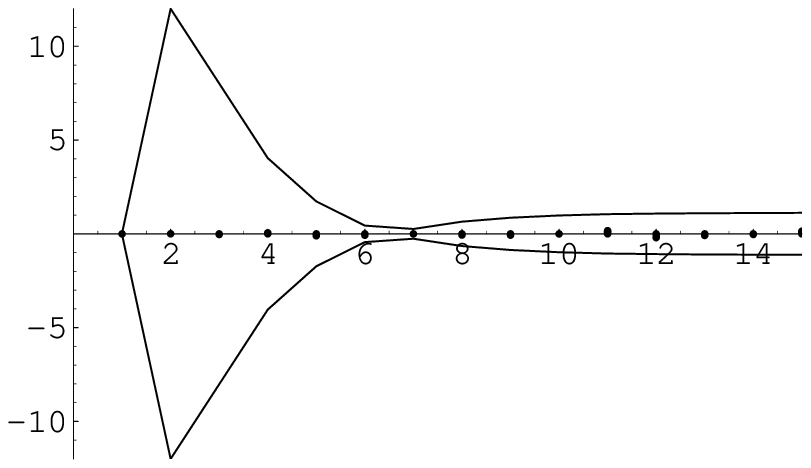} \protect \includegraphics{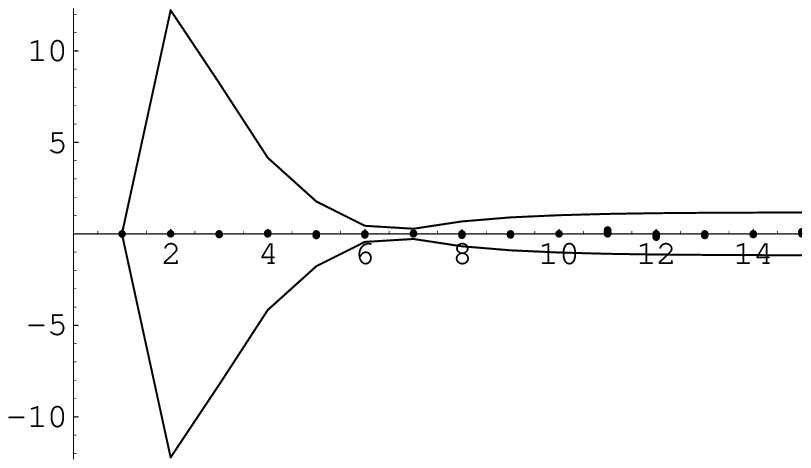} \caption{\label{autocovolsk}
{\footnotesize The same as in Figure \ref{autocorrolsk} but for $\hat{R}_{OLS}^{21}(h)$ (on the left) and $\hat{R}_{OLS}^{12}(h)$ (on the right).}}
\end{figure}

\begin{figure}[h]\!\!\!\!\!\!\!\!\!\!
\vspace*{0.5cm} \hspace*{0.3cm}$\hat{R}_{OLS}^{11}(h)$
\hspace*{5.2cm}$\hat{R}_{OLS}^{22}(h)$

\vspace*{1.5cm} \hspace*{6.3cm}$h$ \hspace*{6.0cm}$h$
\vspace*{1.4cm}

\protect \includegraphics{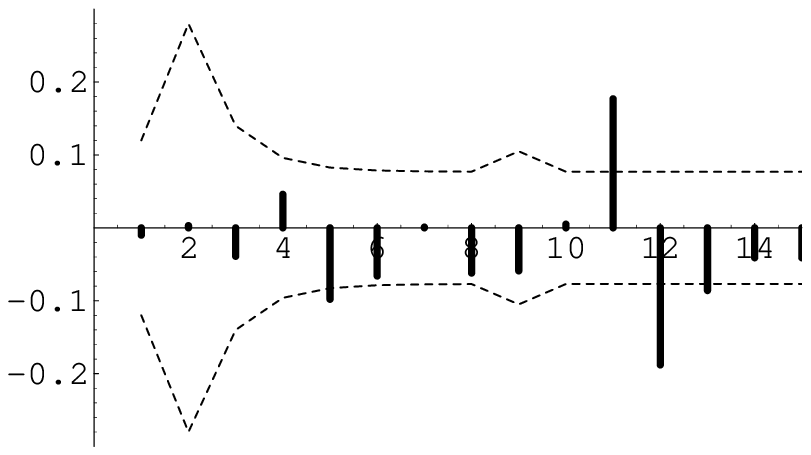} \protect \includegraphics{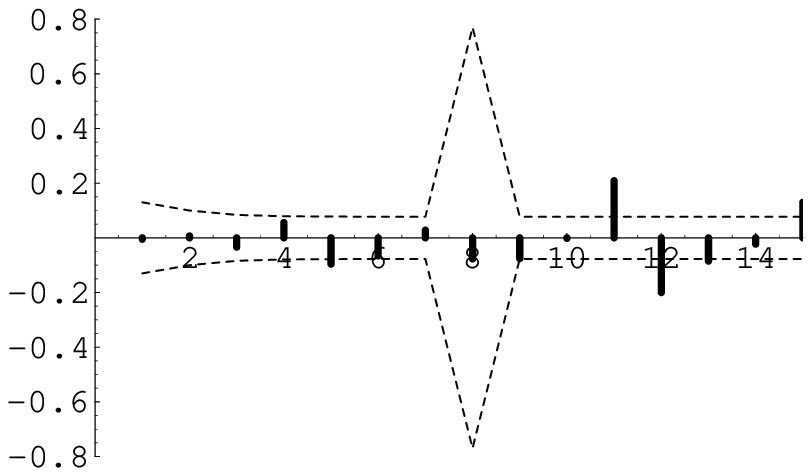} \caption{\label{autocorrstgk}
{\footnotesize The energy-transportation data for the U.S.: the OLS residual autocorrelations $\hat{R}_{OLS}^{11}(h)$ (on the left) and $\hat{R}_{OLS}^{22}(h)$ (on the right). The dotted lines 95\% confidence bounds are obtained using the standard result (\ref{bruges}).}}
\end{figure}

\begin{figure}[h]\!\!\!\!\!\!\!\!\!\!
\vspace*{0.5cm} \hspace*{0.3cm}$\hat{R}_{OLS}^{21}(h)$
\hspace*{5.2cm}$\hat{R}_{OLS}^{12}(h)$

\vspace*{1.5cm} \hspace*{6.3cm}$h$ \hspace*{6.0cm}$h$
\vspace*{1.4cm}

\protect \includegraphics{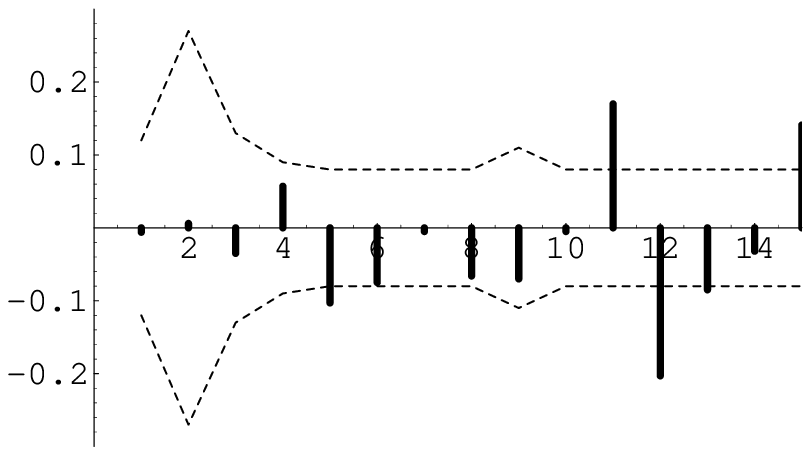} \protect \includegraphics{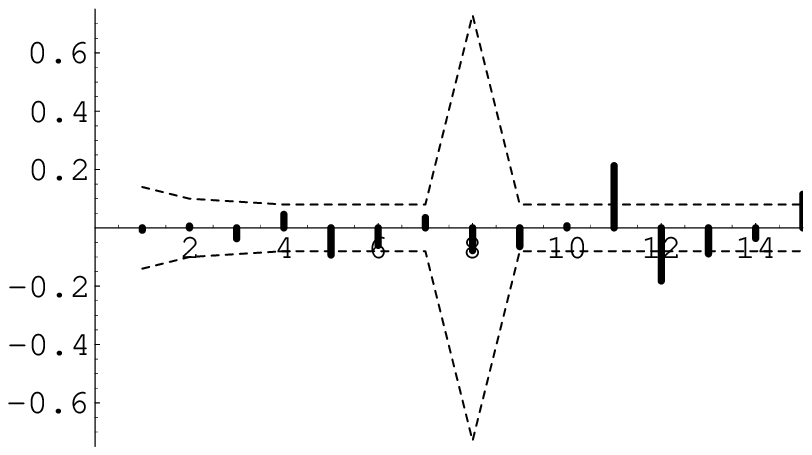} \caption{\label{autocovstgk}
{\footnotesize The same as in Figure \ref{autocorrstgk} but for $\hat{R}_{OLS}^{21}(h)$ (on the left) and $\hat{R}_{OLS}^{12}(h)$ (on the right).}}
\end{figure}

\end{document}